\documentclass[aps, prd, 10pt, twocolumn, superscriptaddress,noshowpacs, preprintnumbers, 
nofootinbib,bibnotes,hyperref,floatfix]{revtex4-1}
\usepackage[colorlinks=true,breaklinks=true]{hyperref}
\usepackage{color}	    
\hypersetup{allcolors=[rgb]{0.0 0.0 1.0},linkcolor=[rgb]{0.75 0.05 0.05}}
\usepackage[dvipsnames]{xcolor}

\usepackage{listings}
\usepackage{mathtools}
\usepackage{amsmath}
\usepackage{amsfonts}
\usepackage{amssymb}
\usepackage{bbold}
\usepackage{epsfig}
\usepackage{graphicx}
\usepackage{bm}
\usepackage{array}
\usepackage{hyperref}
\usepackage{listings}
\usepackage{color}
\usepackage{float}
\usepackage[normalem]{ulem}
\hypersetup{
    urlcolor=purple,
    }

\newenvironment{system}%
{\left\lbrace\begin{array}{@{}l@{}}}%
{\end{array}\right.}

\usepackage{tikz,xcolor,hyperref}

\definecolor{lime}{HTML}{A6CE39}
\DeclareRobustCommand{\orcidicon}{\hspace{-1mm}
	\begin{tikzpicture}
	\draw[lime, fill=lime] (0,0) 
	circle [radius=0.16] 
	node[white] {{\fontfamily{qag}\selectfont \tiny \,ID}};
	\draw[white, fill=white] (-0.0525,0.095) 
	circle [radius=0.007];
	\end{tikzpicture}
	\hspace{-3mm}
}

\foreach \x in {A, ..., Z}{\expandafter\xdef\csname orcid\x\endcsname{\noexpand\href{https://orcid.org/\csname orcidauthor\x\endcsname}
			{\noexpand\orcidicon}}
}

\begin{document}

\title{Transients stemming from collapsing massive stars:\\ The missing pieces to advance joint observations of photons and high-energy neutrinos}

\author{Ersilia Guarini\orcidA{}}
\author{Irene Tamborra\orcidB{}}%
\affiliation{Niels Bohr International Academy \& DARK, Niels Bohr Institute, University of Copenhagen, Blegdamsvej 17, 2100, Copenhagen, Denmark}%
\author{Raffaella Margutti\orcidC{}}
\affiliation{Department of Astronomy, University of California, 501 Campbell Hall, Berkeley, CA 94720, USA}%
\affiliation{Department of Physics, University of California, 366 Physics North MC 7300, Berkeley, CA 94720, USA}
\author{Enrico Ramirez-Ruiz\orcidD{}}
\affiliation{Department of Astronomy \& Astrophysics, University of California, 1156 High Street, Santa Cruz, CA 95064, USA}

\date{\today}
\begin{abstract}
Collapsing massive stars lead to a broad range of astrophysical transients, whose multi-wavelength emission is powered by a variety of processes including radioactive decay, activity of the central engine, and interaction of the outflows with a dense circumstellar medium. These transients are also candidate factories of  neutrinos with energy up to hundreds of PeV. We review  the energy released by such astrophysical objects  across the electromagnetic wavebands as well as neutrinos, in order to outline a strategy to optimize multi-messenger follow-up programs. We find that, while a significant fraction of the explosion energy can be emitted in the infrared-optical-ultraviolet (UVOIR) band, the optical signal alone is not optimal for neutrino searches. Rather, the neutrino emission is strongly correlated with the one in the radio band, if a dense circumstellar medium surrounds the transient, and with X-rays tracking the activity of the central engine.
Joint observations of transients in radio, X-rays, and neutrinos will crucially  complement those  in the UVOIR  band, breaking degeneracies in the transient parameter space.  Our findings call for heightened surveys  in the radio and X-ray bands to warrant multi-messenger detections.
\end{abstract}

\maketitle
\section{\label{sec:intro} Introduction}
 
A number of transients may be linked to the aftermath of collapsing massive stars, ranging from  supernovae (SNe) and gamma-ray bursts(GRBs)~\cite{Janka:2012wk, 1993ApJ...405..273W, MacFadyen:1998vz, Woosley:2006fn, Gehrels:2009qy, Kaneko:2006mt} to exotic transients with puzzling  properties,  e.g.~fast blue optical transients (FBOTs)~\citep{Drout:2014dma, Arcavi:2015zie, Tanaka:2016ncv, DES:2018whm, Ho:2021fyb}, superluminous supernovae (SLSNe)~\cite{Quimby:2009ps,Gal-Yam:2018out} or chameleon SNe~\citep{Milisavljevic:2015bli, Margutti:2016wyh} among the ones detected electromagnetically.  These objects are characterized by a range of  time scales and  peak luminosities~\cite{Villar:2017oya, 2018NatAs...2..324R}, albeit the mechanisms powering their emission remain  uncertain.

In the next future,  the theory behind such sources will progress  through the exponential growth of the number of  astrophysical transients detected across different wavebands with wide field, high cadence surveys, such as the Zwiky Transient Facility (ZTF)~\citep{Dekany:2020tyb}, the All-Sky Automated Survey for SuperNovae~(ASAS-SN)~\cite{2017PASP..129j4502K}, as well as the Panoramic Survey Telescope and Rapid Response System 1 (Pan-STARRS1)~\cite{2016arXiv161205560C}, and the Young Supernova Experiment (YSE)~\citep{YoungSupernovaExperiment:2020dcd}. In addition, while our understanding of the UV emission from explosive transients has already been transformed thanks to \textit{Swift}-UVOT~\citep{Roming:2005hv}, our ability to explore the hot and transient  universe will soon be revolutionized  by the upcoming Vera C. Rubin Observatory~\citep{LSST:2022kad} and   ULTRASAT~\citep{Shvartzvald:2023ofi}.

Such transients are also expected to emit neutrinos with energy between $\mathcal{O}(1)$~TeV and $\mathcal{O}(100)$~PeV, as a result of  particle acceleration~\citep{IceCube:2018dnn, Stein:2020xhk, Reusch:2021ztx, Pitik:2021dyf}, as well as gravitational waves~\citep{LIGOScientific:2017vwq, LIGOScientific:2017ync}. The operating IceCube Neutrino Observatory~\citep{IceCube:2016zyt}, the Baikal Deep Underwater Neutrino Telescope  (Baikal-GVD)~\cite{Baikal-GVD:2020irv} and the ANTARES neutrino telescope~\citep{2011NIMPA.656...11A} routinely search for high-energy neutrinos from transient sources. In particular, neutrinos have been possibly observed in coincidence with a candidate hydrogen-rich SLSN~\citep{Reusch:2021ztx, Pitik:2021dyf} as well as an FBOT~\cite{2018ATel11785....1B, Guarini:2022uyp}. Our potential to explore the transient universe through non-thermal neutrinos will be further enhanced with   upcoming neutrino telescopes such as IceCube-Gen2~\citep{IceCube-Gen2:2020qha}, the Cubic Kilometre Neutrino Telescope (KM3NeT)~\cite{Sanguineti:2023qfa}, the  Giant Radio Array for Neutrino Detection (GRAND200k)~\citep{GRAND:2018iaj}, the orbiting Probe of Extreme Multi-Messenger Astrophysics (POEMMA)~\citep{Venters:2019xwi}, and the Pacific Ocean Neutrino Experiment (P-ONE)~\cite{P-ONE:2020ljt}. 

In order to address  fundamental questions  concerning the physics linked to  high-energy particle emission, efficiency of particle acceleration,  as well as the  mechanisms powering these transients, it is key to exploit multi-messenger observations to break degeneracies in the parameter space of the transient properties otherwise hindering our understanding~\cite{Guarini:2022uyp,Fang:2020bkm,Guepin:2017dfi,Guepin:2022qpl,Pitik:2023vcg}. 

A number of programs are in place to explore transients through multiple messengers and across energy bands; for example, ASAS-SN, ZTF and Pan-STARRS1 carry out  target-of-opportunity searches for  optical counterparts of high-energy neutrino events~\cite{Stein:2022rvc,Pan-STARRS:2019szg,2022MNRAS.516.2455N}, and in turn  the IceCube Neutrino Observatory looks for neutrinos in the direction of the sources discovered by optical surveys, see e.g.~Refs.~\cite{IceCube:2023esf,IceCube:2020mzw}. 
Follow-up searches of (very) gamma-ray counterparts of the high-energy neutrinos observed at the IceCube Neutrino Observatory are also carried out with Fermi-LAT~\citep{2009ApJ...697.1071A, Garrappa:2021ihz} and the Imaging Atmospheric Cherenkov Telescopes (IACTs)~\citep{VERITAS:2021mjg}. 

To capitalize on the  promising  multi-messenger detection prospects, it is necessary and timely to define a strategy to carry out informed follow-up  searches of high-energy neutrinos and electromagnetic emission from transients.
What electromagnetic waveband is better correlated with high-energy neutrinos? What fraction of the bulk of energy released in the collapse of massive stars is deposited across the different electromagnetic wavebands and neutrinos? In this paper,  we address these questions performing   computations of the energy budget of astrophysical transients stemming from collapsing stars. In our analysis, we consider both thermal and non-thermal processes that may power the electromagnetic emission and define a criterion for correlating electromagnetic observations at different wavelengths with the neutrino signal. 

This paper is organized as follows. In Sec.~\ref{sec:model}, we outline the theoretical framework for calculating the energy budget in each electromagnetic waveband for non-relativistic outflows, while we focus on jetted relativistic outflows in Sec.~\ref{app:jet}. In Sec.~\ref{sec:neutrino}, after introducing the distribution of accelerated protons, we outline the channels for neutrino production. Section~\ref{sec:census} presents the energy budget across electromagnetic wavebands and in neutrinos of the astrophysical transients linked to collapsing massive stars. In Sec.~\ref{sec:strategy}, we investigate the most promising strategies to  correlate electromagnetic and neutrino observations depending on the transient properties and the related detection prospects. Finally,  we summarize our findings in Sec.~\ref{sec:conclusion}. In addition, the cooling rates of protons accelerated in the magnetar wind, at CSM interactions as well as in a jetted outflow are discussed in Appendix~\ref{app:A}, while Appendix~\ref{app:B} focuses on radiative shocks. 

\section{\label{sec:model} Modeling of the electromagnetic emission: non-relativistic outflows}

After introducing the one-zone model adopted to compute the bolometric luminosity, in this section we outline the contribution to the electromagnetic emission, across wavebands, from different heating sources. For illustrative purposes we present the results for a benchmark transient in this section, whereas our findings for different transient classes are discussed in Sec.~\ref{sec:census}.

\subsection{Luminosity}\label{sec: luminosity}
We rely on the one-zone model of Refs.~\citep{1980ApJ...237..541A,1980ApJ...237..541A, Chatzopoulos:2011vj, Villar:2017oya} to compute the output bolometric luminosity from collapsing stars. This model only holds for spherical outflows and, since we are interested in the bulk of the emitted radiation, we focus on the properties of the  bolometric light curve around its peak. 

Our model is based on the following assumptions: 1.~the ejecta are spherically symmetric and expand homologously; 2.~the outflow is radiation dominated, namely the radiation pressure is larger than the electron and gas pressure (note that  we do not consider radiation dominated outflows for the production of radio photons and neutrinos when the shock interacts with the CSM; see Secs.~\ref{sec:heatingSource} and~\ref{sec:neutrino}); 3.~a central  heating source is  present~\footnote{The assumption of a central heating source does not hold for all the heating processes, in particular for interactions of the ejecta with a dense CSM surrounding the progenitor. Thus, this simplified model has several limitations, see Ref.~\citep{Chatzopoulos:2011vj} for a discussion. By comparing the analytical model with numerical simulations, Ref.~\citep{Chatzopoulos:2011vj} finds that the approximation of a central heating source reproduces the peak time of the bolometric lightcurve and its normalization within a factor $\simeq 2$ with respect to numerical simulations, which is acceptable for the purpose of this paper.}; 4.~the ejecta propagate with a bulk constant velocity $v_{\rm{ej}}$, i.e.the injected energy is smaller than the kinetic energy of the ejecta.

Because of the hypothesis of homologous expansion, the radius of the ejecta evolves as $R_{\rm{ej}}(t) \simeq v_{\rm{ej}} t$. During the photospheric phase, which can last up to several weeks after the explosion depending on the ejecta mass~\citep{2017hsn..book..769S}, the ejecta are optically thick, i.e.~their optical depth is $\tau_{\rm{ej}} \gg 1$. When and where $\tau_{\rm{ej}} \simeq 2/3$, radiation begins to diffuse  from the outflow~\cite{Chatzopoulos:2011vj}.
Since no significant  kinetic energy is added to the outflow, one can assume that the photosphere  expands with velocity $v_{\rm{ph}} \simeq v_{\rm{ej}}$. 

The first law of thermodynamic can be written as (unless otherwise specified, we carry out our calculations in the reference frame of the expanding outflow):
\begin{equation}
    \frac{{\rm{d}} E}{{\rm{d}} t} + \frac{ {\rm{d}} P}{ {\rm{d}} t} = \dot{q}_{\rm{inj}} - \frac{\partial L}{\partial m} \;
    \label{eq:thermo}
\end{equation}
where $E= a T^4 V$ and $P= a T^4 V/3$ are the specific internal energy and pressure, respectively, $L$ is the output luminosity and $m$ is the mass of the fluid element, $V= \rho^{-1}$ is the specific volume with  $\rho$ being the density and $T$ the temperature. The specific energy injection rate is  $\dot{q}_{\rm{inj}}$.

For a photosphere which homologously expands, the solution of Eq.~\ref{eq:thermo} is~\citep{Chatzopoulos:2011vj}:
\begin{eqnarray}
\label{eq:outLum}
    L(t) & =  &\frac{2}{t_d} e^{-\left( \frac{t^2}{t^2_d} + \frac{2R_0 t}{v_{\rm{ej}} t_d^2} \right)}  \int_0^t  dt^\prime e^{-\left( \frac{t^{\prime 2}}{t_d^2} + \frac{2R_0 t^\prime}{v_{\rm{ej}} t_d^2} \right)}  \\  \nonumber &  & \times \;  \left( \frac{R_0}{v_{\rm{ej}} t_d} + \frac{t^\prime}{t_d} \right) L_{\rm{inj}}(t^\prime) + \rm{HS} \; ,
\end{eqnarray}
where $L_{\rm{inj}}$ is the luminosity injected by the central compact source (linked to $\dot{q}_{\rm{inj}}$ in Eq.~\ref{eq:thermo}), $R_0$ is the initial radius of the source, and $ t_d = \sqrt{2 \kappa M_{\rm{ej}}/ \beta c v_{\rm{ej}}}$ is the time needed for the radiation to diffuse through the ejecta (assumed to be longer than the duration of the energy injection in our model) of mass $M_{\rm{ej}}$ and opacity $\kappa$--- the latter is considered  time-independent and independent on the ejecta composition; the geometrical factor is $\beta \simeq 13.7$ for a variety of diffusion density profiles~\citep{1993ApJ...405..273W}, and HS is the homogeneous solution of Eq.~\ref{eq:thermo} obtained requiring $\dot{q}_{\rm{inj}} =0$.

The homogeneous solution  is only relevant when the outflow expands adiabatically, with no energy source heating the ejecta. Assuming adiabatic expansion, the emitted luminosity is ~\citep{Villar:2017oya}
\begin{equation}
    L_{\rm{ad}}(t) = \frac{E_{k, \rm{ej}}}{t_d} e^{- \left[t^2/t_d^2 + (2 R_0 t)/(v_{\rm{ej}} t_d^2) \right]} \; ,
    \label{eq:HS}
\end{equation}
where $E_{k, \rm{ej}}$ is the kinetic energy content of the ejecta.

When a dense CSM shell surrounds the transient, the outflow crashing with the nearly stationary CSM drives two shocks: one that propagates back in the ejecta and another one which propagates in the CSM. Both these shocks act as heating sources for the ejecta as their kinetic energy is converted into radiation. In this scenario, we   assume that the shock efficiently radiates (i.e.~$t_d=0$), implying $L(t)\equiv L_{\rm{inj}}(t)$~\citep{Ofek:2013afa}. This solution holds as long as the  shock deceleration during the interaction with the CSM is negligible. The full self-similar solution including diffusion through the CSM has been calculated in Ref.~\citep{Chatzopoulos:2011vj}. However, since we are mostly interested in linking the electromagnetic emission to the neutrino one, with the production of the latter  taking place in the optically thin part of the CSM, the simple model outlined in Ref.~\citep{Ofek:2013afa} is a fair approximation for our purposes.
Note that we treat $E_{k, \rm{ej}}$ and $M_{\rm{ej}}$ as free parameters, and the ejecta velocity depends on these two quantities through $v_{\rm{ej}} \simeq \sqrt{2 E_{k, \rm{ej}}/M_{\rm{ej}}}$.

\subsection{Heating  sources}\label{sec:heatingSource}
\begin{figure*}[t]
    \centering
    \includegraphics[width=0.9\textwidth]{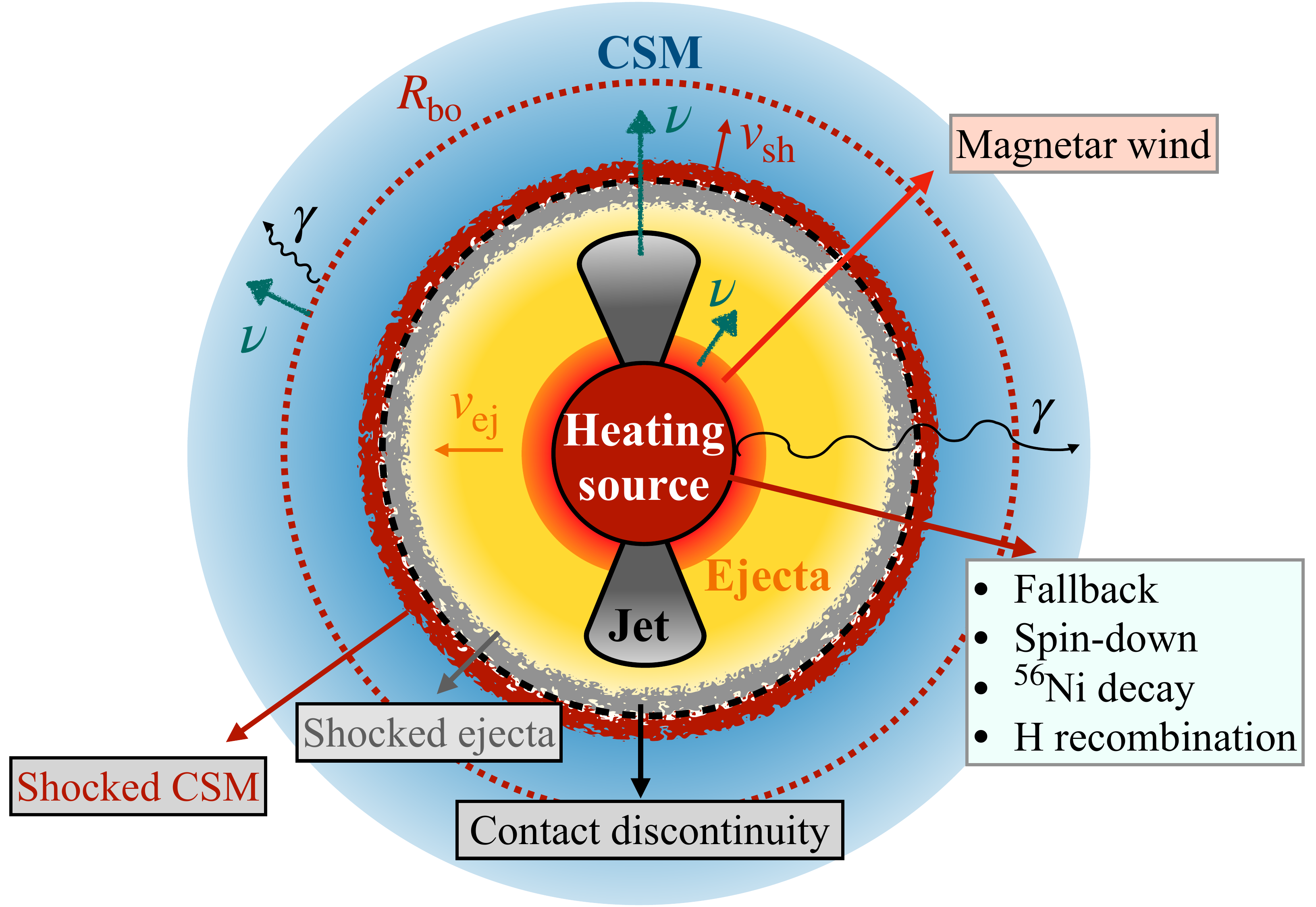}
    \caption{Sketch (not to scale) of the  outflow (orange/yellow region) launched by the collapsing star and powered by a central heating source (red region), moving at velocity $v_{\rm{ej}}$. The heating can be due to  fallback material on the  BH,  spin-down of the magnetar, and/or $^{56}$Ni decay, and hydrogen recombination. A jetted outflow can be harbored  (gray region).
    The outflow  expands and interacts with a dense CSM (blue region), forming a forward shock and a reverse shock. The former propagates outwards and it shocks the CSM  (red outermost shell), the latter propagates inwards and it shocks the ejecta  (light gray shell). The two shocked regions are separated by a contact discontinuity (black dotted line). The forward shock (moving at velocity $v_{\rm{sh}}$) breaks out from the CSM at the breakout radius ($R_{\rm{bo}}$, dotted red line), where non-thermal production of particles starts. Neutrino production can  take place at the forward shock propagating in the CSM and eventually in the magnetar wind and/or in the jet.}
    \label{fig:sketch}
\end{figure*}

For the purposes of this paper, we select the following heating processes~\citep{Villar:2017oya}:
\begin{itemize}
 \item[-] fallback of matter on the black hole (BH);
    \item[-] magnetar spin down;
    \item[-] $^{56}$Ni and $^{56}$Co decay;
    \item[-] hydrogen recombination;
     \item[-] shock breakout from the stellar surface; 
    \item[-] interaction of the outflows with a dense CSM.
\end{itemize}
A sketch of the outflow evolution---including a jetted component---and the heating sources is provided in Fig.~\ref{fig:sketch}. Each process heats the ejecta for a duration $t_{\rm{dur}}$. Unless otherwise specified, we  assume that $t_{\rm{dur}}$ is the timescale such that the bolometric lightcurve luminosity has declined by $90\%$  relative to its peak luminosity. 

\begin{figure*}[t]
    \centering
    \includegraphics[width=0.75\textwidth]{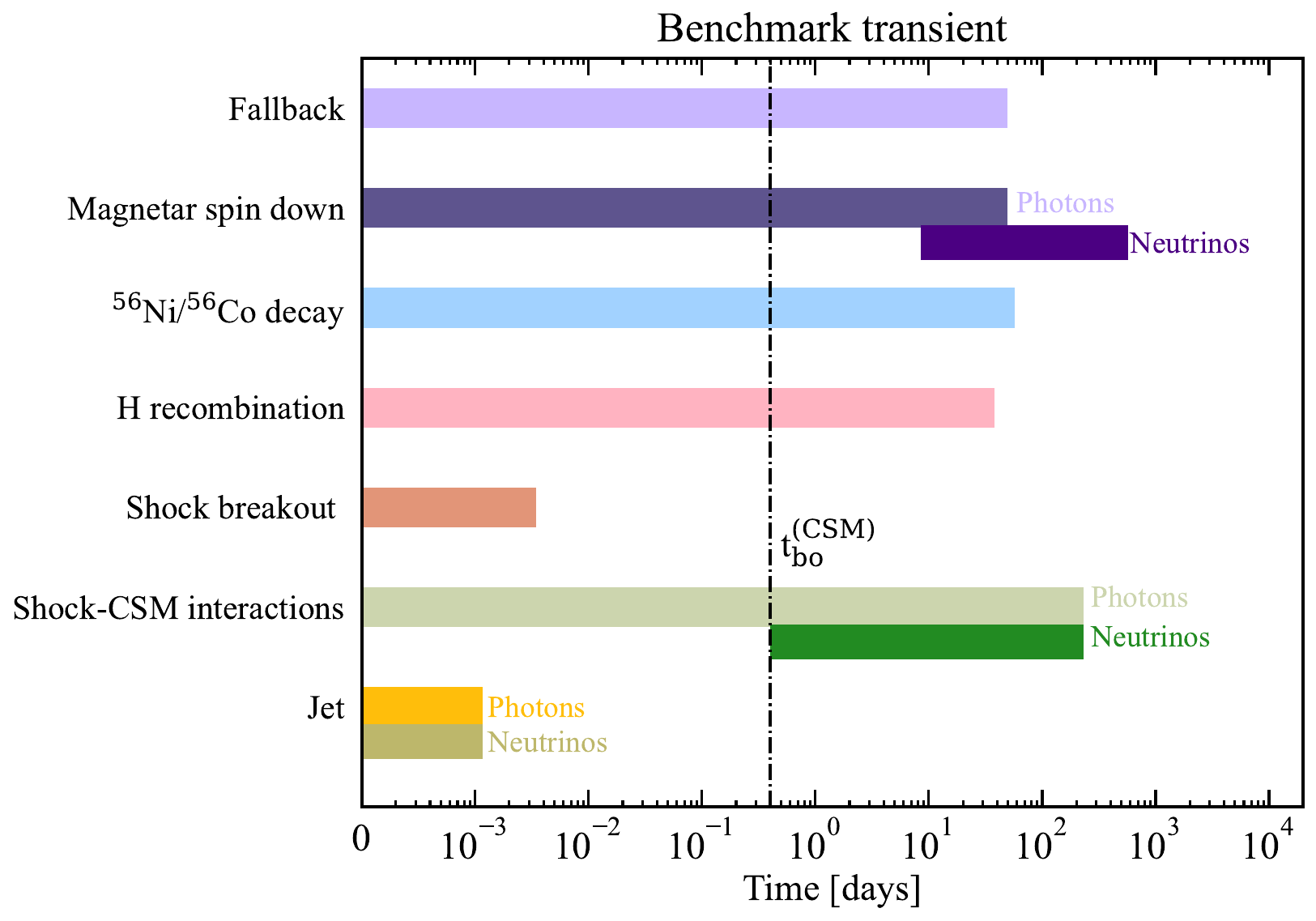}
    \caption{Duration of the bolometric lightcurve powered by different heating processes (Eq.~\ref{eq:outLum}). From top to bottom the observed luminosity is powered by: fallback of material onto a  BH, magnetar spin down, $^{56}$Ni and $^{56}$Co decay, hydrogen recombination, shock breakout from the progenitor star, CSM interactions in the optically thick and thin regimes, and jet. The vertical line marks the time of breakout from the dense CSM shell (Eq.~\ref{eq:breakout}). The time intervals over which neutrino production occurs are  displayed for the magnetar scenario, CSM interactions, and the jet.}
    \label{fig:benchmark}
\end{figure*}

The duration of each heating process is shown in Fig.~\ref{fig:benchmark} for our benchmark transient, whose parameters are listed in Table~\ref{tab:table1}. We assume that the progenitor star of our benchmark transient is a red supergiant. However,  it is unlikely that all considered heating processes simultaneously power the outflow of a collapsing red supergiant. The parameters in Table~\ref{tab:table1} should be interpreted as representative of each  process rather than of a specific transient source.

The total energy radiated by each heating source over the duration of its activity, $t_{\rm{dur}}$, in a specific waveband $\left[ E_{\min}, E_{\max} \right]$ is
\begin{equation}
    E_{\rm{rad}}=  \int_{E_{\rm{min}}}^{E_{\rm{max}}} dE_\gamma\ E_\gamma n_{\gamma}(E_\gamma) \; ,
        \label{eq:totEn}
\end{equation}
where $n_{\gamma}$ is the photon distribution resulting from the  heating process under consideration. Note that  we refer to the total energy radiated after photons diffuse through the  ejecta mass.
Throughout the paper, we consider the following wavebands: 
\begin{itemize}
\item[-] Radio: $ [ E_{\rm{min}}^{\rm{Radio}}, E_{\max}^{\rm{Radio}} ]= [ 4 \times 10^{-15}, 4 \times 10^{-13}]$~GeV =[$1, 100$]~GHz; 
\item[-] Infrared (IR): $ [ E_{\rm{min}}^{\rm{IR}}, E_{\max}^{\rm{IR}} ]= [ 4 \times 10^{-13}, 1.7 \times 10^{-9}]$~GeV = [$0.75, 300$]~$\mu$m;   
\item[-] Optical: $ [ E_{\rm{min}}^{\rm{Opt}}, E_{\max}^{\rm{Opt}} ]= [ 1.7 \times 10^{-9}, 3.3 \times 10^{-9}]$~GeV = [$320,750$]~nm;  
\item[-] Ultraviolet (UV): $ [ E_{\rm{min}}^{\rm{UV}}, E_{\max}^{\rm{UV}} ]= [ 3.3 \times 10^{-9} , 1.2 \times 10^{-7}]$~GeV=[$10, 320$]~nm;  
\item[-] X-ray: $ [ E_{\rm{min}}^{\rm{X-ray}}, E_{\max}^{\rm{X-ray}} ]= [ 3 \times 10^{-7}, 200 \times 10^{-6}]$~GeV= [$0.3, 200$]~keV; 
\item[-] Gamma-ray: $ [ E_{\rm{min}}^{\gamma-\rm{ray}}, E_{\max}^{\gamma-\rm{ray}} ]= [ 200 \times 10^{-6}, 10^3 ]$~GeV. 
\end{itemize} 

Following Ref.~\citep{Villar:2017oya}, we assume that radiation quickly thermalizes and relaxes to a black-body distribution
\begin{equation}
    n^{\rm{BB}}_{\gamma}(E_\gamma) = \int_0^{t_{\rm dur}} dt A_\gamma(t) \frac{E_\gamma^2}{e^{E_\gamma/k_B T_{\gamma}^{\rm{BB}}(t)-1}} \; ,
    \label{eq:blackbody}
\end{equation}
with $k_B$ being the Boltzmann constant, $A_\gamma(t)= L(t) / \left[ \int_0^\infty d E_\gamma E_\gamma n_{\gamma}^{\rm{BB}}(E_\gamma, t)\right]$ the normalization constant and $L$ the emitted luminosity given by Eq.~\ref{eq:outLum}, which depends on the type of heating source. 

The blackbody temperature is
\begin{equation}
    T_{\gamma}^{\rm{BB}}(t) \simeq \left[\frac{L(t)}{4 \pi \sigma_{\rm{SB}} R_{\rm{ph}} (t)^2}\right]^{1/4} \; ,
\end{equation}
where $\sigma_{SB}$ is the Stefan-Boltzmann constant and $R_{\rm{ph}} \simeq R_{\rm{ej}}$ is the photospheric radius in our approximation.  Care should be taken for the photon spectrum resulting from CSM interactions; see Sec.~\ref{sec:csm}.

The black-body assumption holds as long as the outflow is optically thick. Since the bulk of energy is emitted near the lightcurve peak with temperature $\simeq T^{\rm{BB}}_\gamma$, this is a fair approximation.
Note that the total radiated energy in Eq.~\ref{eq:totEn} is calculated in the reference frame of the  star, without considering redshift corrections. For a source at redshift $z$, the observed energy is $E_{\rm{obs}}= E_{\rm{rad}}/(1+z)$.  
\begin{table}[b]
\caption{\label{tab:table1}Characteristic parameters for our benchmark transient. Each parameter is defined in the main text for the corresponding heating source.  {We consider a red supergiant progenitor. These parameters are meant to represent individual heating processes; only a subset of them is expected to be at play for a specific  transient source class.}}
\begin{ruledtabular}
\begin{tabular}{lcr}
\textrm{Parameter}&
\textrm{Symbol} &
\textrm{Value}\\
\colrule
Ejecta energy & $E_{k, \rm{ej}}$ & $10^{51}$~erg \\
Ejecta mass & $M_{\rm{ej}}$ & $ M_\odot $ \\
 {Fallback time} & $t_{\rm fb}$ & $10^7$~s \\
 {Fallback mass} & $M_{\rm fb}$ & $10^{-2} M_\odot$ \\
Jet efficiency & $\epsilon_j$ &  $10^{-2}$\\
Density contributing to $M_{\rm{fb}}$ & $\bar{\rho}$ & $10^{-7}$~cm$^{-3}$ \\
Spin-down period & $P_{\rm{spin}}$ & $10$~ms \\
Magnetar magnetic field & $B$ & $10^{14}$~G \\
Fraction of ejecta mass in $^{56}$Ni & $f_{\rm{Ni}}$ & $0.1$ \\
Progenitor radius & $R_\star$ & $500\ R_\odot$ \\
Progenitor mass & $M_\star$ & $15\ M_\odot$ \\
Mass-loss rate & $M_w$ & $5 \times 10^{-3} \; M_\odot$~yr$^{-1}$ \\
Wind velocity &  $v_w$ & $100$~km s$^{-1}$ \\
CSM radius & $R_{\rm{CSM}}$ & $2 \times 20^{16}$~cm \\
Jet isotropic energy & $E_{\rm{iso}, j}$ & $3.7 \times 10^{54}$~erg \\
Jet lifetime & $t_j $ & 100 s \\
Jet Lorentz factor &  $\Gamma$ & $300$ \\
Jet opening angle & $\theta_j$ & $3^\circ$ \\
\end{tabular}
\end{ruledtabular}
\end{table}

\subsubsection{Fallback}\label{sec:fallback}
 {When a massive star undergoes gravitational collapse its core collapses into a Kerr BH~\citep{MacFadyen:1998vz}, as predicted by the collapsar model. Due to fast rotation, the outer layers of the collapsing star carry too much angular momentum to fall freely into the last stable orbit. Thus, an accretion disk forms, from which both gravitational and rotational energy can be extracted. Energy may also be released through neutrino cooling~\citep{Chen:2006rra}.}

 {Besides the unbound mass ejected during the collapse, a comparable mass (e.g., from tidal tails) could remain bound to the central compact object and fallback onto it.} The rate at which mass falls back onto the BH is~\citep{Lee:2005et, Metzger:2018szx, Metzger:2008av, Metzger:2019zeh, Lopez-Camara:2008trf}:
\begin{equation}
    \dot{M}_{\rm{fb}}(t)=\frac{2}{3} \frac{M_{\rm{fb}}}{t_{\rm{fb}}} \frac{1}{\left( 1+ \frac{t}{t_{\rm{fb}}}\right)^{{5}/{3}}} \; ,
    \label{eq:fallback}
\end{equation}
where $M_{\rm{fb}}$ is the total accreted mass, $t_{\rm{fb}}= \left(  3 \pi/ 32 G \bar{\rho} \right)^{1/2}$ is the free-fall time scale~\citep{kippenhahn1990stellar}, $G$ is the gravitational constant, $\bar{\rho}$ is the mean density of the collapsing layer contributing to $M_{\rm{fb}}$. 
The injected luminosity from this heating process is~\citep{Metzger:2019zeh}
\begin{equation}
    L_{\rm{inj}}^{\rm{fb}}(t)= \epsilon_j \dot{M}_{\rm{fb}} c^2 \; ,
    \label{eq:fallback}
\end{equation}
where $\epsilon_j$ is a constant factor representing the fraction of accreted energy which is used to power the disk wind (or jetted outflow), namely its efficiency. The heating of the spherical ejecta occurs either because of a jet which becomes unstable and looses power~\citep{Bromberg:2015wra}
or  a mildly-relativistic wind which is launched by the inner accretion disk and collides with the more massive outflow emitted at the explosion~\citep{Dexter:2012xk}. In both cases, the energy available to heat the collapsar outflow is given by Eq.~\ref{eq:fallback}; see also the discussion in Ref.~\citep{Metzger:2019zeh}.

\begin{figure*}[t]
    \includegraphics[width=0.75\columnwidth]{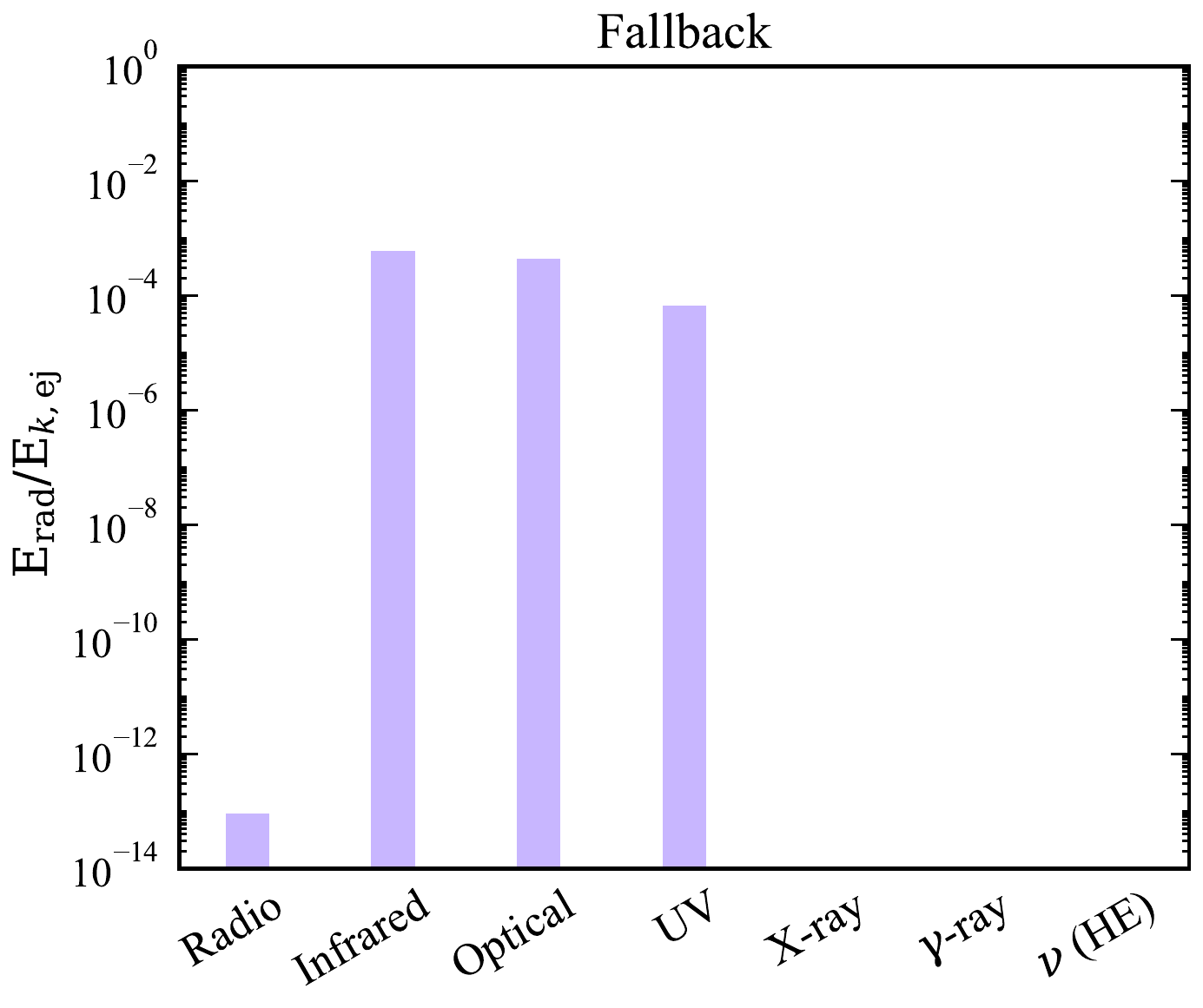}
    \includegraphics[width=0.75\columnwidth]{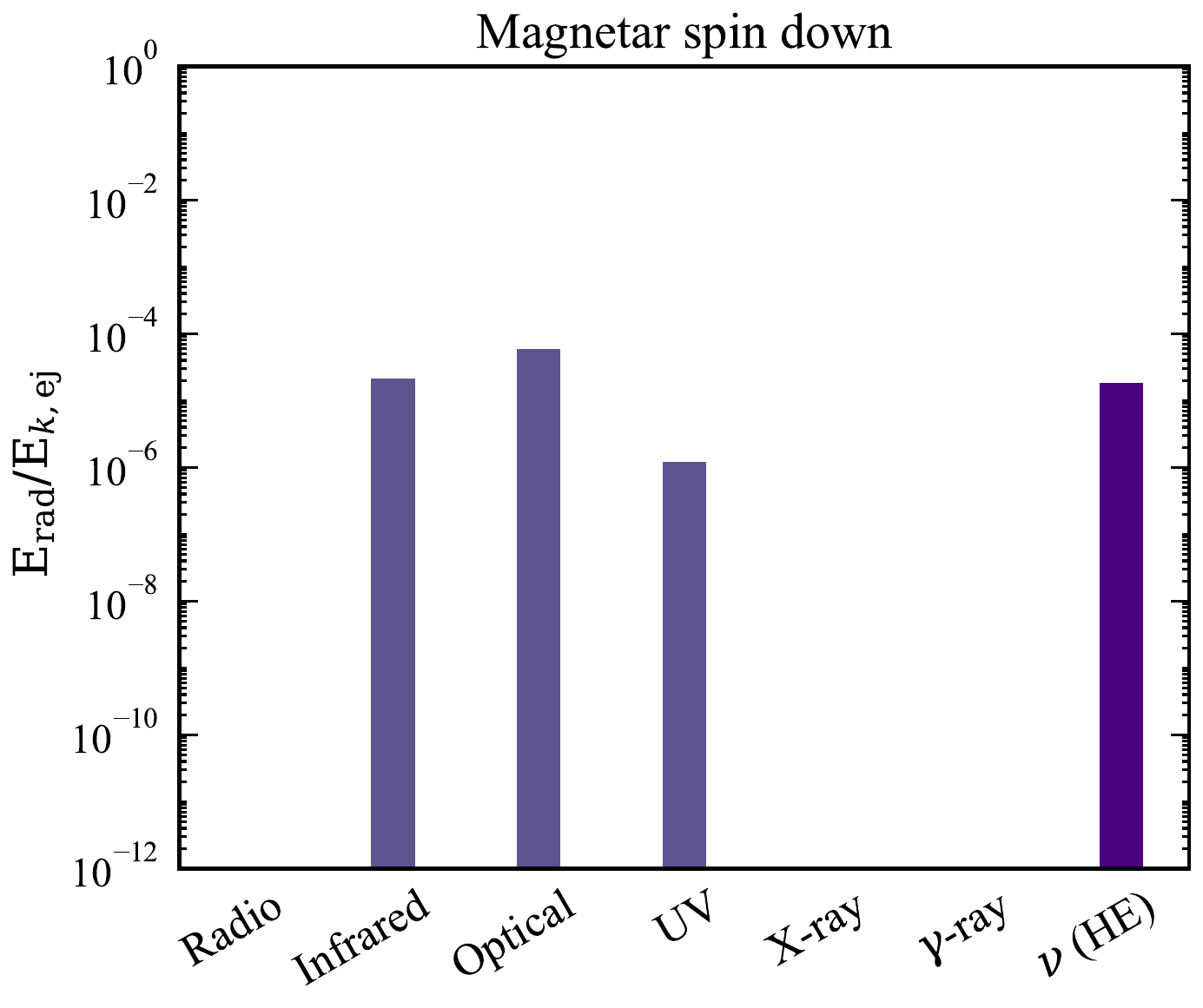}
    \includegraphics[width=0.75\columnwidth]{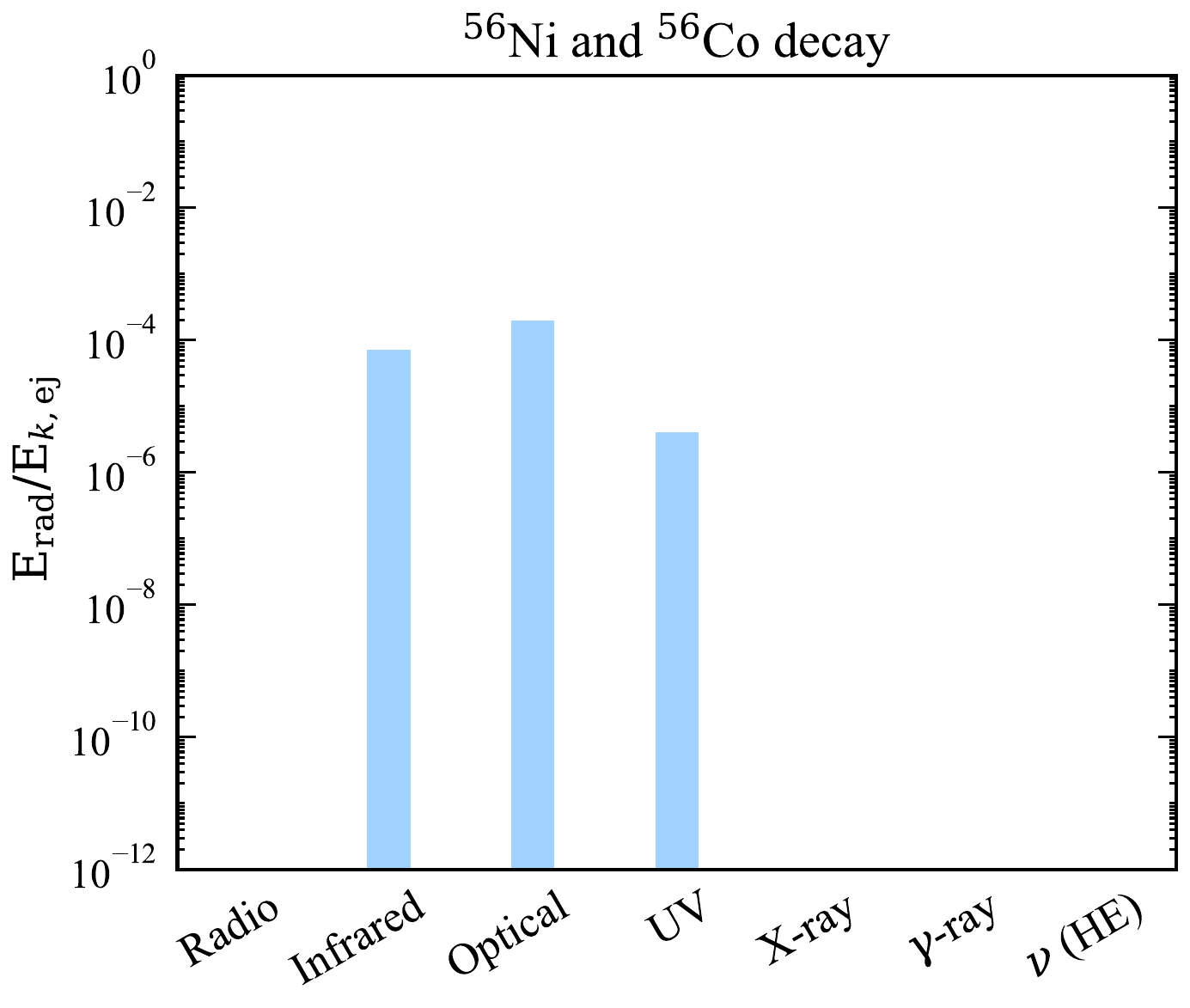}
    \includegraphics[width=0.75\columnwidth]{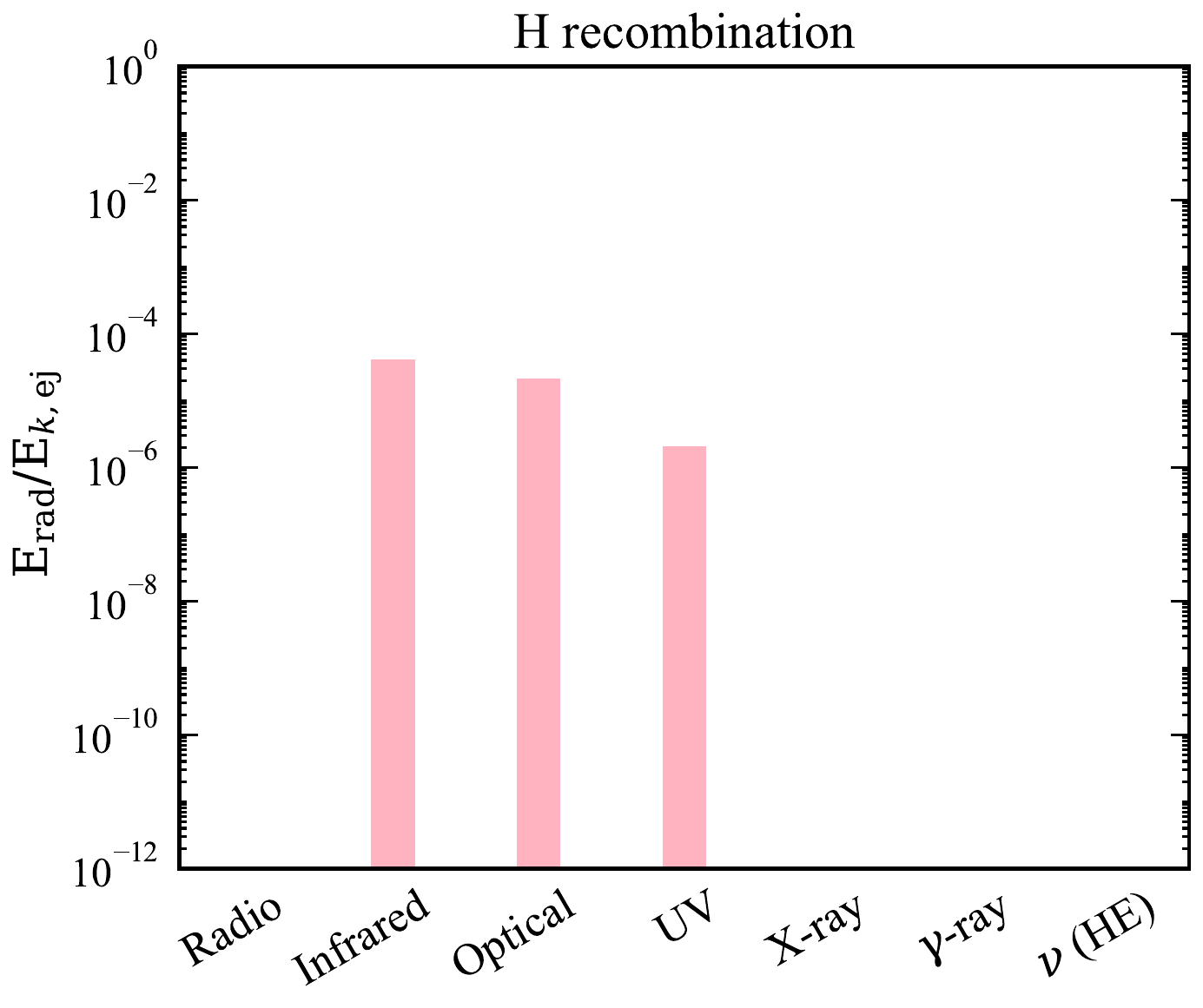}
    \includegraphics[width=0.75\columnwidth]{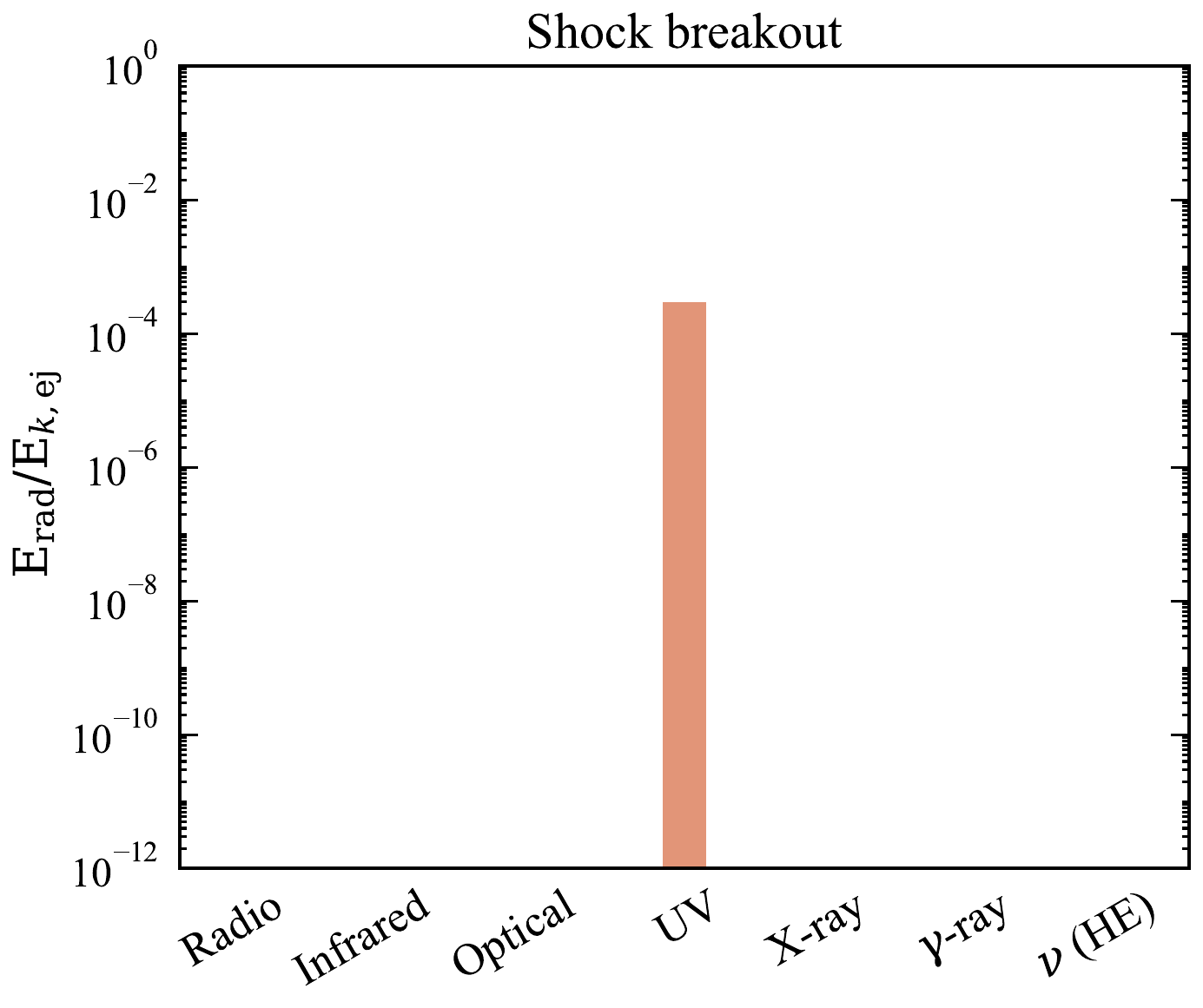}
    \includegraphics[width=0.75\columnwidth]{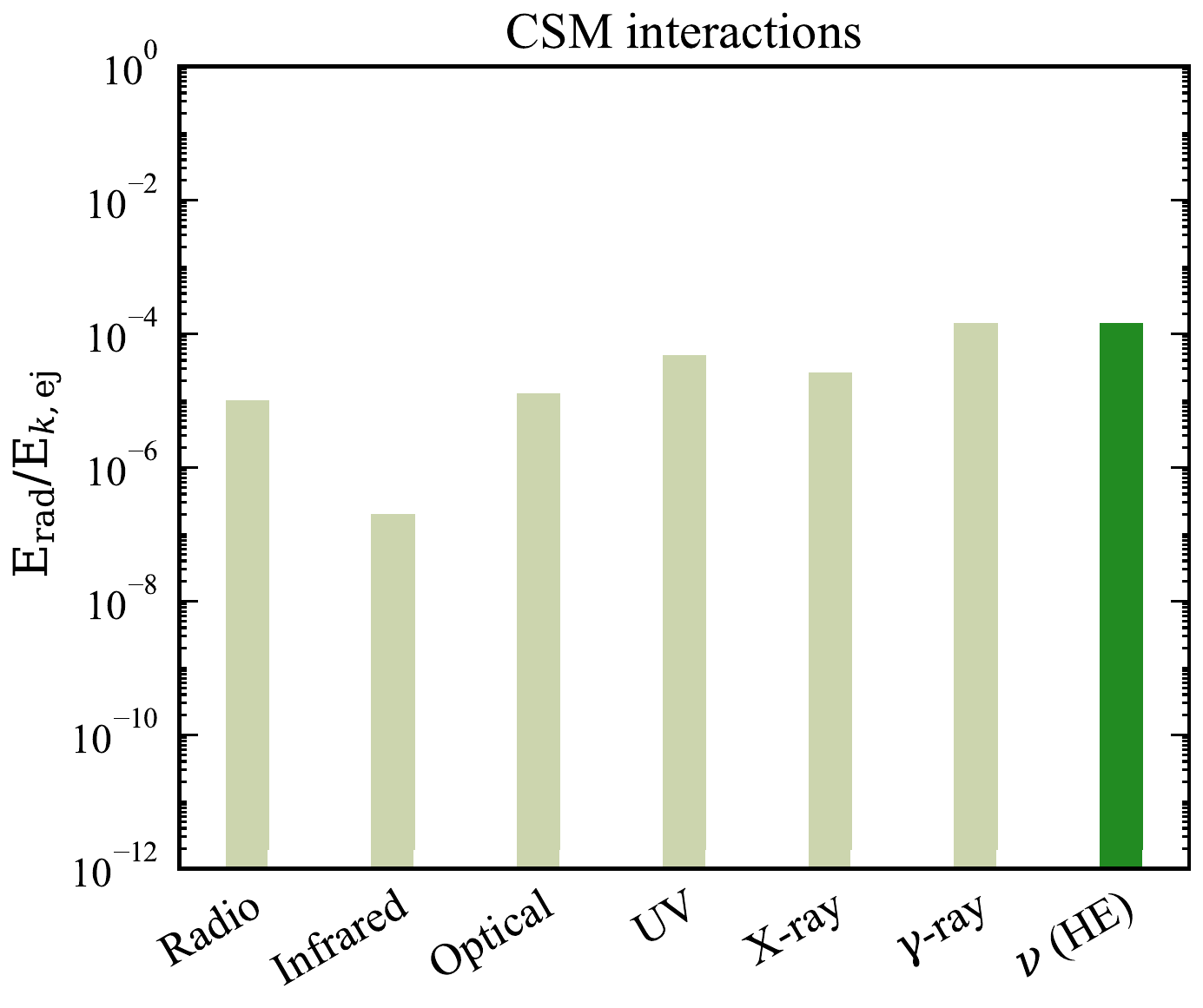}
    \includegraphics[width=0.75\columnwidth]{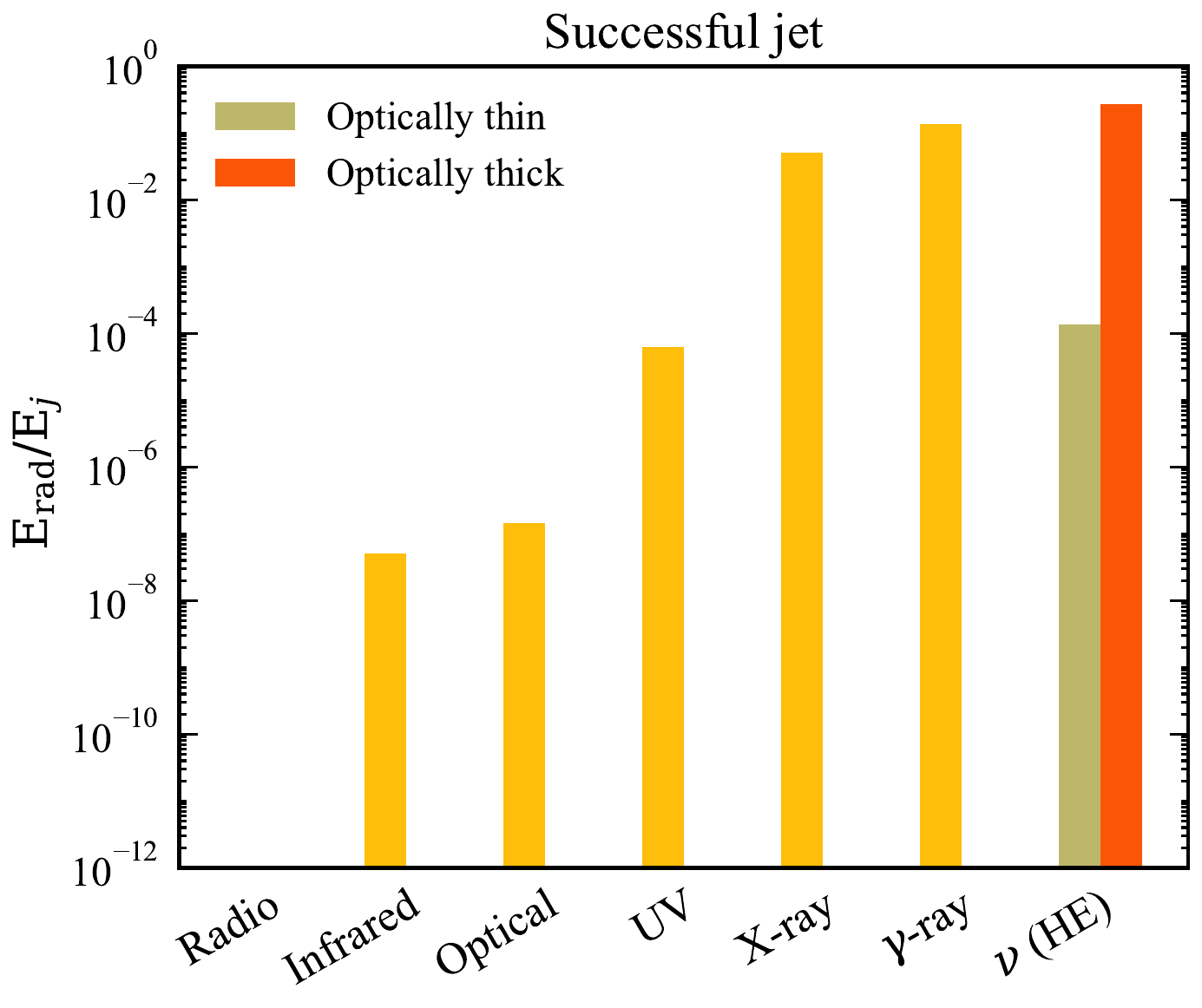}
    \includegraphics[width=0.75\columnwidth]{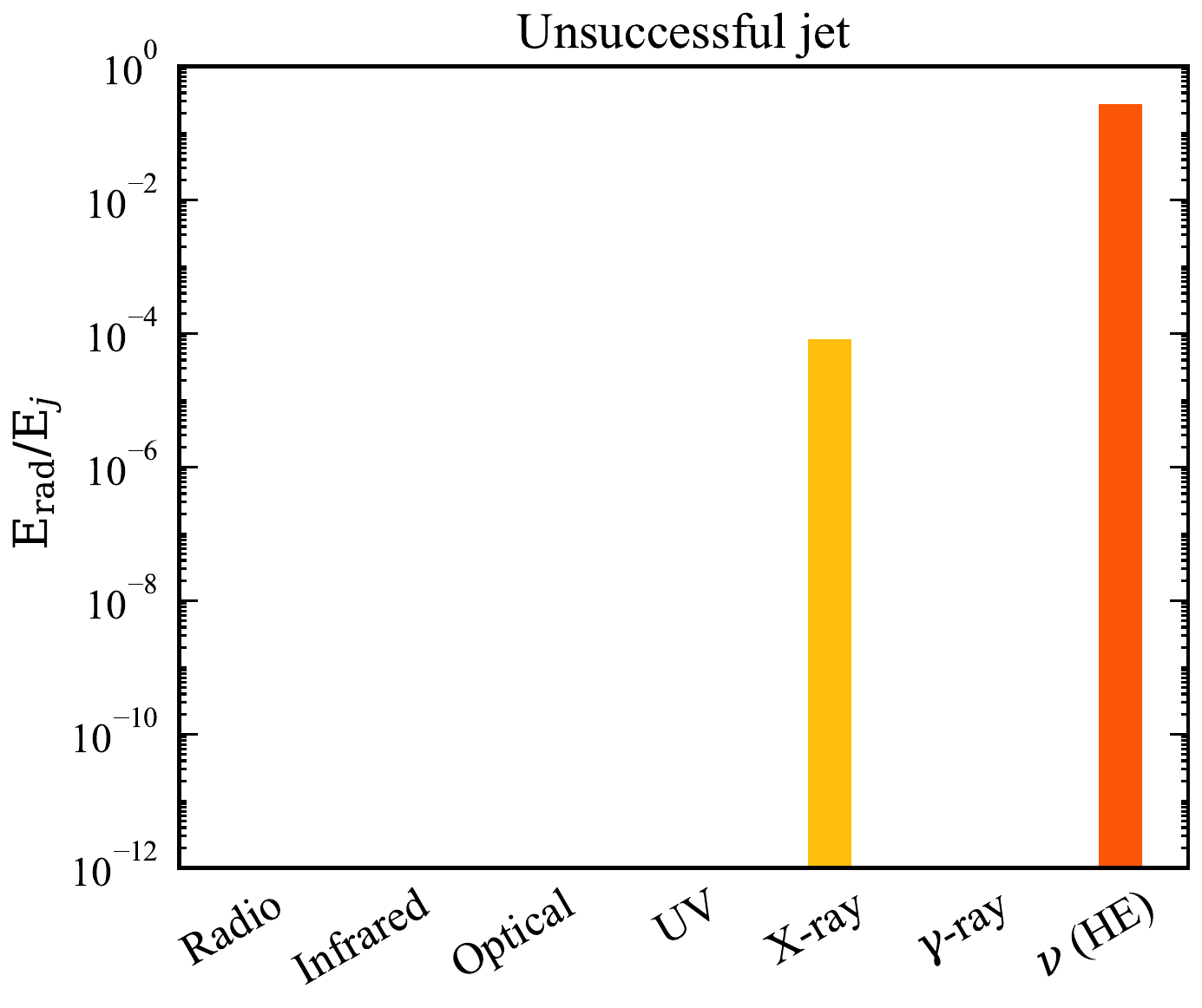}
    \caption{Ratio between the energy radiated  across electromagnetic wavebands as well as  neutrinos  and  the kinetic energy of the ejecta (or the energy of the jet for the bottom panels).
    The results are shown for our benchmark transient (see Table~\ref{tab:table1}). The color code is the same as in Fig.~\ref{fig:benchmark} for each heating process. Heating from fallback material on the BH,  magnetar spin down, $^{56}$Ni, and $^{56}$Co decay and hydrogen recombination lead to  emission of radiation in the UVOIR band. The shock breakout  produces a flash of light in the UV band. CSM interactions in the optically thin regime mostly radiate  in the radio and X-ray bands, with a substantial energy fraction released in gamma-rays and  neutrinos. A successful jet radiates  in the X-ray and gamma-ray bands, whereas the only electromagnetic signature of unsuccessful jets is the flash of light from the shock breakout emitted in the X-ray/gamma-ray band, depending on the outflow and progenitor properties. The jet radiates energy in neutrinos both in the optically thin and thick regimes, the former component only existing for successful jets.}
    \label{fig:tot}
\end{figure*}

 {For a red supergiant progenitor (Table~\ref{tab:table1}), the collapsing layer has mean density $ \bar{\rho} \simeq 10^{-7}$~g cm$^{-3}$, corresponding to the fallback time $t_{\rm fb} \simeq 10^7$~s~\citep{Metzger:2018szx}. The total mass accreted in this case is $M_{\rm fb} \simeq 10^{-2} M_\odot$~\citep{1989ApJ...346..847C}, resulting in a fallback rate $M_{\rm fb} / t_{\rm fb} \simeq 10^{-9} M_\odot$~s$^{-1}$.}
Figure~\ref{fig:tot} (top left panel) shows the energy radiated across the electromagnetic wavebands (Eq.~\ref{eq:totEn}) through fallback of matter on the BH, relative to the kinetic energy of the ejecta $E_{k, \rm{ej}}$.  {For our benchmark transient,} the bulk of radiation powered by fallback onto the BH is emitted in the infrared-optical-ultraviolet (UVOIR) band due to the opacity of the outflow. X-rays may become observable at later times, yet we do not consider this signal in our treatment as it would become relevant  at larger times than the ones considered in this work; see~\citep{Dexter:2012xk} for details.

\subsubsection{\label{sec:magnetar}Magnetar spin down}
Assuming a dipole configuration for the magnetic field, the injected luminosity from the spin down of the compact object is
\begin{equation}
    L_{\rm{inj}}^{\rm{sd}}(t)= \frac{\epsilon_{\rm{sd}} E_{\rm{sd}}}{t_{\rm{sd}} \left( 1+ \frac{t}{t_{\rm{sd}}}\right)^2} \; ,
    \label{eq:sd}
\end{equation}
where $E_{\rm{sd}}= I \Omega^2/2$ is the initial  rotational energy of the magnetar, which depends on the moment of inertia ($I$) and angular velocity of the neutron star ($\Omega$). The spin-down timescale $t_{\rm{sd}}$ is related to the neutron star magnetic field $B_{14}= B/(10^{14} G)$ and the spin period $P_{\rm{spin}}= 2 \pi / \Omega \simeq 10~[E_{\rm{sd}}/(2 \times 10^{50} \; \rm{erg})]^{-0.5}$~ms through~\citep{1969ApJ...157.1395O}
\begin{equation}
    t_{\rm{sd}}= 4 \times 10^7 \frac{P_{\rm{spin, 10}}^2}{B_{14}^2} \; \rm{s} \; .
    \label{eq:sdTime}
\end{equation}
We consider the spin-down injection efficiency to be $\epsilon_{\rm{sd}} = 10\%$, relying on  observations of the Crab Nebula~\citep{Kasen:2009tg}.  {Furthermore, we carry out our calculations for a neutron star with $I = 10^{45}$~g cm$^{-2}$~g cm$^{-2}$~\citep{Lattimer:2004nj}.}

Figure~\ref{fig:tot} (top right panel) shows the energy radiated  (Eq.~\ref{eq:totEn}) through the magnetar spin down. The bulk of radiation powered by the spin down of the magnetar is emitted in the UVOIR band. Note that, during the  the time interval that we consider,  the outflow is optically thick, hence the non-thermal X-rays produced by the compact object are reprocessed in the optical/UV bands~\cite{Fang:2017tla}.

\subsubsection{\label{sec:nichel} Radioactive decay of nickel and cobalt}

Diffusion of radioactive energy produced by newly sinthetized $^{56}$Ni and subsequently  $^{56}$Co was investigated  in Refs.~\citep{1979ApJ...230L..37A, 1980ApJ...237..541A, Wheeler:2014kwa} analytically. The injected luminosity in Eq.~\ref{eq:outLum} can be parametrized as
\begin{equation}
    L_{\rm{inj}}^{\rm{Ni}}(t)=M_{\rm{Ni}} \left[ \epsilon_{\rm{Co}} e^{-t/\tau_{\rm{Co}}} +(\epsilon_{\rm{Ni}}-\epsilon_{\rm{Co}})e^{-t/\tau_{\rm{Ni}}} \right] \;
    \label{eq:nichelDec}
\end{equation}
where $M_{\rm{Ni}}= f_{\rm{Ni}} M_{\rm{ej}}$ is the fraction of the ejecta mass  that goes  into $^{56}$Ni, $\epsilon_{\rm{Ni}}=3.9 \times 10^{10}$~erg s$^{-1}$ g$^{-1}$ ($\epsilon_{\rm{Co}}=6.8 \times 10^{9}$~erg s$^{-1}$ g$^{-1}$) and $\tau_{\rm{Ni}}=8.8$~days ($\tau_{\rm{Co}}=111.3$~days) are the energy generation rates and the decay rates of $^{56}$Ni ($^{56}$Co), respectively. 

Figure~\ref{fig:tot} (second row, left panel) displays the  energy radiated across different wavebands (Eq.~\ref{eq:totEn}) through radioactive decay of $^{56}$Ni and $^{56}$Co. The bulk of radiation powered by these processes is emitted in the UVOIR band; also in this case, the resulting bulk of radiation depends on the assumption of optically thick ejecta.

\subsubsection{\label{sec:recom} Hydrogen recombination}
When the collapsing massive star retains its hydrogen layer, the latter can be ionized by the SN shock. Hydrogen recombination  takes place as the outflow cools to $ \approx 5000$~K, which is the ionization temperature of neutral hydrogen and it has been invoked to explain  the plateau observed in the lightcurve of some SNe~\citep{Hamuy:2002qx, Smartt:2008zd, Sanders:2014uva, Rubin:2015beb}. An analytical model for hydrogen recombination was presented in Refs.~\citep{1993ApJ...405..273W, 2009ApJ...703.2205K, Sanders:2014uva, Villar:2017oya, 2017ApJ...835..282M, 2010ApJ...714..155K}. 

The  luminosity $L_p$ and  duration $t_p$ of the plateau  are~\citep{1993ApJ...405..273W}
\begin{eqnarray}
    L_p &=& 1.64\ \times 10^{42} \frac{R_{\star, 500}^{2/3} E_{k, 51}^{5/6}}{M_{\rm{ej}, 10}^{1/2}} \; \rm{erg} \; \rm{s}^{-1} \; , \\
    t_p &=& 99\ \frac{M_{\rm{ej, }10}^{1/2} R_{\star, 500}^{1/6}}{E_{k, 51}^{1/6}} \; \rm{days} \, ,
\end{eqnarray}
where $R_{\star, 500} = R_\star/(500\ R_\odot)$, $E_{k, 51}= E_{k, \rm{ej}}/(10^{51}$~erg) and $M_{\rm{ej}, 10}= M_{\rm{ej}}/(10\ M_\odot)$ are the kinetic energy and the mass of the ejecta, respectively, with $R_\odot$ and $M_\odot$ being the solar radius and mass. 

The injected luminosity from hydrogen recombination is~\citep{Sanders:2014uva}
\begin{equation}
    L_{\rm{inj}}^{\rm{H}}(t)= \frac{L_p}{e^{-(13.1+0.47\ M_p)t}} \; ,
\end{equation}
where $M_p$ is the peak magnitude, linked to the peak luminosity ($L_p$). 

The energy radiated across different wavebands  through hydrogen recombination is shown in Fig.~\ref{fig:tot} (second row, right panel). The bulk of radiation powered by hydrogen recombination is emitted in the UVOIR band.

\subsubsection{Shock breakout}
A flash of light is expected when the forward shock driven by the outflow breaks out from the progenitor star.
When the CSM surrounding the collapsing star is very dense, the shock breakout may however take place when the shock crosses the CSM. 

The shock breakout theory has been developed in Refs.~\citep{Nakar:2010tt, Nakar:2011mq, Waxman:2016qyw} for non-relativistic and (mildly-)relativistic shocks. The former is the regime expected  for standard core-collapse SNe, while the latter is relevant for  engine-driven SNe. 
The models of Refs.~\citep{Nakar:2010tt, Nakar:2011mq} are challenged by observations, as they cannot reproduce the duration and luminosity of the candidate SNe possibly displaying shock breakout, see e.g.~Refs.~\citep{Soderberg:2008uh, Alp:2020uco, Novara:2020pwy}. Yet, an advanced shock breakout theory does not exist to date. In the light of these uncertainties, we do not adopt any spectral energy distribution for the shock breakout emission. Rather, we assume that photons with temperature $T_{\rm{bo}}$ are emitted over the time $t_{\rm{bo}}$, with  total energy release $E_{\rm{bo}}$. These quantities depend on the stellar progenitor radius ($R_\star$) and mass ($M_\star$), as well as on the energy of the ejecta. For instance, in the case of a non-relativistic shock breakout from a red supergiant one has~\citep{Nakar:2010tt}:
\begin{eqnarray}
\label{eq:sboNakar}
    T_{\rm{bo}} & \simeq & 25 \; \rm{eV} \ M_{\star, 15}^{-0.3}\ R_{\star, 500}^{-0.65}\ E_{k, \rm{ej}, 51}^{0.5} \; ; \\
    t_{\rm{bo}} & \simeq & 300 \; \rm{s} \ M_{\star, 15}^{0.21}\ R_{\star, 500}^{2.16}\ E_{k, \rm{ej}, 51}^{-0.79} \; ; \\
    E_{\rm{bo }} & \simeq & 9 \times 10^{47} \; \rm{erg} \ M_{\star, 15}^{-0.43}\ R_{\star, 500}^{1.74}\ E_{k, \rm{ej}, 51}^{0.56} \; ,
\end{eqnarray}
where $M_{\star, 15} = M_\star /(15 M_\odot)$, $R_{\star, 500}= R_\star / (500 R_\odot)$ and $E_{k, \rm{ej}, 51}= E_{k, \rm{ej}}/ (10^{51} \; \rm{erg})$.
The analytical expressions of these parameters for other stellar progenitors are listed in Appendix A of Ref.~\citep{Nakar:2010tt} for non-relativistic shocks  and  Eq.~29 of Ref.~\citep{Nakar:2011mq} for (mildly-)relativistic shocks. Note that the flash of light produced at the breakout from the stellar surface should be followed by the cooling of the envelope~\citep{Nakar:2010tt, Nakar:2011mq}. However, we neglect this contribution, as it is not correlated with neutrino emission  and thus not of relevance  for the purposes of this work.

Figure~\ref{fig:tot} (third row, left panel)  shows the energy radiated across different wavebands $E_{\rm{rad}}$ (Eq.~\ref{eq:totEn}) through shock breakout. For our benchmark transient, the non-relativistic shock breakout produces a burst of photons in the UV band.   

\subsubsection{Interaction with the circumstellar medium}\label{sec:csm}
Towards the end of their life, massive stars can undergo  eruptive episodes,  polluting the surrounding environment. As a consequence, the collapsing star could have  a  dense CSM shell. We assume that  the CSM density follows  a wind profile
\begin{equation}
    \rho_{\rm{CSM}} (R) = \frac{\dot{M}_w}{4 \pi R^2 v_w f_\Omega} \; 
    \label{eq:rhoCSM}
\end{equation}
where $\dot{M}_w$ is the mass-loss rate of the star, $v_w$ is the wind speed, and $f_\Omega$ is the fraction of the solid angle with dense  CSM. Unless otherwise specified, we assume a spherically symmetric CSM ($f_\Omega = 1$), extended up to the external radius $R_{\rm{CSM}}$, where its density is assumed to drop sharply.  
As the outflow expands in the CSM,  two shocks form: the forward shock, propagating outward and shocking the CSM material, and the reverse shock propagating backward and shocking the ejecta  (in mass coordinates)~\citep{Chevalier:2016hzo}. The forward shock is  the main site of dissipation of kinetic energy, whereas the contribution of the reverse shock is expected to be subleading at the epochs considered in this work and for non-relativistic shocks~\cite{Ellison:2007bga, Patnaude:2008gq, Schure:2010zx, Suzuki:2020qui, 2014ApJ...783...33S, 2018ApJ...853...46S}. The slow deceleration of the forward shock during its interaction with the CSM is not relevant to our purposes, as it would not affect substantially the neutrino emission; we assume that the interaction with the CSM has a total duration $t_{\rm{dur}} \simeq R_{\rm{CSM}}/ v_{\rm{sh}}$, where  $v_{\rm{sh}}$ is the velocity of the forward shock.

The forward shock breaks out from the CSM at the breakout radius $R_{\rm{bo}}$, defined through the following relation
\begin{equation}
    \tau_{\rm{CSM}}(R_{\rm{bo}})= \int_{R_{\rm{bo}}}^{R_{\rm{CSM}}} dR \;  \rho_{\rm{CSM}}(R) \kappa_{\rm{CSM}} = \frac{c}{v_{\rm{sh}}} \; ,
    \label{eq:breakout}
\end{equation}
As the forward shock interacts with the CSM, its kinetic energy is converted into radiation. Within the approximation of constant shock velocity and efficient shock radiation, the injected and emitted luminosity coincide~\citep{Ofek:2013afa}:
\begin{equation}
     \label{eq:csm}
    L_{\rm{inj}}^{\rm{CSM}}  \equiv L(t)=2 \pi \epsilon_{\rm{eff}} \rho_{\rm CSM}(t) R_{\rm{sh}}(t)^2 v_{\rm{sh}}^3, 
\end{equation}
where $\epsilon_{\rm eff}$ is the efficiency conversion factor of kinetic energy into radiation, $R_{\rm{sh}} = v_{\rm{sh}} t$ is the shock radius, and $\rho_{\rm{CSM}}$ is given by Eq.~\ref{eq:rhoCSM} and evaluated at $R_{\rm{sh}}(t)$. As the bulk of radiation from CSM interactions is radiated around $R_{\rm{bo}}$~\citep{Pitik:2023vcg}, we assume that the total  luminosity emitted in the range $R_{0} \leq R \leq R_{\rm{bo}}$ is $L \simeq L_{\rm{bo}}$.

Within our simple framework, the effective temperature of the black-body distribution emerging at $R_{\rm{bo}}$ is~\citep{Chevalier:2011ha, Ofek:2013afa}:
\begin{equation}
    T_{\gamma}^{\rm{BB}}= \left( \frac{18}{7 a} \rho_{\rm{CSM}}(R_{\rm{bo}})  v_{\rm{sh}}^2\right)^{1/4} \; .
\end{equation}

Once the forward shock breaks out from the dense CSM, namely when Eq.~\ref{eq:breakout} is fulfilled, it becomes collisionless.
In this regime, photons are mainly produced through bremsstrahlung and emitted in the X-ray band for  $v_{\rm{sh}} \gtrsim 10^4$~km s$^{-1}$~\citep{Margalit:2021bqe}. The total emitted luminosity produced by the forward shock for $R_{\rm{bo}} \leq R \leq R_{\rm{CSM}}$ is given by~\citep{Margalit:2021bqe, Fang:2020bkm}
\begin{equation}
    L^{\rm{brem}}(R_{\rm{CSM}}) = \min \left( 0.3 \frac{t_{\rm{dyn}}}{t_{\rm{ff}}} , 1 \right) L_{\rm{sh}} \; ,
    \label{eq:bremm}
\end{equation}
where $t_{\rm{dyn}}$ and $t_{\rm{ff}}$ are the dynamical and free-free electron cooling times defined as  in Appendix~\ref{app:B}. The shock kinetic power $L_{\rm{sh}}$ is also defined in Appendix~\ref{app:B}. Note that Eq.~\ref{eq:bremm} is estimated at the edge of the CSM shell ($R_{\rm{CSM}}$).

After shock breakout from the CSM, the radiation due to CSM interactions no longer relaxes to a black-body distribution, hence the non-thermal photon spectrum is
\begin{equation}
    n^{\rm{brem}}_{\gamma}(E_\gamma)= L^{\rm{brem}} \frac{E_\gamma}{k_B T_e} e^{- E_\gamma/ k_B T_e} \; ,
\end{equation}
where $L^{\rm{brem}}$ is the total emitted luminosity given by Eq.~\ref{eq:bremm} and $T_e$ is the post-shock temperature of electrons, defined in Appendix~\ref{app:B}. 

In the optically thin region of the CSM, particle acceleration leads to production  of relativistic electrons. This case is particularly relevant when shocks are not radiative. As the forward shock expands in the CSM, it converts the kinetic energy of the blastwave into internal energy. The internal energy density is
\begin{equation}
    u_{\rm{int}}(R)= \frac{9}{8} v_{\rm{sh}}^2\   \rho_{\rm{CSM}} \ , 
    \label{eq:enInt}
\end{equation}
where $\rho_{\rm{CSM}}$ is given by Eq.~\ref{eq:rhoCSM}. A fraction $\epsilon_B$ of the internal energy density is stored in the post shock magnetic field $B_{\rm{CSM}} = \sqrt{8 \pi \epsilon_B u_{\rm{int}}}$. 

A fraction $\epsilon_e$ of Eq.~\ref{eq:enInt} is  given to accelerated electrons. 
The latter mostly cool through synchrotron radiation~\citep{1998ApJ...499..810C}, whose spectrum for the non-relativistic and mildly-relativistic blastwave is provided in Ref.~\citep{Margalit:2021bqe}. 

In Fig.~\ref{fig:tot} (third row, right panel) we show the total energy radiated through CSM interactions. 
 We also display the relative energy emitted in gamma-rays (see  Sec.~\ref{sec:neutrino}). 
The bulk of energy is radiated in the UVOIR band, whereas bremsstrahlung and synchrotron processes radiate energy mostly in the radio and X-ray bands.

\subsubsection{Multiple heating sources}
If more than one source contributes to heat the outflow as it expands,  the total radiated luminosity is given by the sum of all contributions:  $L^{\rm{tot}} = \sum_{i} L^{i} (t)$, where $L^{i}$ corresponds to the luminosity radiated from the $i$-th heating source.

If the outflow propagates in a dense CSM, then the radiation produced by other heating sources (e.g., $^{56}$Ni decay)  has to propagate through the total mass $M_{\rm{tot}} =  M_{\rm{ej}} + M_{\rm{CSM, th}}$, where $M_{\rm{CSM, th}} = \int_{R_\star}^{R_{\rm{bo}}} dR 4 \pi R^2 \rho_{\rm{CSM}}(R)$ is the mass of the optically thick CSM.

\section{Modeling of the electromagnetic emission: jetted relativistic outflows}\label{app:jet}
In this section, we focus on the modeling of the electromagnetic emission in jetted relativistic outflows, which differs from the treatment outlined in Sec.~\ref{sec:model} for the non-relativistic outflows. 
A bipolar jet may be harbored in the collapsing star and  launched a few ms after the collapse. Given the jet luminosity ${L}_j$ (assumed to be constant) and lifetime ${t}_j$, its  injected energy  is ${E}_j = {L}_j {t}_j$. The jet dynamics only depends on the jet isotropic equivalent energy ${E}_{\rm{iso}, j}={E}_j/ ({\theta}_j ^2 /4 ) $ and  Lorentz factor $\Gamma$~\citep{Piran:1999kx, Kumar:2008dr}, where $\theta_j$ is the jet opening angle. We parameterize  the energy budget of the jet in terms of its energy $E_j$~\citep{Kaneko:2006mt}, rather than $E_{k, \rm{ej}}$ as we have considered for the non-relativistic outflows (see Fig.~\ref{fig:tot}). Furthermore, our results refer to a jet observed on-axis (for a discussion on jets observed off-axis see, e.g., Ref.~\cite{Ramirez-Ruiz:2004gvs}).

Short living engines or progenitor stars which retain the hydrogen envelope, such as partially stripped SNe, are likely to produce unsuccessful jets~\citep{Lee:2005et, Izzard:2003uz, Nakar:2015tma, Gilkis:2021uht, Sobacchi:2017wcq}. 
In this case, the jetted outflow does not  breakout from the stellar mantle or it is choked. If the jet is instead powered for  sufficiently long time and is energetic enough, it breaks out from the star and  produces a GRB.

\subsection{Successful jets}\label{sec:grb}
The mechanism responsible for energy dissipation and shaping the observed non-thermal emission  is still under debate, with particle acceleration possibly due to internal shocks~\cite{Rees:1994nw, Daigne:1998xc, Kobayashi:1997jk} or magnetic reconnection~\citep{Zhang:2010jt, Drenkhahn:2001ue, Drenkhahn:2002ug,Giannios:2007yj}. In both cases,  the observed electromagnetic signal may originate both in  the optically thick and  thin regions of the jet.  
Following Ref.~\citep{Pitik:2021xhb}~\footnote{Note that the calculations of Ref.~\citep{Pitik:2021xhb} are carried out relying on isotropic equivalent quantities. In order to connect isotropic quantities with the observed ones, we correct the total isotropic energy by the beaming factor of the jet ($\theta_j^2 /4$).}, Fig.~\ref{fig:tot} (bottom left panel) shows the total energy radiated by a successful jet across the electromagnetic  wavebands, assuming $t_{\rm{dur}}=100$~s~\citep{Tarnopolski:2015pba, 2009ARA&A..47..567G}. We show the largest energy radiated among the GRB models considered in Ref.~\citep{Pitik:2021xhb}, in order to obtain an upper limit for the energy budget. Note that the relativistic component of the outflow moves with constant Lorentz factor $\Gamma$, hence the observed energy is $E_{\rm{obs}}= E_{\rm{rad}} \Gamma/(1+z)$. 

We  do not consider the deceleration phase of the relativistic jetted component of the outflow. This is motivated by the fact that the neutrino emission during the afterglow is negligible with respect to the prompt one~\cite{Guarini:2021gwh}. 

\subsection{Unsuccessful jets} 
As the jet propagates through the stellar envelope, it inflates a high pressure region  of shocked jet and  stellar material, the cocoon~\citep{Bromberg:2011fg, Ramirez-Ruiz:2002szz, Lazzati:2005xv, Zhang:2002yk}. 
The jet dynamics is highly non-linear due to the mixing with the cocoon, which slows down the jet while increasing its baryon density~\citep{Gottlieb:2022tkb} (see Ref.~\citep{Bromberg:2011fg} for the analytical modeling of the propagation of a relativistic jet in the stellar mantle). Independently on the fate of the jet, the cocoon always breaks out from the star~\citep{Bromberg:2011fg, Lazzati:2005xv}. 

If the jet is smothered within the stellar mantle, the only observable electromagnetic counterpart would be the shock breakout of the cocoon from the collapsing star. The breakout is expected to occur with mildly-relativistic velocities, with signatures of asymmetries in the outflow~\citep{Gottlieb:2020raq, Maeda:2023vfp}. The fraction of energy radiated from an unsuccessful jet is shown in Fig.~\ref{fig:tot} (bottom right panel), for the parameters used in Ref.~\citep{Guarini:2022hry}.

\section{Neutrino emission}\label{sec:neutrino}
In this section, we summarize the processes leading to neutrino production, namely photo-hadronic ($p \gamma$) and hadronic ($pp$) interactions. Furthermore, we outline the  methods adopted to calculate the neutrino signal.

\subsection{Proton spectral energy distribution}\label{sec:protons}
The regions of the outflow where protons can be co-accelerated with electrons are the magnetar wind, the forward shock resulting from CSM interactions and the jet. We now introduce the resulting spectral energy distributions of protons.

\paragraph{Magnetar wind.} The injected proton energy distribution is [in units of GeV$^{-1}$ cm$^{-3}$]~\citep{Fang:2017tla}
\begin{equation}
    {n}_{{p}}({E}_{p}) \equiv \frac{{\rm{d}}^{2} {N}_{{p}}}{{\rm{d}} {E}_{{p}} {\rm{d}}{V}}  =  A_p E^{-1}_p\ ,
    \label{eq:protMagn}
\end{equation}
where the normalization constant is $A_p =1.08 \times 10^{-5} B_{14}^{-1} t_{5.5}^{-3} M_{\rm{ej}, -2}^{3/2} P_{\rm{spin}, -3}^{3}$, with $t_{5.5} = t/10^{5.5}$~s, and the other quantities are defined as in   Sec.~\ref{sec:heatingSource}. Note that the spectrum of protons accelerated in the magnetar wind is  expected to be hard [$n_p(E_p) \propto E^{-1}_p$]. 

\paragraph{CSM interactions.} Protons can be accelerated at the forward shock as the  SN ejecta cross the CSM. Efficient acceleration starts at $R \simeq R_{\rm{bo}}$~\citep{Murase:2010cu, Petropoulou:2017ymv, Pitik:2021xhb, Katz:2011zx, Sarmah:2022vra} and it proceeds over a wide range of radii, up to the outer radius $R_{\rm{out}}= \min \left[ R_{\rm{CSM}}, R_{\rm{dec}} \right]$. Here, $R_{\rm{dec}}=M_{\rm{ej}} v_w / \dot{M}_w$ is the deceleration radius, corresponding to the distance from the center of explosion where the outflow has swept-up a CSM mass comparable to $M_{\rm{ej}}$. 

The injected proton energy distribution at the forward shock is [in units of GeV cm$^{-3}$]:
\begin{equation}
	{n}_{{p}}({E}_{p}) \equiv \frac{{\rm{d}}^{2} {N}_{{p}}}{{\rm{d}}{E}_{{p}}{\rm{d}} {V}}  = {A}_p {E}_{{p}}^{-k_p} \Theta({E}_{{p}} - {E}_{p, \rm{min}}) \Theta({E}_{p, \rm{max}} - {E}_{\rm{p}})\ ,
\label{eq:protonInjection}
\end{equation}
where $k_p = 2$ is the proton index for non-relativistic collisionless shocks~\citep{Matthews:2020lig}. The minimum proton energy for non-relativistic shocks is ${E}_{p, \rm{min}} = m_p c^2$ (for mildly-relativistic shocks, the minimum proton energy is ${E}_{p, \rm{min}}= \Gamma_{\rm{sh}} m_p c^2$, where $\Gamma_{\rm{sh}}= 1/ \sqrt{1- (v_{\rm{sh}}/c)^2}$ is the shock Lorentz factor; since for mildly-relativistic shocks $\Gamma_{\rm{sh}} \lesssim 2$, the correction to the minimum proton energy does not affect  our results for the neutrino signal substantially), while the maximum energy ${E}_{p, \rm{max}}$ is obtained by the condition ${t}_{p, \rm{acc}}^{-1} = {t}^{-1}_{p, \rm{cool}}$, where ${t}^{-1}_{p, \rm{acc}}$ is the proton acceleration rate and ${t}^{-1}_{p, \rm{cool}}$ is the proton total cooling rate; see Appendix~\ref{app:A} for the proton cooling rates. 

The normalization constant is ${A}_p = {9 \epsilon_{{p}} n_{{p, \rm{CSM}}}}(R) m_p c^2/[{8  {\rm{ln}} ({E}_{p, \rm{max}}/ {E}_{p, \rm{min}})}] ({v_{\rm{sh}}}/c)^2 $.  Here, $\epsilon_p$ is the fraction of the blastwave internal energy expressed by Eq.~\ref{eq:enInt} which is stored into accelerated protons. Finally, $n_{p, \rm{CSM}}= \rho_{\rm{CSM}}/m_p$ is the CSM proton number density. 

\paragraph{Jetted outflows.} Protons accelerated in the jet follow a power-law spectrum~\cite{Sironi:2013ri}. The proton distribution in the comoving frame of the jet (we denote quantities in this frame as primed: $X^\prime$) reads [in units of GeV cm$^{-3}$]
\begin{eqnarray}
n^\prime_p(E^\prime_p) = \frac{{\rm{d}}^{2}N^\prime_{\rm{p}}}{{\rm{d}}E^\prime_{\rm{p}}{\rm{d}}V^\prime} & = &  A^\prime_p E^{^\prime -k_p}_p  \exp \left[-\left( \frac{E^\prime_p}{E^\prime_{p, \rm{max}}} \right)^{\alpha_p} \right] \nonumber \\   &    \times &
\Theta(E^\prime_p - E^\prime_{p, \rm{min}}) \ ,
\label{eq:protonJET}
\end{eqnarray}
where $\alpha_p$ mimics an exponential cutoff~\citep{2001RPPh...64..429M} and $\Theta$ is the Heaviside function. The minimum energy of accelerated protons is $E^\prime_{p, \min} = m_p c^2$, while their maximum energy is obtained equating the proton acceleration rate with the proton cooling rate, namely $t^{\prime -1}_{p, \rm{acc}}= t^{\prime -1}_{p, \rm{cool}}$; see Appendix~\ref{app:A} for the  proton cooling rates in the jet. 
$A^\prime_p = \epsilon_{j, d} \epsilon_{j, p} E^{\prime}_{\rm{iso}, j}/(4 \pi R_j^2 c t^\prime_j)$ is the normalization constant, where $\epsilon_{j, p}$ is the fraction of the dissipated isotropic energy of the jet $\epsilon_{j, d} E^\prime_{\rm{iso}}$ which is stored in accelerated electrons; $R_j$ is the position along the jet where proton acceleration takes place, while $t^\prime_j= {t}_j \Gamma$ is the comoving jet lifetime. 

The microphysical parameters $\epsilon_{j, d}$ and $\epsilon_{j, p}$ depend on the process assumed to be responsible for energy dissipation along the jet. The spectral index is $k_p = 2.2$, if acceleration occurs at relativistic collisionless internal shocks or sub-shocks~\citep{Sironi:2013ri, Beloborodov:2016jmz}, while it depends on the magnetization of the jet if protons are accelerated through magnetic reconnection~\citep{Werner:2016fxe}. 

\subsection{Neutrino production channels}
\paragraph{Proton-photon ($p\gamma$) interactions. \label{sec:pg}} 
Electrons co-accelerated with protons cool producing a photon distribution which serves as a target for accelerated protons. Neutrinos are mainly produced through the $\Delta^{+}$ resonance~\citep{Kelner:2008ke, Hummer:2010vx}:
\begin{equation}
   p+ \gamma \longrightarrow \Delta^+ \longrightarrow
   \begin{system}
   n + \pi^+ \; \; \; \; \; \; \; 1/3 \; \rm{\; of\; all\; cases} \\
   p+   \pi^0  \; \; \; \; \; \; \;  \;   2/3 \; \rm{\; of \; all\; cases }  \ .
   \end{system}
    \label{reaction_channel}
\end{equation} 
Subsequently, neutral pions decay into gamma-rays $\pi^0 \longrightarrow 2 \gamma$, while charged pions decay producing neutrinos $\pi^+ \longrightarrow \mu^+ + \nu_\mu \longrightarrow \bar{\nu}_\mu + \nu_e + e^+$. 
Unless otherwise specified, we do not distinguish between neutrinos and antineutrinos in the following. 

\paragraph{\label{sec:pp}Proton-proton ($pp$) interactions.} Accelerated protons can interact with a target of non-relativistic protons, producing neutral and charged pions~\citep{Kelner:2006tc}. Subsequently, pions decay as detailed above for $p \gamma$ interactions. Throughout the paper, we consider the energy radiated in gamma-rays both through the electromagnetic processes discussed in Sec.~\ref{sec:model} and through $pp$ interactions. 

\subsection{Expected neutrino emission}
Both $p \gamma$ and $pp$ interactions can take place in the magnetar wind, at the external shock driven by the outflow in the CSM and in the jet. The duration of the expected neutrino signals in the wind of a central magnetar and at CSM interactions is summarized in  Fig.~\ref{fig:benchmark} for our benchmark transient, whose parameters are listed in Table~\ref{tab:table1}. Along the jet, neutrino production takes place throughout the whole jet lifetime $t_j$.

Both neutrinos and photons at CSM interactions are produced through the dissipation of kinetic energy of the blastwave as the forward shock expands within the optically thin CSM. Consequently, the duration of the electromagnetic and neutrino signals in Fig.~\ref{fig:benchmark} is similar. On the contrary, neutrino production in the magnetar wind starts when photopion production becomes efficient and it ceases when pion production freezes out~\citep{Fang:2017tla}; these times are defined in Appendix~\ref{app:A}. As the processes producing photons and neutrinos in the magnetar wind are not correlated, their duration in Fig.~\ref{fig:benchmark} is different.

\paragraph{Magnetar wind.} Protons accelerated in the wind of the magnetar can undergo both $p \gamma$ and $pp$ interactions. The injected proton energy distribution is given by Eq.~\ref{eq:protMagn}, while thermal optical photons and non-thermal X-ray photons produced in the wind nebula serve as  targets for $p \gamma$ interactions.

We calculate the total energy emitted in neutrinos in the magnetar wind following Ref.~\citep{Fang:2017tla}: 
\begin{eqnarray}
    \label{eq:nuMax}
    {E}^{\nu}_{\rm{tot}} & \simeq & 1.7  \times 10^{41} \eta_{-1}^{2/3} B_{14}^{-4/3} M_{\rm{ej}, -2}^{-1/4} P_{\rm{spin}, -3}^{-1/2} \\  \nonumber &  & \times \;  \epsilon_{\rm{mag}, -2}^{-1/6} f^{p}_{\rm{sup}} f^{\pi}_{\rm{sup}}  f^{\mu}_{\rm{sup}} \; \rm{erg} \; , 
\end{eqnarray}
where $\eta_{-1}= \eta/10^{-1}$ is the acceleration efficiency in the magnetar wind nebula  normalized to its nominal value, $\epsilon_{\rm{mag}, -2}= \epsilon_{\rm{mag}}/10^{-2}$ is the nebula magnetization parameter, whose nominal value is motivated by observations of the Crab Nebula~\citep{1984ApJ...283..710K}. 
Finally, $f^{p}_{\rm{sup}}$ is the suppression factor for pion creation, while $f^{\pi}_{\rm{sup}}$ and $f^{\mu}_{\rm{sup}}$ are the suppression factors for neutrino creation from $\pi^{\pm}$ and $\mu^{\pm}$ decays, respectively (see  Appendix~\ref{app:B}). The fraction of the ejecta kinetic energy emitted in neutrinos in the magnetar wind is shown in Fig.~\ref{fig:tot} (top right panel) for our benchmark transient.

\paragraph{CSM interactions.} Accelerated protons follow the input energy distribution in Eq.~\ref{eq:protonInjection} and  can interact  with the photon spectrum produced at the forward shock. Furthermore, accelerated protons  undergo $pp$ interactions with the non-relativistic CSM protons.

In most cases, $p \gamma$ interactions at the forward shock are subleading for non-relativistic and mildly-relativistic shocks~\cite{Pitik:2021dyf, Guarini:2021gwh, Guarini:2022uyp, Murase:2010cu, Petropoulou:2017ymv, Sarmah:2022vra}. This result also holds when the shocks are radiative, as the energy threshold for $p \gamma$ interactions can be reached only when the CSM covers a small fraction of the solid angle ($f_\Omega \ll 1$), which is not the case for SNe~\citep{Fang:2020bkm,Brethauer:2022nag}. 
Therefore, we only consider $pp$ interactions as a viable neutrino production channel at the forward shock. 
We calculate the total energy emitted in neutrinos through $pp$ interactions following  Refs.~\citep{Kelner:2006tc, Petropoulou:2017ymv}. The fraction of the ejecta kinetic energy radiated in neutrinos from CSM interactions for our benchmark transient is shown in Fig.~\ref{fig:tot} (third row, right panel).

\paragraph{Jetted outflows.}
In a magnetized jet, neutrino production begins in the optically thick part of the outflow~\citep{Gottlieb:2021pzr, Guarini:2022hry}. 
Hereafter we rely on the results of Ref.~\citep{Guarini:2022hry} for the expected neutrino signal. In particular, we consider the case with initial jet magnetization $\sigma_0=200$ of Ref.~\citep{Guarini:2022hry}.

In the absence of jet magnetization, neutrino production below the jet photosphere may take place only if the jet is smothered in an extended envelope surrounding the progenitor core. We refer the interested reader to Refs.~\citep{Senno:2015tsn, Guarini:2022uyp} for the neutrino signal expected in this scenario, and we explicitly include it in our calculations in Sec.~\ref{sec:census}. However, in Fig.~\ref{fig:tot} (bottom right panel) we only show the case of a jet smothered in a Wolf-Rayet progenitor star. 

In the optically thin region of the jet, the input proton distribution is given by Eq.~\ref{eq:protonJET}. The non-thermal photon distribution that serves as target for $p \gamma$ interactions  depends on the mechanism assumed  for energy dissipation. We rely on Ref.~\citep{Pitik:2021dyf} and  take the maximum energy radiated in neutrinos across the different GRB models considered in the aforementioned reference.
In the optically thin part of the jet, the baryon density is not large enough for $pp$ interactions to be efficient~\cite{Pitik:2021dyf}. Therefore, we only consider $p \gamma$ interactions as the viable channel for neutrino production. 

In the bottom left panel of Fig.~\ref{fig:tot} we show the energy radiated by a successful jet in neutrinos both in the optically thick and thin regimes. However, we warn the reader that the results for the optically thick regime outlined in Ref.~\citep{Guarini:2022hry} are obtained for a jet with isotropic luminosity larger than the one assumed for the optically thin component in Ref.~\citep{Pitik:2021xhb}; the comparison in the bottom left panel of Fig.~\ref{fig:tot} is intended to be representative.

\begin{table*}[t]
\caption{\label{tab:table2} Characteristic parameters for each class of transients originating from the collapse of massive stars considered throughout this work.}
\begin{ruledtabular}
\begin{tabular}{lcccccr}
\textrm{Parameter}&
\textrm{SNe Ib/c}&
\textrm{SNe Ib/c BL with jet} &
\textrm{SNe IIP} &
\textrm{SNe IIn} &
\textrm{SLSNe} & 
\textrm{LFBOTs} \\
\colrule
$E_{k, \rm{ej}}$~[erg]  & $10^{51}$ & $10^{52}$ & $10^{51}$ & $10^{51}$ & $10^{52}$ & $10^{52}$ \\
$M_{\rm{ej}}$ [M$_\odot$] & 1  & 1 & 5 & 2 & 5 & $10^{-1}$ \\ 
$M_{\rm{fb}}/ t_{\rm{fb}}$ [M$_\odot$ s$^{-1}$] &  N/A & $5 \times 10^{-4}$ &  N/A &  N/A &  N/A & $1.5 \times 10^{-8}$ \\
$\epsilon_j$ & N/A & 0.01 & N/A &  N/A &  N/A & 0.01 \\
$\bar{\rho}$ [cm$^{-3}$] & N/A & 100 &  N/A &  N/A &  N/A & $10^{-7}$ \\
$P_{\rm{spin}}$ [ms] &  N/A & N/A &  N/A &  N/A & $5$ & 1 \\
$B$ [G] &   N/S & N/A &  N/A &  N/A & $10^{15}$ & $10^{15}$ \\
$f_{\rm{Ni}}$ & 0.1 & 0.15 & $10^{-3}$ &  0.01 & 0.01 & 0.01 \\
$R_\star$ [R$_\odot$] &  4 & 4 &  500 & 434 & 434 & 434 \\
$M_w$ [M$_\odot$ yr$^{-1}$] & $10^{-5}$ & $10^{-5}$ & $10^{-3}$ & $10^{-2}$ & $10^{-2}$ & $10^{-3}$ \\
$v_w$ [km s$^{-1}$]  & $1000$ & $1000$ & 15 & 100 & 100 & 1000  \\
$\epsilon_{\rm{eff}}$ & 0.1 & 0.1 & 0.1 & 0.1 & 0.1 & 0.1 \\
$\epsilon_e$ &  $10^{-1}$ & $10^{-1}$ & $10^{-1}$ & $10^{-1}$ & $10^{-1}$ & $10^{-1}$ \\
$\epsilon_B$ &   $10^{-1}$ & $10^{-2}$ & $10^{-2}$ & $10^{-2}$ & $10^{-2}$ & $10^{-2}$ \\
$\epsilon_p$ & $10^{-1}$ & $10^{-1}$ & $10^{-1}$ & $10^{-1}$ & $10^{-1}$ & $10^{-1}$ \\
\colrule
$E_{\rm{iso}, j}$ [erg] & N/A & $3.7 \times 10^{54}$ & N/A & N/A & N/A & $2.5 \times 10^{53}$ \\
$\Gamma$ & N/A & $300$ & N/A & N/A & N/A & $100$ \\
$\theta_j $ & N/A & $3^\circ$ & N/A & N/A & N/A & $6^\circ$ \\
$\epsilon_{j, d}$ & N/A & 0.2 & N/A & N/A & N/A & 0.2 \\
$\epsilon_{j, p}$ & N/A & 0.1 & N/A & N/A & N/A & 0.1 \\
\end{tabular}
\end{ruledtabular}
\end{table*}

\section{\label{sec:census}Transients from collapsing massive stars}

In this section, we present the energy radiated through the mechanisms outlined in Sec.~\ref{sec:model} across the electromagnetic wavebands as well as in neutrinos, for the transients originating from collapsing stars:  SNe Ib/c as well as SNe Ib/c broad line (BL) and GRBs, SNe IIP, SNe IIn, SLSNe, and LFBOTs. The considered transient categories together with the characteristic parameters adopted for each of the heating processes are listed in Table~\ref{tab:table2}.
While a range of parameters should be considered~\citep{Villar:2017oya}, we aim to compute ballpark figures for the source energetics  to gauge the best multi-messenger detection strategies.

One should also consider neutrinos  from the shock breakout from the progenitor star. A calculation of the neutrino signal arising from the breakout of a (quasi) spherical outflow has been attempted in Ref.~\citep{Gottlieb:2021pzr}, which concluded that other dissipation mechanisms taking place within the outflow dominate the time integrated neutrino signal. Furthermore, the photon spectrum emerging from shock breakout is highly uncertain and it is challenging to  reproduce observations. 
In the light of such uncertainties, we neglect neutrinos in the energy budget of shock breakout and leave this task to  future work.

\subsection{Supernovae of Type Ib/c, Ib/c broad line and gamma-ray bursts}
Type Ib/c SNe and GRBs are thought to be linked to massive and compact hydrogen-depleted stars, which experience reduced mass loss ($\dot{M}_w \simeq 10^{-7}$--$10^{-4} \; M_\odot$yr$^{-1}$)~\citep{Smith:2014txa,YoungSupernovaExperiment:2021fur, Jung:2021pjj}. The wind velocities are typically $v_w \simeq 10^3$~km s$^{-1}$~\citep{Margutti:2016wyh}. For  Type Ib/c SNe, $^{56}$Ni decay, CSM interactions and shock breakout of a non-relativistic outflow from a Wolf-Rayet star can contribute to heat the outflow. 

\begin{figure*}[t]
    \includegraphics[width=\columnwidth]{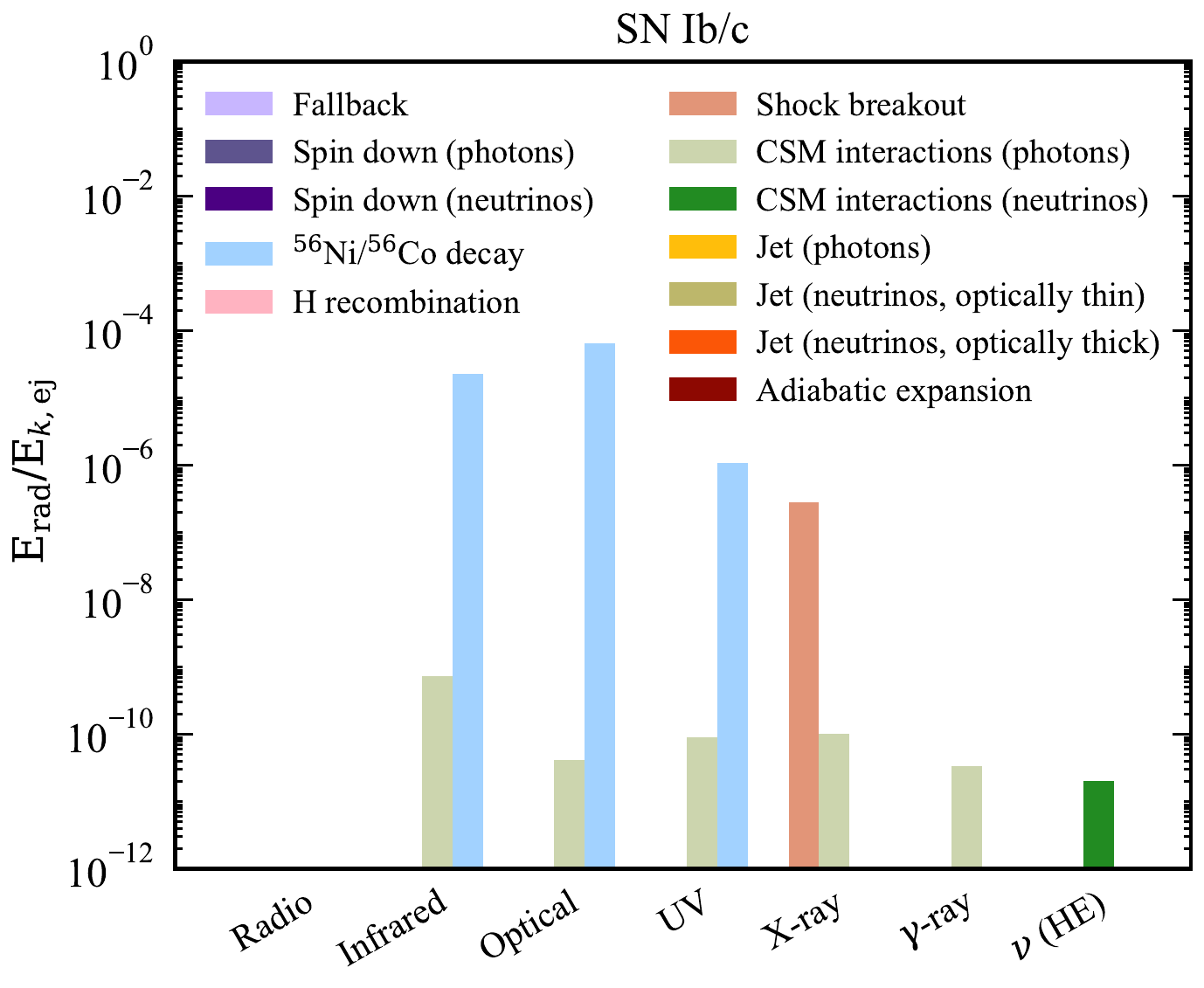}
    \includegraphics[width=\columnwidth]{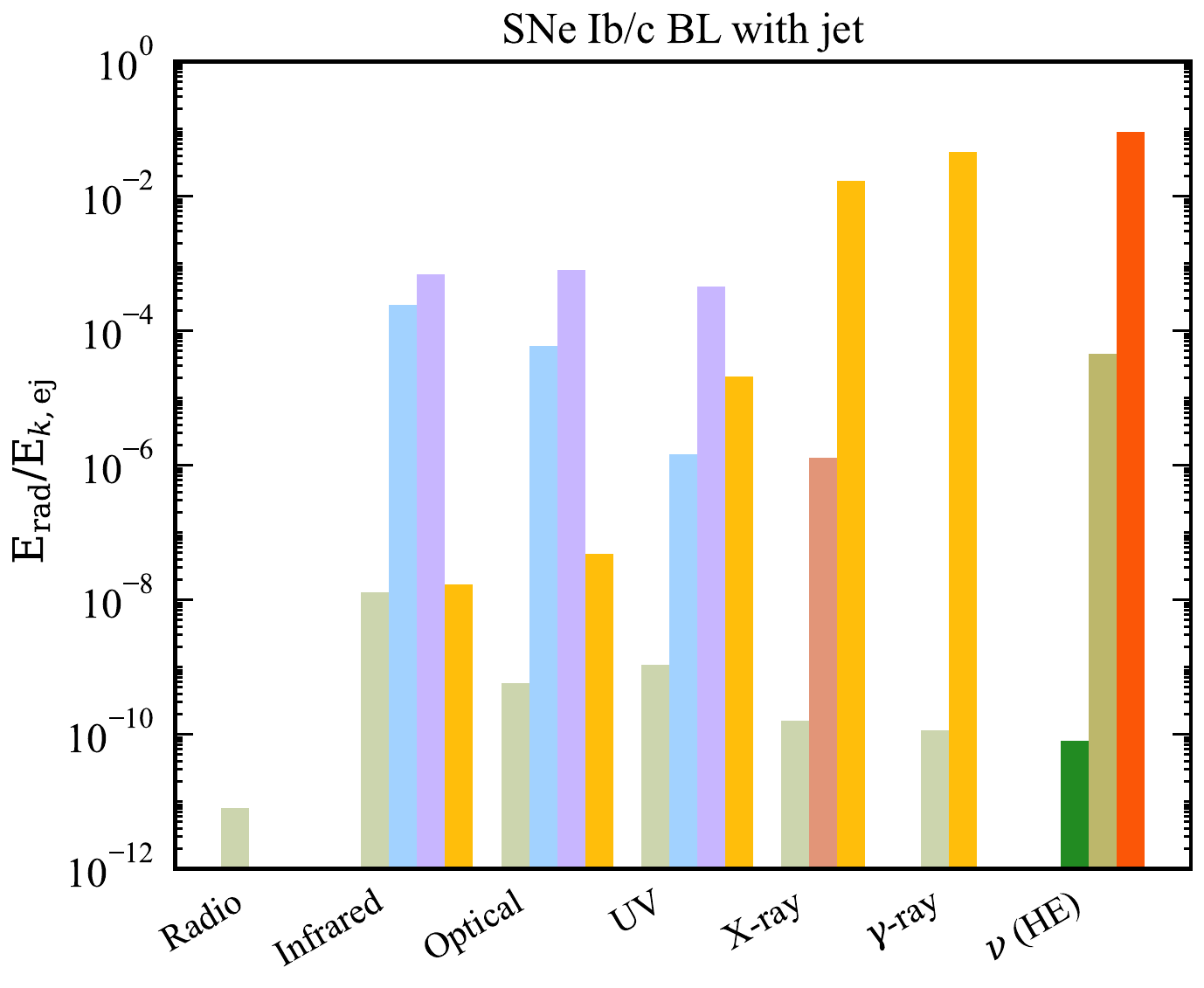}
    \includegraphics[width=\columnwidth]{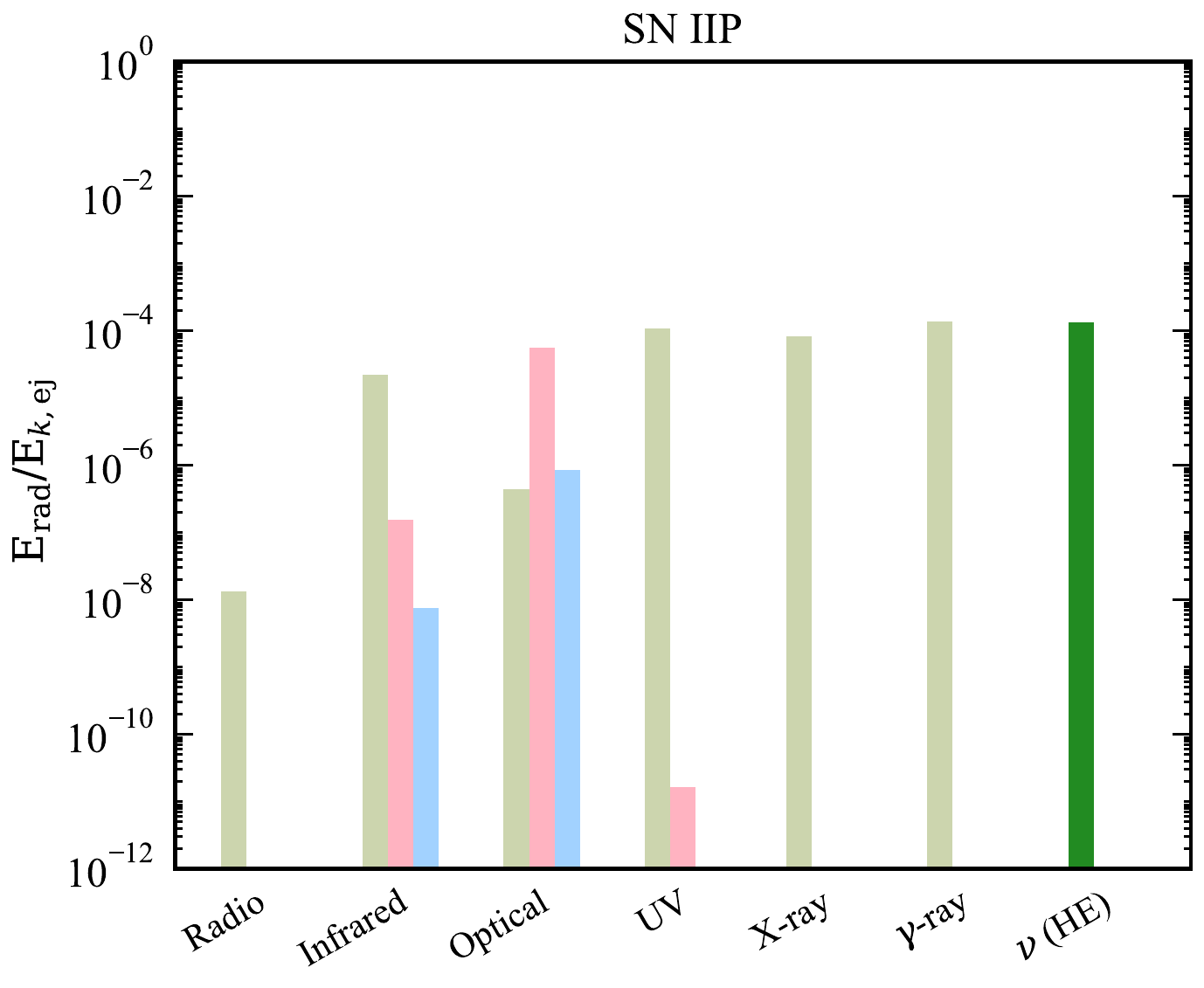}
    \includegraphics[width=\columnwidth]{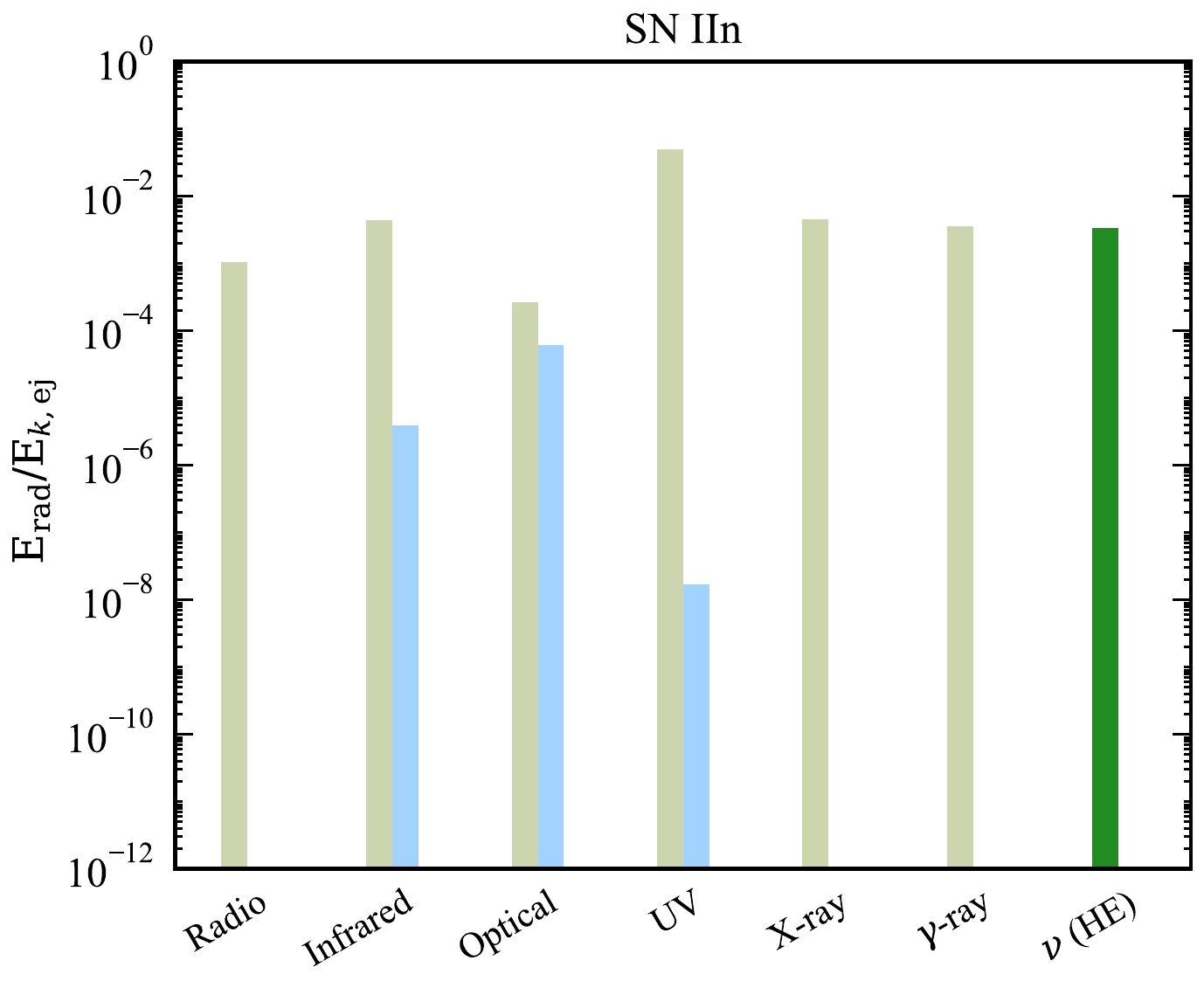}
    \includegraphics[width=\columnwidth]{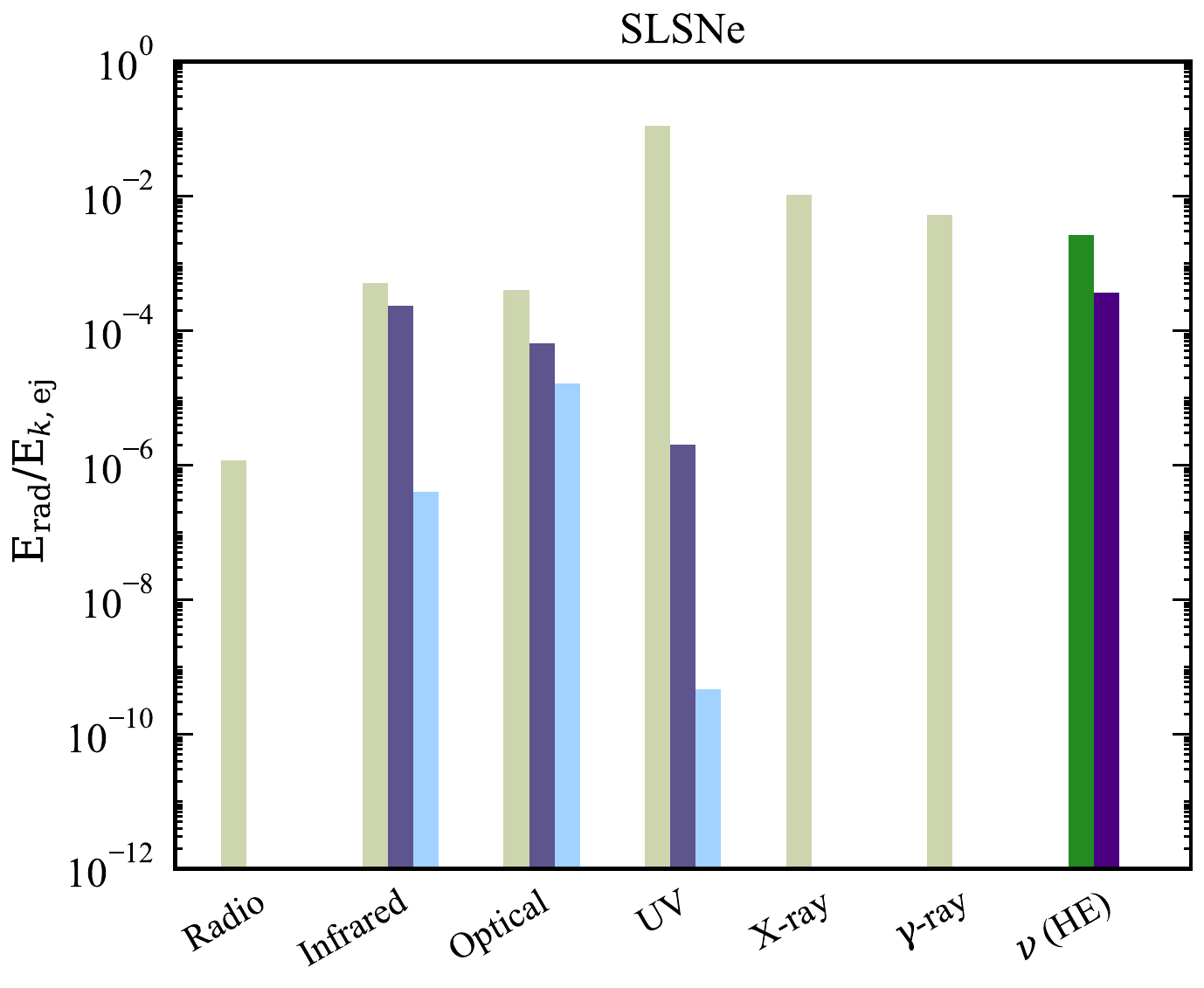}
    \includegraphics[width=\columnwidth]{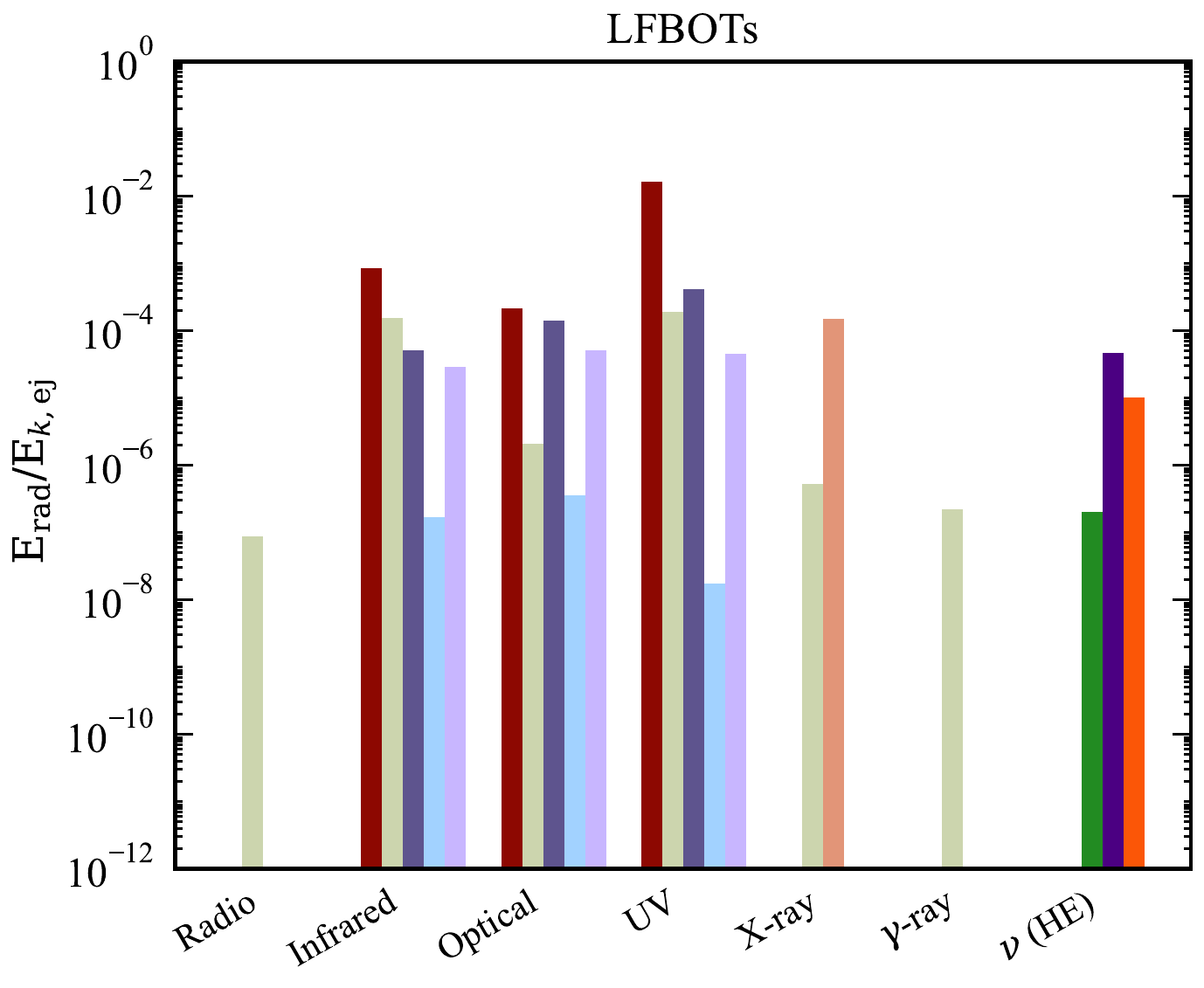}
    \caption{Ratio of the energy radiated across electromagnetic wavebands and in neutrinos  (Eq.~\ref{eq:totEn}) and the kinetic energy of the ejecta (or energy of the jet) for SNe Ib/c, SNe Ib/c BL with jet, SNe II-P, SNe IIn, SLSNe and LFBOTs (from top left to bottom right, respectively).  {The color code for each process is the same as in Fig.~\ref{fig:benchmark}}. For each transient, we assume the fiducial parameters  in Table~\ref{tab:table2}. If the transient is engine driven, then the bulk of radiation is emitted in the UVOIR band through either fallback of matter onto the BH or the magnetar spin down. In the case of successful jet (GRB), most of the  energy is emitted in the X-ray/gamma-ray bands, whereas a dimmer flash of light in the same bands resulting from shock breakout is expected for unsuccessful jets. If a dense CSM surrounds the collapsing star, then a significant fraction of energy is radiated in the UVOIR, radio and X-ray bands. In this case, also bright gamma-ray and neutrino signals are expected. Finally, when the heating source is either radioactive $^{56}$Ni decay or H recombination, the outflow radiates energy in the UVOIR band.}
    \label{fig:transients}
\end{figure*}

Figure~\ref{fig:transients} (top left panel) shows the fraction of energy radiated  across different electromagnetic wavebands and in neutrinos  for SNe Ib/c.
Radioactive decay of $^{56}$Ni is the most relevant heating source for SNe Ib/c and it radiates the bulk of energy in the UVOIR band, with $E^{\rm{UVOIR}} / E_{k,\rm{ej}} \simeq 10^{-4}$.

The forward shock mediating CSM interactions is the only site of neutrino production for SNe Ib/c, as detailed in Sec.~\ref{sec:neutrino}. 
Due to the small mass-loss rates of Wolf-Rayet stars~\citep{Ramirez-Ruiz:2004ucz}, this class of SNe is not expected to radiate a bright neutrino signal ($E^{\nu}/ E_{k,\rm{ej}} \lesssim 10^{-11}$), consistently with the findings of  Ref.~\citep{Sarmah:2022vra}. 
However, about $10 \% $ of SNe Ib/c  show signs of late time interaction with a dense CSM~\citep{Margutti:2016wyh}, starting $\gtrsim 1$~year after the explosion [SNe Ib/c late time (LT)].  
SNe Ib/c LT can release an amount of energy in neutrinos  larger than standard SNe Ib/c. An investigation of the  neutrino production due to CSM interactions for this class of SNe can be found in Refs.~\citep{Sarmah:2022vra, Sarmah:2023sds}. 

The number of SNe observed with  broad spectral features similar to the ones of SN 1998bw---dubbed SNe Ib/c broad line (BL)---is growing~\citep{Berger:2002je, Milisavljevic:2014caa, Pignata:2010ap}. Many of these SNe are not observationally linked to GRBs~\cite{Berger:2002je}, even though their ejecta move with mildly-relativistic velocity ($v_{\rm{ej}} \gtrsim 0.1 c$), hinting that the explosion mechanism may be different from the one of standard core-collapse SNe. It has been suggested that the explosion of  SNe Ib/c BL is not spherical, but  either accompanied by an off-axis GRB~\cite{Kawabata:2002yf} or a jet that barely fails to break out from the stellar mantle~\cite{Margutti:2014gha}.  
Due to the very high energies, SNe Ib/c BL and GRBs are usually modeled by considering  a spinning BH~\citep{Woosley:1993wj, Kumar:2008dr, Hayakawa:2018uxm} or a magnetar~\citep{Duncan:1992hi, Mazzali:2006tk, Metzger:2010pp}  that powers the outflow.

For the class of SNe Ib/c BL, the contribution of fallback material onto the central compact object should be included as an energy source. 
The fraction of energy radiated  across different electromagnetic wavebands and in neutrinos for SNe Ib/c BL is shown in Fig.~\ref{fig:transients} (top right panel).  
Fallback of matter on the BH constitutes the most important heating source for SNe Ib/c BL, with $E^{\rm{UVOIR}} / E_{k,\rm{ej}} \simeq 2 \times 10^{-4}$. If the central engine is not efficient then radiation is powered by $^{56}$Ni decay only.
Similarly to SNe Ib/c, CSM interactions are not an efficient neutrino production mechanism for SNe Ib/c BL ($E^{\nu}/ E_{k, \rm{ej}} \lesssim 10^{-10}$).

Assuming that SNe Ib/c BL harbor an unsuccessful jet, shock breakout of the cocoon from a Wolf-Rayet star  produces a burst of radiation in the X-ray band, with $E^{\rm{X-ray}} / E_{j} \simeq 10^{-6}$. 
A bright neutrino signal ($E^{\nu}/E_{j} \simeq 10^{-1}$) is expected only if the unsuccessful jet is magnetized and points towards Earth, as detailed in Sec.~\ref{sec:neutrino}. 
If the jet is successful, as in the case of GRBs, the bulk of energy is emitted in the X-ray/gamma-ray band [$(E^{\rm{X-ray}} + E^{\gamma-\rm{ray}})/ E_{j} \simeq 5 \times 10^{-2}$], as shown in Fig.~\ref{fig:transients} (bottom left panel). In this case, the expected neutrino ($E^{\nu} / E_{j} \simeq 5 \times 10^{-5}$) and electromagnetic signals are observable on Earth only if the jet is on-axis. 

\subsection{Supernovae of Type IIP}
Type IIP SNe  originate from red supergiants, massive stars which retain the extended hydrogen envelope. The abundance of hydrogen in their progenitor may cause the plateau of variable duration observed in the light curve of these  SNe due to hydrogen recombination~\citep{Hamuy:2002qx, Smartt:2008zd, Sanders:2014uva, Rubin:2015beb}.

Typical values for the mass-loss rates of red supergiant stars are $\dot{M}_w \simeq 10^{-6}$--$10^{-5} \; M_\odot$~yr$^{-1}$, with  wind velocity $v_w \simeq 10$~km s$^{-1}$~\cite{Margutti:2016wyh}. Nevertheless, larger CSM densities are inferred from the observation of  SNe IIP, with $\dot{M}_w \simeq \mathcal{O}(10^{-3}) \; M_\odot$~yr$^{-1}$~\citep{Nakaoka:2018nfl, Yaron:2017umb, Bullivant:2018tru}. Such large densities can be explained  invoking eruptive mass loss of the progenitor star  $\simeq \mathcal{O}(1)$~year before the SN explosion~\citep{Morozova:2016efp, Wagle:2019aqr, Nakaoka:2018nfl}. 
Besides hydrogen recombination, $^{56}$Ni decay can heat the SN outflow, together with CSM interactions. Recent work  shows that $f_{\rm{Ni}} \equiv M_{\rm{Ni}}/M_{\rm{ej}} \lesssim 0.05$~\cite{Muller:2017bdf}, thus the contribution from the radioactive decay of $^{56}$Ni is expected to be subleading.

The total energy radiated across all electromagnetic wavebands and the neutrino energy budget are shown in Fig.~\ref{fig:transients} (middle left panel) for the  parameters in Table~\ref{tab:table2}. The bulk of energy radiated in the UVOIR band is produced through CSM interactions ($E^{\rm{UVOIR}} / E_{k,\rm{ej}} \simeq 10^{-4}$) and hydrogen recombination ($E^{\rm{UVOIR}} / E_{k, \rm{ej}} \simeq 6 \times 10^{-5}$). Significant X-ray emission is also expected due to bremmsthralung as the ejecta propagate in the optically thin CSM ($E^{\rm{X-ray}}/ E_{k,\rm{ej}} \simeq 7 \times 10^{-5}$). 
These results depend on the assumption of  eruptive mass-loss episodes prior to the stellar collapse. If typical mass-loss rates of red supergiants were adopted,  the  energy in the UVOIR band would be radiated through hydrogen recombination  and the  X-ray energy would be a negligible fraction of the  explosion energy. This may be the case for most of the SNe IIP, as suggested by the lack of X-ray bright SNe IIP~\citep{Dwarkadas:2014jqa}. Due to the large CSM density, neutrinos are produced with $E^{\nu} / E_{k, \rm{ej}} \simeq 10^{-4}$.

\subsection{Supernovae of Type IIn}
Type IIn SNe show clear signs of strong CSM interactions and some of them may linked to luminous blue variable, red supergiants or yellow hypergiant stars~\cite{Smith:2016dnb, Taddia:2013nga}. The mass-loss rate of the surrounding CSM ranges between $\dot{M} = 10^{-4}$--$10 \; M_\odot$~yr$^{1}$~\citep{Smith:2014txa, Moriya:2014cua}, with wind velocity $v_w \simeq 30$--$600$~km s$^{-1}$. As a result of the dense CSM, this class of SNe exhibits signs of strong CSM interactions. 

We consider CSM interactions and $^{56}$~Ni decay as the main processes contributing to the heating of the outflow. The results are shown in Fig.~\ref{fig:transients} (middle left panel) for the  parameters in Table~\ref{tab:table2}. The bulk of energy is emitted in the UVOIR band and it is produced by CSM interactions, with $E^{\rm{UVOIR}}/ {E_{k,\rm{ej}}} \simeq 4 \times 10^{-2}$. A significant amount of energy is also emitted through non-thermal processes in the radio ${E}^{\rm{Radio}}/ {E_{k, \rm{ej}}} \simeq 10^{-3}$ and X-ray bands ${E}^{\rm{X-ray}}/ {E_{k,\rm{ej}}} \simeq 5 \times 10^{-3}$. Due to the large CSM density,  $E^{\nu} / E_{k, \rm{ej}} \simeq 3.3 \times 10^{-3}$.

\subsection{Superluminous supernovae}

SLSNe are an emerging class of SN explosions whose optical luminosity is ten or more times larger than standard core-collapse SNe~\cite{Gal-Yam:2018out}. They can be broadly classified as H-poor (Type I) and H-rich (Type II) SLSNe; the lightcurve of many H-rich SLSNe is  consistent with the interaction of the SN outflow with a dense CSM~\citep{Moriya:2018sig}, similarly to the case of SNe IIn.
The mechanism powering Type I SLSNe is not clear, even though observations suggest that these transients may be powered  by a magnetar~\citep{Kasen:2009tg, Dexter:2012xk}, which would explain the observed large kinetic energy of the outflow and radiation output~\citep{Quimby:2009ps, Gal-Yam:2012ukv}. On the contrary, Type II SLSNe exhibit signs of strong CSM interactions, like SNe IIn, and they are thought to be powered by CSM interactions~\citep{2010ApJ...709..856S}. Since hybrid mechanisms invoking  magnetar spin down, CSM interactions and $^{56}$Ni decay can also be considered  for this class of transients, we include all these  heating sources~\cite{Chen:2016rkk, Inserra:2017uyc}. 

The  energy radiated across the electromagnetic wavebands and  neutrinos  is displayed in Fig.~\ref{fig:transients} (bottom left panel) for the  parameters  in Table~\ref{tab:table2}.
Most of the energy is radiated in the UVOIR band, thanks to interactions with the CSM (${E}^{\rm{UVOIR}}/{E}_{k, \rm{ej}} \simeq 10^{-1}$) and  spin down of the magnetar (${E}^{\rm{UVOIR}}/ {E}_{k, \rm{ej}} \simeq 2.4 \times 10^{-4}$). A significant amount of energy is also emitted in  X-rays through bremmstrahlung (${E}^{\rm{X-ray}}/ {E}_{k,\rm{ej}} \simeq 10^{-2}$).
Due to the large CSM density, a bright neutrino counterpart is expected. Furthermore, neutrinos can be produced in the magnetar wind. The fraction of energy radiated in neutrinos is $E^{\nu}/ E_{k, \rm{ej}} \simeq 2 \times 10^{-3}$ ($E^{\nu}/ E_{k, \rm{ej}} \simeq 4 \times 10^{-4}$) for CSM interactions (for the magnetar wind). 

\subsection{Luminous fast blue optical transients}

Luminous FBOTs (LFBOTs, namely FBOTs with optical luminosity $L_{\rm{opt}} \gtrsim 10^{44}$~erg s$^{-1}$) are an emerging SN-like class reaching  peak luminosity in less than $10$~days~\citep{Drout:2014dma, Arcavi:2015zie, Tanaka:2016ncv, DES:2018whm}, whose observed outflow asymmetry and variability of the X-ray light curve hint towards the presence of a  compact object~\citep{Ho:2018emo, Margutti:2018rri, Gottlieb:2022old}. The latter should be responsible for the ejection of the observed asymmetric and fast outflow~\citep{Margutti:2018rri, Maund:2023jkd}. 

One of the  scenarios proposed to explain LFBOT observations invokes the collapse of a massive star, followed by the launch of a jet which inflates the cocoon~\citep{Gottlieb:2022old}. The star may not be completely depleted of hydrogen, thus the jet may fail in breaking out and be choked in the stellar mantle. This scenario would explain the lack of direct association between gamma-rays and LFBOTs~\citep{Bietenholz:2019ptf}, as well as the asymmetric outflow and the hydrogen lines observed in the spectra of some LFBOTs~\citep{Perley:2018oky, Margutti:2018rri,Coppejans:2020nxp}. 

Radio observations suggest that a fast blastwave drives the shock moving with $v_{\rm{sh}} \gtrsim 0.1 c$ in the dense CSM, extended up to $R_{\rm{CSM}} \gtrsim 10^{16}$~cm. Even though observations reveal an asymmetric CSM,  using the normalization in Eq.~\ref{eq:rhoCSM}, $M_w \simeq 10^{-4}$--$10^{-3}$ M$_\odot$ yr$^{-1}$ is inferred, for a wind velocity $v_w \simeq 1000$~km s$^{-1}$~\citep{Ho:2018emo, Margutti:2018rri, Coppejans:2020nxp}.

The energy radiated across the electromagnetic wavebands and in neutrinos is shown in Fig.~\ref{fig:transients} (bottom right panel). We rely on the benchmark parameters  in Table~\ref{tab:table2} and consider CSM interactions, $^{56}$Ni decay,  magnetar spin down, matter fallback,  and shock breakout from a massive star that is not completely hydrogen stripped star. Additionally, we consider the possibility that radiation is emitted through the adiabatic expansion of the ejecta~\citep{Gottlieb:2022old}, whose output luminosity is described by the homogeneous solution  in Eq.~\ref{eq:HS}. However, we warn the reader that the mechanism powering LFBOTs is still uncertain and that they   may not be linked to  collapsing massive stars, see e.g.~Ref.~\citep{Metzger:2022xep}. 

Following Ref.~\citep{Guarini:2022uyp}, we consider CSM interactions and a jet choked in an extended envelope surrounding the progenitor core as sites of neutrino production. We show the most optimistic scenario considered in Ref.~\citep{Guarini:2022uyp} as a representative case, however the results are model dependent.
The assumed total energy of the explosion only holds if LFBOTs originate from the core collapse of a massive star, whereas different origin (e.g.~cf.~Ref.~\citep{Metzger:2022xep}) may affect the energy budget considered in this work.

From Fig.~\ref{fig:transients}, we deduce that most of the energy is emitted in the UVOIR band, with ${E}^{\rm{UVOIR}}/ {E}_{k, \rm{ej}} \simeq 1.6 \times 10^{-2}$ (${E}^{\rm{UVOIR}}/ {E}_{k, \rm{ej}} \simeq 5 \times 10^{-4}$), through adiabatic expansion of the outflow (magnetar spin down). Radioactive decay of $^{56}$Ni does not contribute significantly to the emitted radiation in the UVOIR band~\citep{Perley:2018oky}. This is consistent with the model outlined in Ref.~\citep{Gottlieb:2021pzr}, where most of the energy is radiated through the cooling of the cocoon inflated as the jet propagates in the stellar envelope. Consistently with observations, synchrotron radiation from accelerated electrons is responsible for the observed radio emission~\citep{Margutti:2018rri, Coppejans:2020nxp, Ho:2018emo}, with ${E}^{\rm{Radio}} / E_{k, \rm{ej}} \simeq 10^{-7}$.  
Neutrinos can be produced through CSM interactions and in the magnetar wind, with  $E^{\nu} / E_{k, \rm{ej}} \simeq 2 \times 10^{-7}$ and $E^{\nu} / E_{k, \rm{ej}} \simeq 5 \times 10^{-5}$, respectively. For the assumed choked jet scenario, neutrinos are produced with $E^{\nu} / E_{k, \rm{ej}} \simeq 10^{-5}$.

\section{Connection between  electromagnetic emission and neutrinos}\label{sec:strategy}
In this section, we investigate the  correlation between electromagnetic radiation and neutrinos from transient sources resulting from massive stars. Since neutrino emission is expected for sources powered by the magnetar spin down, CSM interactions or sources harboring a jet, we focus on these scenarios. The magnetar spin down could be applied to the case of SLSNe and LFBOTs. On the other hand, SNe IIn, IIp, SLSNe as well as LFBOTs may have efficient CSM interactions. Efficient neutrino production is also expected in GRB jets and in jets smothered in an extended envelope, which may be the case for LFBOTs.

If the CSM   is not very dense, a small fraction of the ejecta kinetic energy is radiated in neutrinos and the neutrino counterpart is not bright enough to be detected. This may be the case for non-jetted SNe Ib/c or SNe IIP which do not show signs of strong CSM interactions.  Therefore,  if the observed transient is only powered  by $^{56}$Ni decay or hydrogen recombination and does not show any signs of engine or  CSM interactions,  we expect the corresponding neutrino signal to be negligible and  do not  discuss this case further.

\subsection{Magnetar spin down: Superluminous supernovae and fast blue optical transients}\label{sec:magStrategy}
A magnetar could power the emission of  SLSNe and LFBOTs (see Fig.~\ref{fig:transients}). As the spin down of the magnetar  powers bright UVOIR radiation, we can correlate the neutrino signal with the electromagnetic signal. 
\begin{figure}[t]
\centering
    \includegraphics[width=\columnwidth]{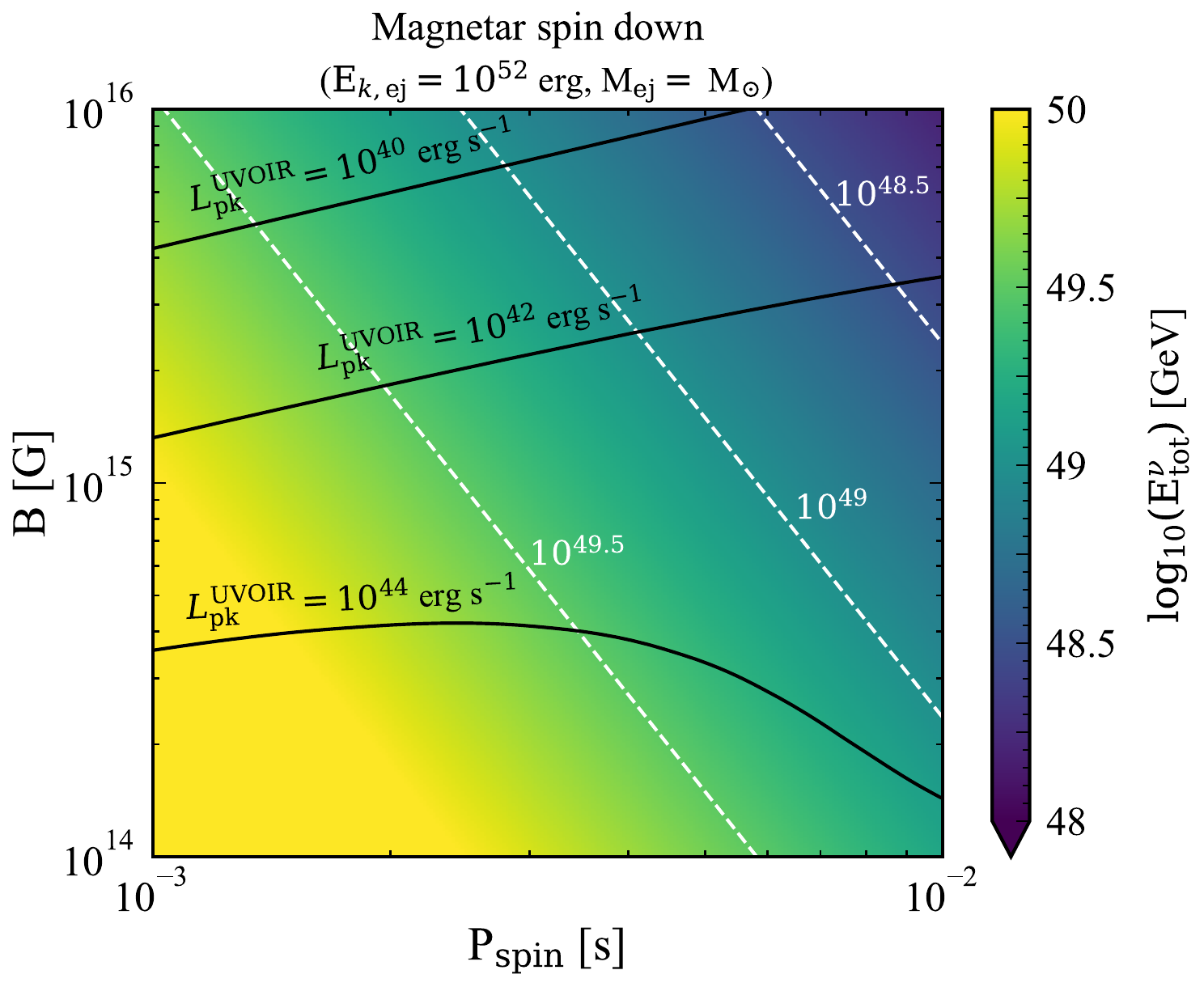}
    \caption{Isocontours of the total energy radiated in neutrinos ${E}_{\rm{tot}}^{\nu}$ (Eq.~\ref{eq:nuMax}) for transients whose UVOIR lightcurve is powered by the magnetar  spin down, in the plane spanned by the magnetar spin $P_{\rm{spin}}$ and magnetic field $B$. The  brown dashed isocontours are displayed to guide the eye. The solid black isocontours mark benchmark values for the peak of the UVOIR luminosity, which is degenerate with respect to ($P_{\rm{spin}}, B$). For a transient whose UVOIR lightcurve is powered by the magnetar spin down, the expected energy radiated in neutrinos can be inferred by  localizing the transient in this plane.}
    \label{fig:magnetarCorr}
\end{figure}    
 Figure~\ref{fig:magnetarCorr}  shows  contours for the total energy radiated in neutrinos from the magnetar wind (Eq.~\ref{eq:nuMax}), in the plane spanned by the magnetar spin $P_{\rm{spin}}$ and the magnetic field $B$. The black solid lines mark the values of the peak bolometric luminosity in the UVOIR band for each ($P_{\rm{spin}}, B$) pair. The results are shown for  ${E}_{k, \rm{ej}}=10^{52}$~erg and $M_{\rm{ej}}=M_\odot$, however Eq.~\ref{eq:nuMax} should be used for a given kinetic energy and mass of the ejecta. These parameters can be inferred from the bolometric lightcurve, which gives information on the photospheric velocity and the rise time; the former scales as $v_{\rm{ph}} \propto \sqrt{E_{k, \rm{ej}}/v_{\rm{ej}}}$, while the latter goes like $t_{\rm{rise}} \simeq t_{\rm{d}} \propto \sqrt{M_{\rm{ej}}}$. 

The peak luminosity ($L_{\rm{pk}}^{\rm{UVOIR}}$) is degenerate with respect to  the  ($P_{\rm{spin}}, B$) pairs. The only way to break this degeneracy is to complement the information from the UVOIR band with the non-thermal signal produced by the compact object  observable in the X-ray band. To a first approximation, the total energy of non-thermal photons is proportional to the magnetic field $E_{\rm{n-th}} \propto B^{-2}$, whereas it is independent on the spin $P_{\rm{spin}}$~\citep{Fang:2017tla}. Note that we have not considered the non-thermal signal in Sec.~\ref{sec:model}, as its modeling is affected by large theoretical uncertainties (see   Ref.~\citep{Fang:2017tla} for details).

The total  energy radiated in neutrinos ($E^{\nu}_{\rm{tot}}$) from a transient powered by the magnetar spin down can be obtained from Eq.~\ref{eq:nuMax}, with the characteristic parameters inferred  combining observations in the UVOIR and X-ray bands. From Fig.~\ref{fig:magnetarCorr} we conclude that sources with a bright UVOIR signal consistent with the spin down of a magnetar are expected to produce a very bright neutrino counterpart. Intriguingly, if  neutrinos should be detected  in coincidence with the UVOIR signal, the total energy emitted in neutrinos can be combined with the peak of the bolometric UVOIR lightcurve to break the degeneracy between $P_{\rm{spin}}$ and $B$, as shown in Fig.~\ref{fig:magnetarCorr}. 

Note that we consider time-integrated quantities, yet  neutrino production in the magnetar wind starts later than the UVOIR radiation, at  $t_{\nu, \rm{in}} \simeq 1.4 \times 10^{5} \eta_{-1}^{8/25} B_{14}^{-18/25} M_{\rm{ej}, -2}^{9/50} P_{\rm{spin}, -3}^{9/25} \epsilon_{B, -2}^{8/25}$~s. The neutrino flux is expected to be maximum at  $t_{\nu, \rm{max}} \simeq 9.3 \times 10^5 \eta^{1/3}_{-1} B^{-2/3}_{14} M_{\rm{ej}, -2}^{1/4} P_{\rm{spin}, -3}^{1/2} \epsilon_{B, -2}^{1/6}$~s. This time does not correspond to the peak of the UVOIR light curve, which is expected around  $10$--$100$~days~\citep{Villar:2017oya}. For example, for the benchmark transient in Fig.~\ref{fig:benchmark}, the neutrino signal peaks at $t \simeq 34$~days when  the production of thermal UVOIR radiation already stopped. Therefore, the search for neutrinos from a magnetar-powered transient should be performed for $ t_{\nu, \rm{in}} \lesssim t \lesssim t_{{\nu}, \rm{max}}$.

\subsection{Circumstellar interactions: Supernovae IIP, IIn, superluminous supernovae, and luminous fast blue optical transients}\label{sec:csmStrategy}
When the observed transient exhibits strong signs of CSM interactions in the UVOIR light curve, bright radio and X-ray counterparts are expected--- modulo  absorption processes taking place in the CSM---together with high-energy neutrinos; see also Refs.~\cite{Katz:2011zx,Murase:2010cu}. 
Here, we focus on the relation existing between the synchrotron radio and neutrino signals produced by the decelerating blastwave.
This case is of relevance for SNe IIP and IIn, SLSNe, and LFBOTs  (see Fig.~\ref{fig:transients}).

For these  transients a direct temporal correlation between the synchrotron radio and neutrino signals can be established, since both signals are produced through non-thermal processes in the proximity of the same blastwave. As the outflow propagates in the dense CSM, the forward shock converts its kinetic energy into internal energy, whose density at each time $t$ is given by Eq.~\ref{eq:enInt}. 
The energy density stored in protons is 
\begin{equation}
    u_{p}(t) \simeq E_p^{2} \frac{{\rm{d}} N_p}{{\rm{d}}E_p {\rm{d}}V} \simeq \epsilon_p \frac{u_{\rm{int}}}{\ln({E_{p, \max}/E_{p, \min}})} \; ,
    \label{eq:enPtot}
\end{equation}
where we  assume the injection spectrum given by Eq.~\ref{eq:protonInjection}.

Neutrinos are produced at the  forward shock trough $pp$ interactions (Sec.~\ref{sec:neutrino}). The neutrino energy density in the blastwave at each radius $R$ can be approximated as~\citep{Fang:2020bkm} 
\begin{equation}
    {u}_{\nu}(R) \approx \frac{1}{2} {u}_{p} \left( 1- e^{-\tau_{pp}} \right) \; ,
    \label{eq:nuCSM1}
\end{equation}
where $u_{p}$ is given in Eq.~\ref{eq:enPtot} and ${\tau}_{pp}$ is the optical depth of relativistic protons. The latter is given by $\tau_{\rm{pp}} \approx \sigma_{\rm{pp}} n_{p, \rm{CSM}} R_{\rm{sh}}$ for $E_{p} = E_{p, \max}$, while $\tau_{pp} = t_{\rm{dyn}} / t_{pp}$ for $E_{p} \ll E_{p, \max}$. Here, $E_{p, \max}$, $t_{\rm{dyn}}$ and $t_{pp}$ are the maximum energy, the dynamical and $pp$ interaction timescales of protons accelerated at the external shock, respectively; see Appendix~\ref{app:A}. Finally, the cross section for $pp$ interactions is assumed to be independent of energy ($\sigma_{pp} \simeq 5 \times 10^{-26}$~cm$^{2}$).  

The total energy emitted in neutrinos from the transient during its interaction with the CSM is
\begin{equation}
    E_{\rm{tot}}^{\nu} = \int_{R_{\rm{bo}}}^{R_{\max}} dR 4 \pi R^2 u_{\nu}(R) \; ,
    \label{eq:nuCSM}
\end{equation}
where $R_{\rm{bo}}$ is the breakout radius  (Eq.~\ref{eq:breakout}) and $R_{\max}$ is the outer edge of neutrino production region defined as indicated in Sec.~\ref{sec:neutrino}. From Eq.~\ref{eq:nuCSM}, we deduce that the total energy emitted by the blastwave in neutrinos is related to the upstream CSM density and  the blastwave velocity at the considered time. The same dependence holds for the flux radiated in the radio band, which is produced through synchrotron losses~\citep{Margalit:2021bqe}. 

While the total energy radiated in neutrinos scales with $\epsilon_p$, the radio signal strongly depends on $\epsilon_B$. Thus, the ratio $E_{\rm{tot}}^{\nu} / E^{\rm{Radio}} \propto \epsilon_p / \epsilon_B$. Typical values inferred from observations are $\epsilon_B \simeq 10^{-3}$-- $10^{-2}$~\cite{2001MNRAS.321..433B}, while the fraction of energy stored in protons accelerated at the forward shock is expected to be $\epsilon_p \lesssim 0.1$~\cite{Aharonian:2008nw, 2009Sci...325..719H}.
Therefore, when a bright radio source whose signal is consistent with synchrotron radiation is detected, its radio flux sets a lower limit on the total energy emitted in neutrinos by the expanding blastwave. 

\begin{figure}
    \centering
    \includegraphics[width=\columnwidth]{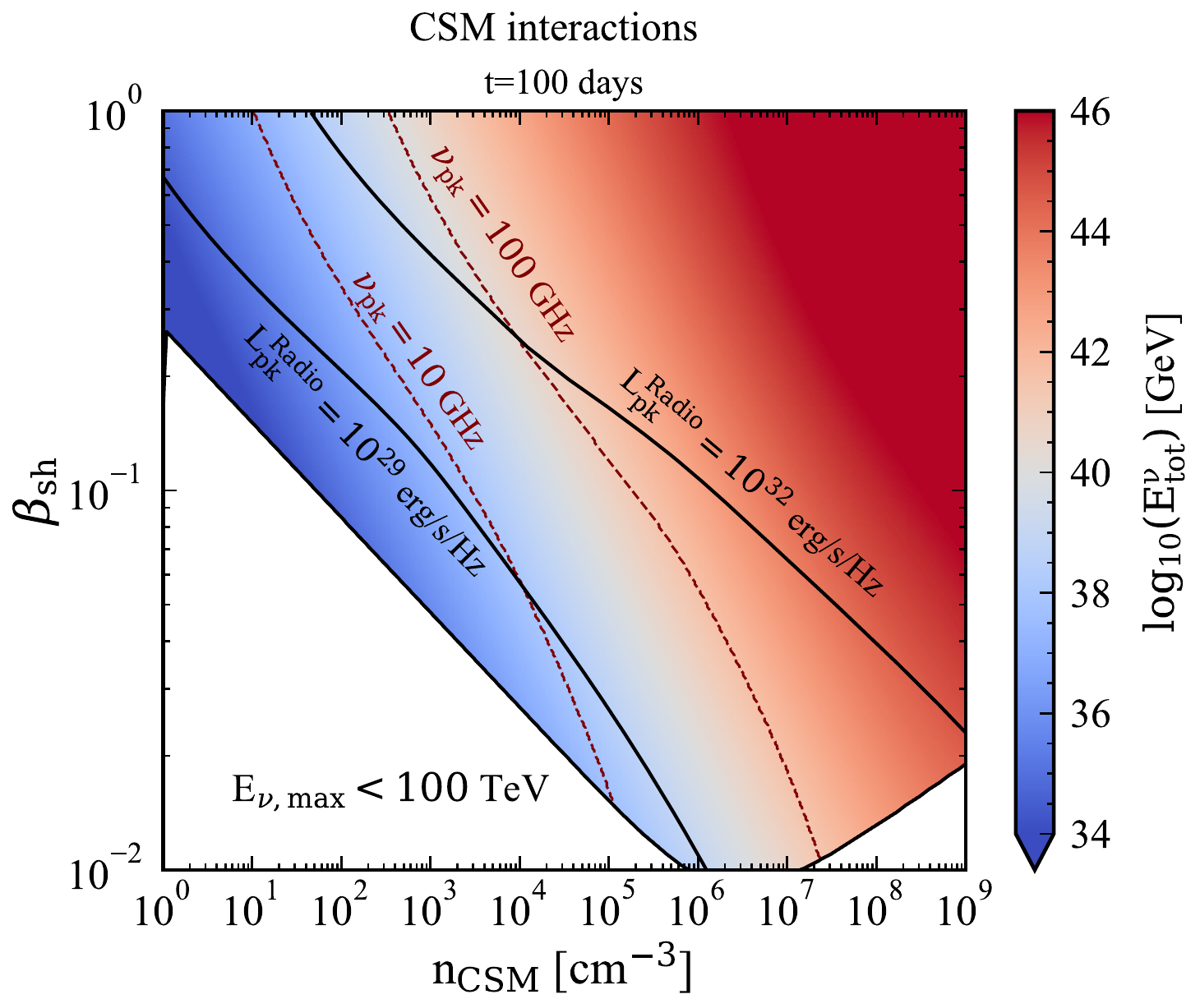}
    \caption{Isocontours of the total energy radiated in neutrinos (Eq.~\ref{eq:nuCSM}) through CSM interactions, in the plane spanned by the upstream density $n_{\rm{CSM}}$ and the shock adimensional velocity $\beta_{\rm{sh}}= v_{\rm{sh}}/c$, both measured at $t=100$~days after the explosion. The solid black (dashed purple) lines mark the peak of the radio flux $L_{\rm{pk}}^{\rm{Radio}}$ (peak frequency $\nu_{\rm{pk}}$) for  $\epsilon_B=0.01$ and $\epsilon_e =0.1$. We exclude the region of the parameter space producing neutrinos with energy $E_{\nu, \max} \lesssim 100$~TeV throughout the  duration of CSM interactions; see main text for details. When a  transient bright in radio is detected and its light curve is consistent with synchrotron radiation, the ($n_{\rm{CSM}}, \beta_{\rm{sh}}$) pair can be inferred  and the expected energy radiated in neutrinos at a fixed time can be estimated from Eq.~\ref{eq:nuCSM}.}
    \label{fig:radioNeut}
\end{figure}

Figure~\ref{fig:radioNeut} shows the contour plot of the total energy radiated in neutrinos, in the plane spanned by the upstream CSM density $n_{\rm{CSM}}$ and the blastwave dimensionless  velocity $\beta_{\rm{sh}}= v_{\rm{sh}}/c$, both measured at $t=100$~days. We use $\epsilon_p=\epsilon_e=10^{-1}$ and $\epsilon_B= 10^{-2}$ in our calculations. Radio data allow to measure the CSM density at the  time $t$, while the velocity of the fastest component of the ejecta $\beta_{\rm{sh}}$ can be inferred  from radio data of the transient~\citep{1994ApJ...420..268C, 1998ApJ...499..810C}. A transient whose radio signal is produced through interactions of the outflow with the CSM can be located in the ($n_{\rm{CSM}}, \beta_{\rm{sh}}$) plane. Once the  ($n_{\rm{CSM}}, \beta_{\rm{sh}}$) pair is fixed, the observed peak radio luminosity $L_{\rm{pk}}^{\rm{Radio}}$ and peak frequency $\nu_{\rm{pk}}$ can be obtained simultaneously only for a specific ($\epsilon_B, \epsilon_e$) pair and vice-versa~\citep{Granot:2004rh}. 

The minimum luminosity radiated  in neutrinos can be inferred from radio observation as $L_{\rm{min}}^{\nu} \simeq L_{\rm{pk}}^{\rm{Radio}} \nu_{\rm{pk}}$. 
The total energy  in neutrinos $E_{\rm{tot}}^{\nu}$ can be estimated  locating the transient in the plane  in Fig.~\ref{fig:radioNeut}. Otherwise, once the  ($n_{\rm{CSM}}, \beta_{\rm{sh}}$) pair is  inferred from radio observations, the corresponding $E_{\rm{tot}}^{\nu}$ can be estimated from Eq.~\ref{eq:nuCSM}. 

In summary,  transients detected with a bright radio counterpart are expected to produce a bright neutrino signal. As neutrinos and radio photons are produced over the same time interval during CSM interactions (see Fig.~\ref{fig:benchmark}), it is fundamental to identify radio sources at early times, in order to quickly initiate follow-up neutrino observations. However, we stress that the neutrino curve is expected to peak at a time likely shifted with respect to  the one when   the radio and optical light curves peak~\citep{Pitik:2021xhb,Pitik:2023vcg}. 
The procedure outlined here can be performed at different times of radio observations. 

We exclude in Fig.~\ref{fig:radioNeut}  the region of the parameter space leading to the production of neutrinos with maximum energy $E_{\nu, \max} \simeq 0.05 E_{p, \max} \lesssim 100$~TeV throughout the  duration of CSM interactions. In fact, the  neutrino events detected below $100$~TeV are contaminated by  the atmospheric background and  astrophysical neutrino detection would be challenging~\citep{Vitagliano:2019yzm}. 

If neutrinos are produced as a result of  CSM interactions, then a  gamma-ray counterpart should be also expected~\cite{Sarmah:2022vra,Sarmah:2023sds}. 
However, gamma-rays undergo $\gamma$-$\gamma$ and Bethe-Heitler processes before reaching Earth, making the correlation with the corresponding neutrino signal less straightforward.

\subsection{Jetted transients}\label{sec:jetStrategy}
The neutrino signal produced in the optically thin part of GRBs is strictly correlated with X-ray/gamma-ray radiation and its detectability has been extensively discussed in Ref.~\citep{Pitik:2021dyf}. We refer the reader to the criterion outlined in Ref.~\citep{Guepin:2017dfi} for the detectability of neutrinos from GRBs whether the bolometric X-ray/gamma-ray light curve is powered by internal shocks. The criterion does not hold if energy is dissipated through magnetic reconnection along the jet, and the correlation between neutrinos and photons is no longer trivial. 

When a GRB is detected electromagnetically, correlated neutrino searches should be carried out also at energies $ 10^{-1} \lesssim E_{\nu} \lesssim 10^5$~GeV, since neutrinos may be produced in this energy range in the optically thick part of the jet~\citep{Guarini:2022hry}. Subphotospheric neutrinos could be easily differentiated from the prompt signal, as the latter peaks at energies $E_{\nu} \simeq 10^5-10^6$~GeV~\citep{Pitik:2021xhb}. We note that neutrinos produced in the optically thick part of the jetted outflow do not have any direct electromagnetic counterpart, yet their detection in the direction of a GRB could be the smoking gun of the jet magnetization.

The only electromagnetic counterpart of unsuccessful jets would be the flash of light in the hard X-ray/soft gamma-ray band~\citep{Nakar:2011mq, Nakar:2015tma} due to the shock breakout of the cocoon, as discussed in Sec.~\ref{app:jet}. 
Neutrinos with energy $ 10^{-1} \lesssim E_{\nu} \lesssim 10^5$~GeV ~\citep{Guarini:2022hry} can be produced below the photosphere, if the jet is magnetized, while a neutrino signal peaking at $E_{\nu} \simeq 10^5$~GeV may exist if the jet is smothered in an extended envelope~\citep{Guarini:2022uyp}.

\section{Detection prospects}\label{sec:detections}

In this section, we explore the detection prospects of neutrinos emitted from the transients considered throughout this paper (all of them already observed electromagnetically). Finally, we discuss the best strategy for follow-up searches of single transient sources and stacking searches.

\subsection{Expected number of neutrino events}
In order to compute the expected number of neutrino events, where suitable, we consider IceCube-Gen2~\citep{IceCube-Gen2:2020qha} for representative purposes because of its large expected rate.
The  number of muon neutrino events expected at IceCube-Gen2~\citep{IceCube:2021xar, IceCube-Gen2:2020qha} for a source at redshift $z$ is $N_{\nu_\mu}(z)= \int_{E_{\nu, \rm{min}}}^{E_{\nu, \rm{max}}} dE_\nu \Phi_{\nu_{\mu}}^{\rm{obs}}(E_\nu, z) \mathcal{A}_{\rm{eff}}(E_{\nu}, \delta)$, where $\mathcal{A}_{\rm{eff}}(E_\nu, \delta)$ is the detector effective area for a source at declination $\delta$~\citep{2022ascl.soft08024V}, $E_{\nu, \min}$ and $E_{\nu, \max}$ are the minimum and maximum neutrino energy, respectively. We fix $E_{\nu, \min} = 100$~TeV, in order to avoid the  background of atmospheric neutrinos, and  choose $\delta=0^\circ$ to maximize the effective area of the detector. The observed fluence of muon neutrinos is $\Phi_{\nu_{\mu}}^{\rm{obs}}(E_\nu, z)$   [in units of GeV$^{-1}$ cm$^{-2}$], calculated as outlined in Sec.~\ref{sec:neutrino} for the model parameters in Table~\ref{tab:table2} and including neutrino flavor conversion~\citep{Anchordoqui:2013dnh,Farzan:2008eg}. 

\begin{figure}
    \centering
    \includegraphics[width=\columnwidth]{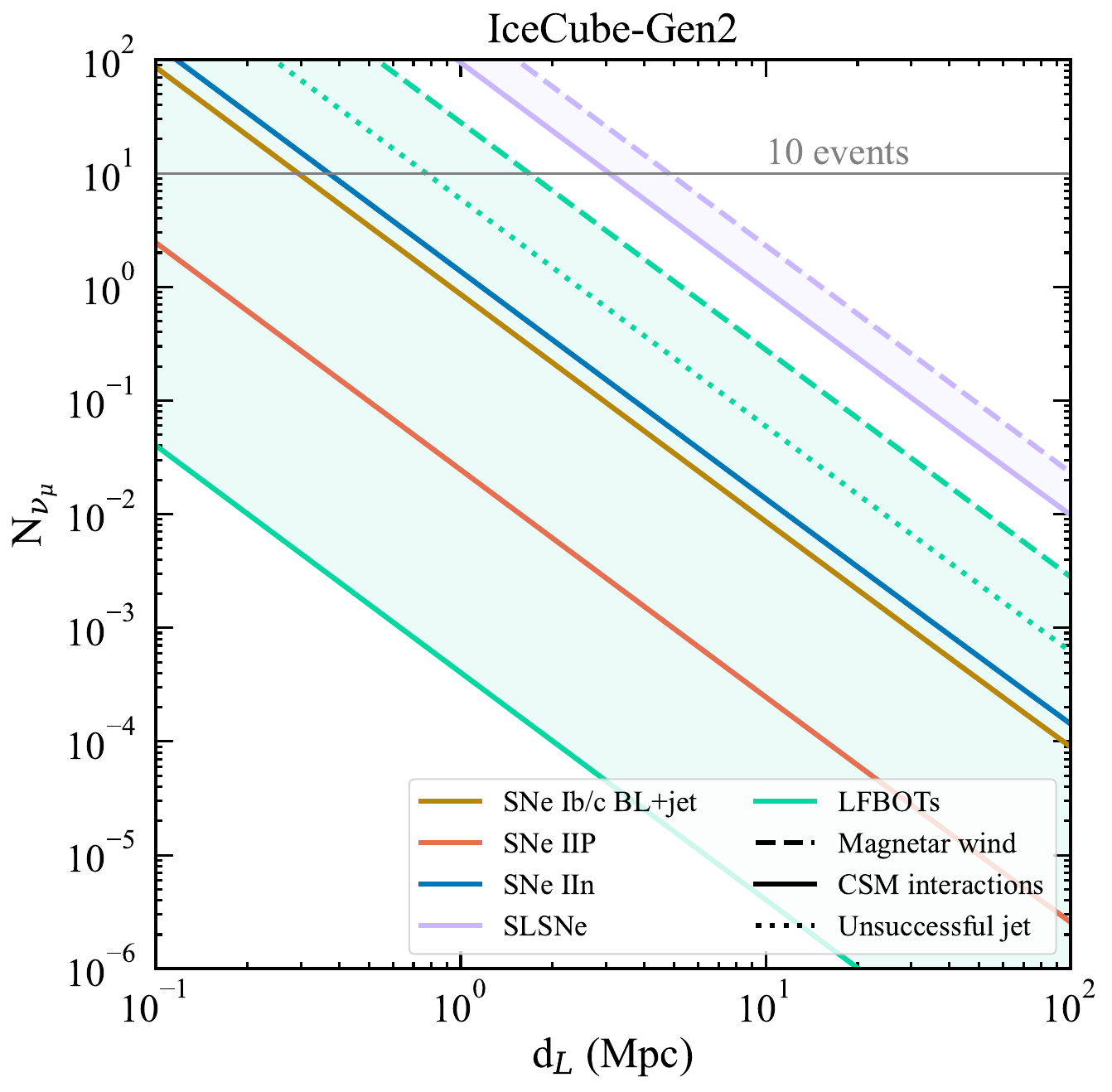}
    \caption{Expected number of muon neutrino and antineutrino events at IceCube-Gen2 as a function of the luminosity distance  for SNe Ib/c BL harboring jets, SNe IIP, SNe IIn, SLSNe and LFBOTs. The gray horizontal line marks $N_{\nu_\mu}=10$.  We consider neutrinos from CSM interactions (solid lines), from the magnetar wind (dashed lines) and from jets (dotted lines). For SNe Ib/c BL harboring a jet, we display the total number of neutrinos given by CSM interactions and the jet; CSM interactions alone produce a number of neutrinos  which falls below the plotted range.  
    CSM interactions can produce $ N_{\nu_\mu} \simeq \mathcal{O}(10)$ at IceCube-Gen2 for SLSNe (SNe IIn) located at $d_L \lesssim 4$~Mpc ($d_L \lesssim 0.6$~Mpc). The magnetar wind can produce $N_{\nu_\mu} \simeq \mathcal{O}(10)$ at IceCube-Gen2 radio for SLSNe (LFBOTs) located at $d_L \lesssim 5$~Mpc ($d_L \lesssim 2 $~Mpc).  $ N_{\nu_\mu} \simeq \mathcal{O}(10)$ is expected at IceCube-Gen2 from LFBOTs harboring unsuccessful jets and placed at $d_L \lesssim 1$~Mpc. Note that the number of neutrino events from jets, both successful and unsuccessful, is model dependent; see main text for details.}
    \label{fig:nuGen2}
\end{figure}

Figure~\ref{fig:nuGen2} shows the  number of muon neutrino events expected at IceCube-Gen2 as a function of the luminosity distance  for SNe Ib/c BL harboring a jet, SNe IIP and IIn, SLSNe, as well as LFBOTs. For all  source classes, we consider neutrino production through CSM interactions. For our fiducial parameters, CSM interactions  produce neutrinos with $E_{\nu, \max} \lesssim 10^{8}$~GeV, in agreement with  previous work~\citep{Murase:2010cu, Murase:2013kda, Cardillo:2015zda, Pitik:2021dyf, Pitik:2023vcg, Petropoulou:2017ymv, Sarmah:2022vra, Guarini:2022uyp}. 

For SLSNe and LFBOTs, we also calculate the number of neutrino events expected from the magnetar wind. 
These neutrinos have energies larger  than the ones produced through CSM interactions, with their signal expected to peak at $E_{\nu} \simeq 10^8$--$10^9$~GeV~\citep{Fang:2017tla}. In this energy range the sensitivity of the radio extension of IceCube-Gen2 is better than its optical component~\citep{IceCube-Gen2:2020qha}, thus we estimate the detection perspectives of neutrinos from the magnetar wind at IceCube-Gen2 radio.
In our simplified model, we assume that  $N_{\nu_\mu} \simeq E_{\rm{rad}}^{\nu, \max} / 10^{8.5} \; \rm{GeV} \mathcal{A}_{\rm{eff}}(10^{8.5} \; \rm{GeV})$, where $\mathcal{A}_{\rm{eff}}(10^{8.5} \; \rm{GeV})$ is the effective area of the radio extension IceCube-Gen2 at $\simeq 10^{8.5}$~GeV~\citep{2022ascl.soft08024V}. This is an  approximation due to the fact that we do not  consider the energy distribution of neutrinos from the magnetar.

As for SNe Ib/c BL harboring jets,  we show the total number of events expected at IceCube-Gen2 in Fig.~\ref{fig:nuGen2} from a successful jet, whereas the neutrino signal from CSM interactions only would be too small to be detected (see Sec.~\ref{sec:census}).
If the jet is smothered in the Wolf-Rayet star progenitor, neutrinos with  $E_{\nu} \lesssim 10^5$~GeV may be produced; the related detection prospects of subphotospheric neutrinos havew been explored in  Ref.~\citep{Guarini:2022hry}.  

As outlined in Sec.~\ref{sec:census}, LFBOTs may harbor jets which are smothered in the extended envelope surrounding the progenitor core~\citep{Gottlieb:2022old}. In this scenario, a signal peaking at  $E_{{\nu}} \simeq 10^5$~GeV may be produced in the unsuccessful jet~\citep{Guarini:2022uyp} (see also Ref.~\citep{Senno:2015tsn} for neutrino production in jets smothered in an extended envelope). In Fig.~\ref{fig:nuGen2} we show the corresponding expected number of neutrino events, obtained by relying on the most optimistic model of Ref.~\citep{Guarini:2022uyp}.

From Fig.~\ref{fig:nuGen2}, we deduce that the expected number of neutrino events from CSM interactions is $ N_{\nu_\mu} \simeq \mathcal{O}(10)$ for SLSNe (SNe IIn) located at $d_L \lesssim 4$~Mpc ($d_L \lesssim 0.6 $~Mpc). Large CSM densities may be possible around SLSNe and SNe IIn, with $M_w \lesssim 10 M_\odot$ yr$^{-1}$~\citep{Smith:2014txa, Moriya:2014cua}; in this case, the expected number of neutrino events from  SLSNe and SNe IIn could be larger than  considered here~\citep{Pitik:2023vcg}. Neutrinos from magnetar winds show promising detection perspectives at IceCube-Gen2 radio, with $ N_{\nu_{\mu}} \simeq  \mathcal{O}(10)$ for SLSNe and LFBOTs located at $d_L \lesssim 5$~Mpc and $d_L \lesssim 2$~Mpc, respectively.
Unsuccessful jets in LFBOTs may produce $ N_{\nu_{\mu}} \gtrsim  \mathcal{O}(10)$, if the source is  at $d_L \lesssim 1$~Mpc. However, we note that the neutrino signal from the choked jet peaks at energies $E_{\nu} \simeq 10^5$~GeV and it quickly drops at larger energies~\citep{Guarini:2022uyp}, where the sensitivity of IceCube-Gen2 increases~\citep{IceCube-Gen2:2020qha}. Thus, the most promising detection prospects for LFBOT sources are obtained with IceCube, due to its sensitivity range~\citep{IceCube:2021xar} (see Fig.~6 in Ref.~\citep{Guarini:2022uyp} for the expected number of neutrinos in this case). We stress that the results for both successful and smothered jets are model dependent and the number of events is calculated assuming that the jet is observed on-axis. 

In order to assess the likelihood of finding such transients and contrast the local rate with the expected number of muon neutrino events, we assume that all these sources follow the star formation rate as a function of redshift. 
The local rates of the sources considered throughout this paper ($\mathcal{R}_0$) relative to the one of core-collapse SNe---$\mathcal{R}_0^{\rm{CCSN}} = (1.02 \times 10^{-4})^{+ 70\% }_{-30 \%} $~Mpc$^{-3}$ yr$^{-1}$~\citep{Margutti:2014gha, Vitagliano:2019yzm}---are listed in Table~\ref{tab:table3}.

\begin{table}[t]
\caption{\label{tab:table3}Local rate ($\mathcal{R}_0$) of the sources considered throughout this paper including their error bands, relative to the local rate of core-collapse SNe [$\mathcal{R}_0^{\rm{CCSN}} = (1.25 \times 10^{-4})^{+ 70\% }_{-30 \%} $~Mpc$^{-3}$ yr$^{-1}$]~\citep{Yuksel:2008cu, Vitagliano:2019yzm}. Note that the local rate of GRBs refers to GRBs beamed towards us. For reference, we also show the rate of low-luminosity (LL) GRBs as they are more abundant and might also be related to choked jets and/or shock breakouts.}
\begin{ruledtabular}
\begin{tabular}{lcr}
\textrm{Source}&
\textrm{$\mathcal{R}_0/ \mathcal{R}_0^{\rm{CCSNe}}$} &
\textrm{Reference}\\
\colrule
SNe Ib/c & $(26\%)^{+5.1 \%}_{-4.8 \%}$ & \citep{Smith:2010vz} \\
SNe Ib/c BL & $\lesssim 13 \% $ & \citep{2021NewAR..9201614C} \\
SNe Ib/c BL with choked jet & Unknown & \citep{2021NewAR..9201614C} \\
GRBs & $\lesssim 10^{-5}$ & \citep{Lien:2013qja} \\
LL GRBs & $\lesssim 10^{-3}$ & \citep{Virgili:2008gp} \\
SNe IIP & $(48.2 \%)^{+5.7 \%}_{-5.6}$ & \citep{Smith:2010vz} \\
SNe IIn & $(8.8 \%)^{+3.3 \%}_{-2.9 \%}$ & \citep{Smith:2010vz} \\
SLSNe & $\lesssim 2.8 \times 10^{-3}$ & \citep{Quimby:2013jb} \\
LFBOTs & $\lesssim 10^{-3}$ & \citep{Coppejans:2020nxp}
\end{tabular}
\end{ruledtabular}
\end{table}

SLSNe and LFBOTs display the most promising chances of neutrino detections if powered by a magnetar, however these sources are the least abundant in the local universe. Using Table~\ref{tab:table3},  $ \simeq 4 \times \mathcal O(10^{-7})$~Mpc$^{-3}$ yr$^{-1}$ [ $ \simeq 2 \times \mathcal O(10^{-7})$~Mpc$^{-3}$ yr$^{-1}$ ] SLSNe (LFBOTs) are expected at $d_L = 10$~Mpc (note that we consider the central values of the rates).
On the contrary, SNe IIP are the most abundant sources locally, with $\mathcal{R}_0^{\rm{SN \; IIP}} = 1.1 \times 10^{-4}$~Mpc$^{-3}$ yr$^{-1}$. Nevertheless, their neutrino signal is too weak to be detected at IceCube-Gen2.  Jetted outflows are also expected to produce a significant number of neutrinos. Yet the probability that the jet points towards us is $\theta_j^2/4 \simeq \mathcal{O}(10^{-3})$ for typical opening angles (see Table~\ref{tab:table2}). The beaming factor and the small local rate of GRBs, LFBOTs and SNe Ib/c BL which may harbor jets (Table~\ref{tab:table3}) challenge the associated
neutrino detection.

\subsection{Combining multi-messenger signals}

On the basis of our findings, we now outline a possible strategy to carry out multi-messenger observations of transients originating from collapsing massive stars.
As outlined in Sec.~\ref{sec:csmStrategy}, radio sources whose signal is consistent with synchrotron radiation are expected to have a bright neutrino counterpart. SLSNe, SNE IIn, LFBOTs and SNe IIP with eruptive episodes fall in the category of transients with strong CSM interactions, as shown in  Figs.~\ref{fig:transients} and~\ref{fig:nuGen2}.
The synchrotron signal is the signature of a collisionless shock expanding in a dense CSM and it plays a crucial role for multi-messenger searches. 
First, as neutrinos and radio photons are produced over the same time interval from CSM interactions (Fig.~\ref{fig:benchmark}),  early detection of the radio signal will be crucial  to swiftly initiate follow-up neutrino searches. The latter can be guided by the criterion outlined in Sec.~\ref{sec:csmStrategy}. Since gamma-rays are also expected to be produced together with neutrinos~\citep{Kelner:2008ke} (see Fig.~\ref{fig:transients}), radio detection should also guide gamma-ray follow-up searches, e.g.~with Fermi-LAT~\citep{2009ApJ...697.1071A} or the Cherenkov Telescope Array (CTA)~\citep{Mazin:2019ykz}.

Sources emitting in X-rays due to bremsstrahlung emission are also hosted in a dense CSM, although this signal is produced through radiative shocks and may hint towards the existence of an asymmetric CSM~\citep{Brethauer:2022nag}. Neutrinos produced at the same site of bremmsthralung radiation have energies below the sensitivity range of IceCube and IceCube-Gen2~\citep{Fang:2020bkm} and we have not considered them throughout this work. Yet, X-ray data from bremmsthralung can be combined with synchrotron radio data to infer the CSM properties, that  affect the expected neutrino signal~\citep{Fang:2020bkm, Sarmah:2023sds,Pitik:2023vcg}. 

If the UVOIR lightcurve shows signs of central magnetar activity, as it may be the case for SLSNe and LFBOTs, X-ray telescopes should look for a non-thermal and time variable signal. The latter may emerge at later times than the UVOIR light, due to the opacity of the outflow~\citep{Margutti:2018rri}. As detailed in Sec.~\ref{sec:magStrategy}, the non-thermal X-ray signal is key to disentangle the degeneracies plaguing the UVOIR lightcurve.
Neutrino searches from this class of transients should start later than the UVOIR observations, and they should be carried out in the time window [$t_{\nu, \min}, t_{\nu, \max}$] defined in Sec.~\ref{sec:magStrategy}, e.g.~with IceCube-Gen2 radio. 

Intriguingly, SLSNe and LFBOTs may be powered either by CSM interactions or  magnetar spin down. While neutrinos from the former have energies $E_{\nu} < 10^8$~GeV, a signal peaking at $E_{\nu} \gtrsim 10^8$~GeV is expected from the latter. The time window during which neutrinos are radiated is different and it depends on the mechanism responsible for their production (see Fig.~\ref{fig:benchmark}).
Thus, the energies and the detection time of neutrinos in the direction of the transient source can be combined with electromagnetic observations to disentangle the dominant mechanism powering the lightcurve.    

Some sources, such as LFBOTs and SNe Ib/c BL, may harbor a choked jet pointing towards us. The resulting outflow has an asymmetry observable in the UVOIR and radio bands and it moves with middly-relativistic velocity, otherwise unreachable through symmetric explosions~\citep{Maeda:2023vfp}.
The electromagnetic signature of the choked jet would be a flash of light in the X-ray band~\citep{Nakar:2011mq}; see Fig.~\ref{fig:transients}. Improving X-ray detection techniques to unambiguously detect shock breakouts will be crucial to model the associated neutrino signal.

If a mildly-relativistic outflow is inferred from radio observations, one should search for neutrinos in the direction of the transient hundred to thousand seconds before and after the first observation in the UVOIR band (see also Fig.~\ref{fig:benchmark}). Indeed, if an unsuccessful jet is hidden in the source, neutrinos may be produced while the outflow is still optically thick and for a time $t_{\rm{dur}}^{\nu} \simeq t_j$.
IceCube and IceCube-Gen2 could potentially detect neutrinos from a jet smothered in a red supergiant progenitor star, whereas IceCube DeepCore~\citep{IceCube:2011ucd} is needed to observe neutrinos from a jet choked in Wolf-Rayet stars~\citep{Guarini:2022hry, Guarini:2022uyp}. A combined search may be promising  for neutrinos from mildly-relativistic sources.

Finally, if the UVOIR lightcurve should mostly exhibit signs of $^{56}$Ni decay or hydrogen recombination, the corresponding neutrino emission would be a negligible fraction [$\lesssim \mathcal{O}(10^{-13})$] of the ejecta kinetic energy. Searches of neutrinos in the direction of sources only powered through these processes would not be successful. 
\begin{table*}
\caption{\label{tab:table4} Summary of our results. We list the luminosity distance ($d_L$) where $N_{\nu_{\mu}} = 10 $ for our benchmark transients (Fig.~\ref{fig:nuGen2}), the number of transients expected per year within $d_L$ [$N_{\rm{trans}}(\leq d_L) $] and the best wavelength to correlate with neutrinos. The bands reflect the uncertainty on the local core-collapse SN rate~\citep{Mathews:2014qba} and on the fraction of SNe belonging to each class~\citep{Smith:2010vz}. Note that for SNe Ib/c BL with a choked jet we calculate the number of neutrinos expected at IceCube DeepCore~\citep{IceCube:2011ucd}, by relying on the results in Ref.~\citep{Guarini:2022hry}.}
\begin{ruledtabular}
\begin{tabular}{lcccr}
\textrm{Source}&
\textrm{Model} &
\textrm{$d_{L}$} [Mpc] &
\textrm{$N_{\rm{trans}}(\leq d_L)$ [yr$^{-1}$]} &
\textrm{Best correlated wavelength}\\
\colrule
SLSNe & CSM interactions  &4  & $\lesssim 10^{-3}$ & Radio \\
SLSNe & Magnetar wind  & 5  & $\lesssim 2 \times 10^{-3}$ & UVOIR+X-ray \\ 
SNe IIn & CSM interactions &  0.6 & $3.5 \times 10^{-3}-2 \times 10^{-2} $ & Radio \\
SNe IIP & CSM interactions & $0.05 $ & $6 \times 10^{-4}- 2 \times 10^{-3}$ & Radio \\
LFBOTs & Magnetar wind & 2 & $ \lesssim 2 \times 10^{-6}$ & UVOIR+X-ray \\
LFBOTs with jet & Choked jet in extended envelope & 1 & $ \lesssim 5 \times 10^{-7}$ & X-ray/gamma-ray \\
GRBs & Envelope of more models (Ref.~\citep{Pitik:2021dyf}) & $0.2$ & $\lesssim 2 \times 10^{-8}$ & X-ray/gamma-ray \\ 
SNe Ib/c with choked jet & Choked jet in Wolf-Rayet star & 90 & Unknown  & X-ray/gamma-ray
\end{tabular}
\end{ruledtabular}
\end{table*}
\subsection{Follow-up searches for  selected sources and stacking  searches for a source class}
The detection prospects for follow-up searches of a selected source together with the best wavelength to correlate with neutrinos for each transient are summarized in Table~\ref{tab:table4}. 
We list the luminosity distance ($d_L$) where $N_{\nu_\mu}=10$ for our benchmark transients in Fig.~\ref{fig:nuGen2}, and the number of transients expected per year within $d_L$ [$N_{\rm{trans}}(\leq d_L) $]. The bands reflect the uncertainty on the local core-collapse SN rate~\citep{Mathews:2014qba} and  the fraction of SNe belonging to each class~\citep{Smith:2010vz}. We do not include SNe Ib/c as the number of expected neutrinos from CSM interactions only is too low to be detected. For completeness, we also show the expected distance where $N_{\nu_\mu}=10$ at IceCube DeepCore~\citep{IceCube:2011ucd} for jets choked in Wolf-Rayet star progenitors, by relying on the results of Ref.~\citep{Guarini:2022hry}.

In order to carry out stacking searches of neutrinos from radio-bright transients, one can search through  archival all-sky neutrino data for  clusters of a few neutrino events  in the direction of identified radio transients. To this purpose, it would be useful to  compile  catalogues of transients detected in the radio band, e.g.~relying on data from the Very Larger Array Sky Survey (VLASS)~\citep{Lacy:2019rfe}.
Additional radio catalogues will be available in the near future, through  the Square Kilometer Array Observatory (SKA), which will cover the  Southern hemisphere~\citep{Macquart:2015uea}.
Note, however, that an appropriate weighting of the sources relative to each other is recommended in order to optimize neutrino searches~\cite{Pitik:2023vcg}. 

Another important factor in the search for neutrinos from radio sources is the time window. As extensively discussed in this work and shown in Fig.~\ref{fig:benchmark}, the neutrino and radio signals are produced over the same window. The peak of the neutrino signal is expected to occur not too far from the breakout time of the forward shock from the CSM, or anyway around the peak of the optical lightcurve~\citep{Pitik:2023vcg}. The same results do not hold for the radio signal, whose peak can occur much later than the optical one, depending on the properties of the CSM and the forward shock. Thus, it is crucial to combine UVOIR and radio data to optimize the time window for neutrino searches.

The atmospheric neutrino background increases when a long time window is chosen. Yet the criterion presented in Fig.~\ref{fig:radioNeut} excludes the parameter space contaminated by atmospheric neutrinos, considering only the  $(n_{\rm{CSM}}, \beta_{\rm{sh}})$ pairs which allow for the production of neutrinos with $E_{\nu} \gtrsim 10^5$~GeV. Our findings provide guidance to identify the ideal time window to carry out radio and neutrino stacking searches. We also encourage to initiate radio follow-up observations of neutrino alerts~\citep{IceCube:2016cqr}. 
In order to better assess the CSM properties, follow-up observations in the X-ray bands are needed to break the degeneracies in the $(n_{\rm{CSM}}, \beta_{\rm{sh}})$ space~\citep{Margalit:2021bqe}.

\section{Discussion and conclusions}\label{sec:conclusion}
In this work,  we consider SNe Ib/c, SNe Ib/c BL harboring jets, SNe II-P, SNe IIn, SLSNe, as well as LFBOTs and compute
the  energy radiated across the observable electromagnetic wavebands and  neutrinos. 
Our findings reveal that most of the energy is radiated in the UVOIR band. However, a significant fraction of the outflow kinetic energy can be emitted either in the radio or the X-ray bands through synchrotron or bremsstrahlung processes, when a dense CSM engulfs the collpasing star. Since the UVOIR light curve is degenerate with respect to the transient model parameters, a correlation of neutrino observations with this band alone is not sufficient, in agreement with the findings of Ref.~\cite{Pitik:2023vcg}. However, one could combine UVOIR observations with radio data to infer upper and lower limits, respectively, on the ejecta energy $E_{\rm{ej}}$ and mass $M_{\rm{ej}}$~\citep{Coppejans:2020nxp, Ho:2018emo, 1998ApJ...499..810C}. 

While the peak of the UVOIR luminosity of transients powered by the spin down of a magnetar is degenerate with respect to the spin period and magnetic field of the pulsar, multi-wavelength observations are fundamental to break these degeneracies. In particular, X-ray/non-thermal data can be combined with the thermal UVOIR ones to infer the spin and magnetic field~\citep{Fang:2017tla} and  allow to forecast the neutrino number of events. Neutrino observations could be instrumental to break the degeneracy between the spin period and the magnetic field. As the neutrino production starts (ends) when photopion processes become efficient (inefficient), neutrino searches should be carried out in a time window uncorrelated with the UVOIR lightcurve.

Our findings reveal that bright radio sources are promising high-energy neutrino factories. Opposite to the UVOIR signal, a  correlation between the radio and optical signals exist. Radio photons and neutrinos are produced over the same time interval and therefore neutrino searches should be performed over the  duration of the radio emission.  The radio counterpart allows to infer the CSM density $n_{\rm{CSM}}$ and the  shock velocity $\beta_{\rm{sh}}= v_{\rm{sh}}/c$ at the observed time~\citep{1998ApJ...499..810C}. The minimum neutrino luminosity expected at each emission time from the transient can be computed considering the product of the radio peak luminosity and peak frequency $E_{\rm{rad}, \min}^{\nu} \simeq L_{\rm{pk}}^{\rm{Radio}} \nu_{\rm{pk}}$, and the total energy radiated in neutrinos can be localized in the plane spanned by $n_{\rm{CSM}}$ and $\beta_{\rm{sh}}$.

For our fiducial parameters, IceCube-Gen2 will be able to detect neutrinos from SLSNe at $d_L \lesssim 4$~Mpc,  when neutrinos are produced from CSM interactions. If SLSNe (LFBOTs) harbor a central magnetar,  $10$~neutrino events produced in the magnetar wind are expected in IceCube-Gen2 radio for sources at $d_L \lesssim 5 (2)$~Mpc. 

While transients linked to massive stars are routinely detected in the UVOIR band, our findings urge to optimize the detection opportunities  in the radio and X-ray bands to swiftly identify CSM and magnetar powered transients. Furthermore, neutrino searches would be useful for mildly-relativistic transients, as neutrinos may  signal the presence  of a choked jet. Improving observational techniques in the UV/X-ray will be fundamental to detect the shock breakout light and model the corresponding neutrino signal. Neutrino searches from mildly-relativistic sources should be performed $\mathcal(10$--$1000)$~s  before and after the first UVOIR signal.  

In summary, in order to optimize the chances of joint detection of electromagnetic radiation and neutrinos from transients stemming from collapsing massive stars, follow-up programs solely based on  UVOIR observations are not optimal. UVOIR data should be  complemented by radio data tracing CSM interactions or X-ray data carrying imprints of the activity of the central engine, if any. Only exploiting multi-wavelength and neutrino data can we explore the physics powering these fascinating sources and properly guide  multi-messenger follow-up  programs.


\begin{acknowledgments}
In Copenhagen, this project has received funding from the  Villum Foundation (Project No.~37358), the Carlsberg Foundation (CF18-0183), and the Deutsche Forschungsgemeinschaft through Sonderforschungsbereich
SFB~1258 ``Neutrinos and Dark Matter in Astro- and Particle Physics'' (NDM). 
R.~M.~and the TReX team at UC Berkeley are supported in part by the National Science Foundation under Grant No.~AST-2221789 and AST-2224255, and by the Heising-Simons Foundation under grant \# 2021-3248. 
E.~R-R.~acknowledges support from the Heising-Simons Foundation. 

\end{acknowledgments}

\appendix

\section{Interaction rates of accelerated protons}\label{app:A}
In this appendix we summarize the interaction rates of  protons accelerated in the magnetar wind as well as at CSM interactions or in a jetted outflow.

\subsection{Magnetar wind}
The energy deposited by the spin down is partially deposited into kinetic energy of the outflow, with the remaining energy being used to produce thermal and non-thermal radiation. Therefore, protons accelerated in the magnetar wind  interact through $p \gamma$ interactions both with thermal and non-thermal photons in the wind nebula. 
The corresponding interaction rates are~\citep{Fang:2017tla}:
\begin{eqnarray}
    t^{-1}_{p \gamma, \rm{th}} & = & 7.7 \times 10^{-5} t_{5.5}^{-15/8} B_{14}^{-3/4} \beta_{w}^{-15/8} \; \rm{s} \; , \\ 
    t^{-1}_{p \gamma, \rm{nth}} & = & 2.4 \times 10^{-5} t_{5.5}^{-27/8} B_{14}^{-7/4} \beta_w^{-19/8} \; \rm{s} \; ,  
\end{eqnarray} 
where $\beta_w \simeq 1 M_{\rm{ej}, -2}^{-1/2} P_{\rm{spin}, -3}^{-1}$ is the wind velocity after its acceleration, at times $t \gg t_{\rm{sd}}$, where $t_{\rm{sd}}$ is the spin-down time defined as in Eq.~\ref{eq:sdTime}. 

The interaction rate for $pp$ interactions is
\begin{equation}
    t_{pp}^{-1}=6.25 \times 10^{-9} M_{\rm{ej}, -2}^{-1} t_{5.5}^3 \beta_w^3 \; \rm{s} \; .
\end{equation}

Pions are  created in the wind nebula at a rate
\begin{equation}
    t^{-1}_{\pi, \rm{cre}}= t^{-1}_{p \gamma, \rm{th}} + t^{-1}_{p\gamma, \rm{nth}} + t^{-1}_{pp} \; .
\end{equation}

The only proton cooling process competing with pion production is the synchrotron cooling, whose characteristic time is
\begin{equation}
    t_{p, \rm{rad}}= 5.6 \times 10^{-6} \eta^{-1}_{-1} t_{5.5}^5 B^3_{14} \beta_w^3 \epsilon_{B, -2}^{-1} \; \rm{s} .
\end{equation}
Thus, pion creation in the wind nebula is suppressed by a factor
\begin{equation}
    f^{p}_{\rm{sup}}= \min \left(1, \frac{t_{\rm{ad}}}{t_{\pi, \rm{cre}}},\frac{t_{p, \rm{rad}}}{t_{\pi, \rm{cre}}} \right) \,
\end{equation}
where $t_{\rm{ad}} \simeq R / \beta_w c$ is the adiabatic expansion time scale of the wind nebula.

The onset of neutrino production corresponds to the time when efficient photopion production starts, namely when $t_{p, \rm{rad}}^{-1} \equiv t_{\pi, \rm{cre}}^{-1}$. Similarly, neutrino production ends at the time when photopion processes are no longer efficient, i.e. $t_{\pi, \rm{cre}}^{-1} \equiv t_{\rm{cross}}^{-1}$.

Before decaying, secondary pions and muons also cool. Their cooling affects the neutrino signal through  the following suppression factors: 
\begin{eqnarray}
    f^\pi_{\rm{sup}} & = & 0.3 \eta_{-1}^{-2} B_{14}^4 \epsilon_{B, -2}^{-1} t_{5.5}^6 \; , \\ 
    f^\mu_{\rm{sup}} & = & 1.5 \times 10^{-3} \eta_{-1}^{-2} B_{14}^4 \beta_w^3 \epsilon_{B, -2}^{-1} t_{5.5}^6 \; .
\end{eqnarray}

\subsection{CSM interactions and jets}
In the case of non-relativistic and mildly relativistic shocks---such as the external shock driven by the outflow as it expands in the CSM---the proton acceleration rate is obtained from the Bohm limit~\citep{protheroe_clay_2004}
\begin{equation}
 t^{\prime -1}_{\rm{acc}} \simeq \frac{3 e B v_{\rm{sh}}^{2}}{20 \gamma_p m_{p} c^3 } \; ,
 \end{equation}
where $e = \sqrt{\hbar \alpha c}$ is the elementary electric charge, $\alpha = 1/137$ is the fine structure constant and $\hbar$ is the reduced Planck constant; for non-relativistic shocks, the comoving frame of the outflow and the center of explosion frame coincide---we  carry out the calculations in the comoving frame of the outflow and we denote quantities with $X^\prime$. The magnetic field $B$ is defined as in Sec.~\ref{sec:model} and $\gamma_p$ is the proton Lorentz factor. 

In the case of relativistic and mildly relativistic outflows, the proton acceleration rate is 
\begin{equation}
t^{\prime -1}_{\rm{acc}} = \frac{c e B^{\prime}}{\xi E^{\prime}_p} \; ,
\end{equation}
where $\xi$ represents the number of gyroradii needed for accelerating protons. We assume~$\xi = 10$~\citep{Gao:2012ay}. Finally, $B^\prime$ is the magnetic field along the jet, which  depends on the energy dissipation mechanism~\citep{Pitik:2021xhb}.

The total cooling rate of accelerated protons is
\begin{equation}
t^{^\prime -1}_{p, \rm{cool}} = t^{^\prime -1}_{\rm{ad}} + t^{^\prime -1}_{p, \rm{sync}} +t^{^\prime -1}_{p \rm{\gamma}} + t^{^\prime -1}_{pp} + t^{^\prime -1}_{p, \rm{BH}} + t^{^\prime -1}_{p, \rm{IC}}\ ,
\label{eq:total_cooling}
\end{equation}
where $t^{^\prime -1}_{\rm{ad}}$, $t^{^\prime -1}_{p, \rm{sync}}$, $t^{^\prime -1}_{p \gamma}$, $t^{^\prime -1}_{{pp}}$, $t^{^\prime -1}_{p, \rm{BH}}$, $t^{^\prime -1}_{p, \rm{IC}}$ are the adiabatic, synchrotron, photo-hadronic ($p \gamma$), hadronic ($pp$), Bethe-Heitler (BH, $p \gamma \rightarrow p e^+ e^-$) and inverse Compton (IC) cooling rates, respectively,  defined as follows~\citep{dermer_book, Gao:2012ay, Razzaque:2005bh}: 
\begin{eqnarray}
 t^{\prime -1}_{\rm{ad}} &=& \frac{v}{R}\ , \label{eq:adiabatic_time} \\
 t^{\prime -1}_{p, \rm{sync}} &=& \frac{4 \sigma_T m_e^2 E^\prime_p B^{\prime 2}}{3 m_p^4 c^3 8 \pi}\ , \\
 t^{\prime -1}_{p \gamma} &=& \frac{c}{2 \gamma^{\prime 2}_p} \int_{E_{\rm{th}}}^\infty dE^\prime_\gamma \frac{n^\prime_{\gamma}(E^\prime_\gamma)}{E^{\prime 2}_\gamma} \\ \nonumber & & \; \times \int_{E_{\rm{th}}}^{2 \gamma^\prime_p E^\prime_\gamma} dE_r E_r  \times \sigma_{p \gamma}(E_r) K_{p \gamma}(E_r)\ ,  \\
 t^{\prime -1}_{{pp}} &=& c n^\prime_{p} \sigma_{pp} K_{pp}\ ,  \\
 t^{\prime -1}_{p, \rm{BH}} &=& \frac{7 m_e \alpha \sigma_T c}{9 \sqrt{2} \pi m_p \gamma^{\prime 2}_p} \int_{\gamma_p^{\prime -1}}^{\frac{E^\prime_{\gamma, \rm{max}}}{m_e c^2}} d\epsilon^\prime \frac{n^\prime_{\gamma} (\epsilon^\prime)}{\epsilon^{^\prime 2}} \\ \nonumber & & \; \times \biggl\{ (2 \gamma^\prime_p \epsilon^\prime)^{3/2} \biggl[\ln(\gamma^\prime_p \epsilon^\prime) -\frac{2}{3} \biggr]+ \frac{2^{5/2}}{3} \biggr\}\ ,  \\
 t^{\prime -1}_{p, \rm{IC}} &=& \frac{3 (m_e c^2)^2 \sigma_T c}{16 \gamma_p^{\prime 2}( \gamma^\prime_p-1) \beta^\prime_p} \int_{E^\prime_{\gamma, \rm{min}}}^{E^\prime_{\gamma, \rm{max}}} \frac{dE^\prime_\gamma}{E_\gamma^{^\prime 2}} \\ \nonumber & & \; \times F(E^\prime_\gamma, \gamma^\prime_p) n^\prime_{\gamma}(E^\prime_\gamma)\ , 
\end{eqnarray}
where $v= 2 c \Gamma$ for the jetted outflow and $v = v_{\rm{sh}}$ for CSM interactions, $\gamma_p = E^\prime_p/m_p c^2$ is the proton Lorentz factor, $\epsilon^\prime = E^\prime_\gamma /m_e c^2$, $E_{\rm{th}}=0.150$~GeV is the energy threshold for photo-pion production, and $\beta^\prime_p \approx 1$ for relativistic particles. The function $F(E^\prime_\gamma, \gamma^\prime_p)$ is defined in Ref.~\cite{PhysRev.137.B1306},  with the replacement $m_e \rightarrow m_p$. The cross sections for $p \gamma$ and $pp$ interactions, $\sigma_{p \gamma}$ and $\sigma_{pp}$ are taken from Ref.~\cite{ParticleDataGroup:2020ssz}.
The function $K_{p\gamma}(E_r)$ is the inelasticity of $p\gamma$ interactions defined in Eq.~9.9 of~\cite{dermer_book}:
\begin{equation}
K_{p\gamma}(E_r) = 
\begin{system}
0.2 \; \; \; \; \; \;  \; \; \;  E_{\rm{th}} < E_r < 1~\rm{GeV}\\
0.6 \; \; \; \; \; \;  \; \; \;  E_r > 1~\rm{GeV}\ ,
\end{system} \
\end{equation}
with $E_r = \gamma^\prime_p E^\prime_\gamma (1 - \beta^\prime_p \cos\theta^\prime)$ being the relative energy between a proton with Lorentz factor $\gamma^\prime_p$ and a photon with energy $E^\prime_\gamma$, which move in the comoving frame of the interaction region along directions which form an angle $\theta^\prime$. The comoving proton density is $n^\prime_{p} = {E}_{\rm{iso}}/(4 \pi R_j^2 c {t}_j \Gamma^2)$ for the jetted outflow, and $n^\prime_{p} = 4 n_{p, \rm{CSM}} m_p c^2$ for CSM interaction. The inelasticity of  $pp$ interactions is $K_{pp} = 0.5$ and $n^\prime_{\gamma}(E^\prime_\gamma)$ is the photon target for accelerated protons, defined in the main text for the jetted and spherical outflow. Before decaying, also secondary pions and muons cool through synchrotron, adiabatic and hadronic energy losses. Their cooling rates are defined as for protons, but replacing $m_p \rightarrow m_x$, with $m_x$ being the mass of the $x$ secondary particle. 

\section{Radiative shocks}\label{app:B}
When the gas behind the shock immediately cools  in a thin shell, the shocks are radiative. This happens when the CSM density is very large~\citep{Fang:2020bkm, Margalit:2021bqe}.
Also in the radiative regime the dynamical timescale for the forward shock driven by the outflow in the optically thin CSM is $t_{\rm{dyn}} = R/ v_{\rm{sh}}$.

When the forward shock breaks out from the dense CSM shell, bremmstrahlung becomes the leading mechanisms for photon production and electrons mainly cool through free-free emission. The timescale for this process reads
\begin{equation}
t_{\rm{ff}} = \frac{3 n_e k_B T_e}{ 2 \Lambda_{\rm{ff}}(n_e, T_e) } \; ,
\end{equation}
where $\Lambda_{\rm{ff}}$ is the free-free cooling rate~\citep{2011piim.book.....D}, $k_B T_e = 3/16 \mu_p m_p v_{\rm{sh}}^2$ is the post-shock temperature of the gas, with $\mu_p \simeq 0.62$ being the mean molecular weight for a fully-ionized gas. 
The post-shock electron density is $n_e = 4 \rho_{\rm{CSM}} m_p / \mu_e$  and $\mu_e \simeq 1.18$.
Finally, $L_{\rm{sh}}= 2 \pi R_{\rm{CSM}}^2 \rho_{\rm{CSM}}(R_{\rm{CSM}}) v_{\rm{sh}}^3$ is the total kinetic shock power computed at the CSM edge.

\bibliography{main.bib} 

\end{document}